\documentclass{iopart}
\usepackage{graphicx}
\eqnobysec
\graphicspath{{pictures/caseNVN/}{pictures/caseNVN/10/}{pictures/caseNVN/20/}{pictures/caseNVN/01/}
{pictures/caseNVN/02/}
{pictures/caseNVN/11/}{pictures/caseVN/}{pictures/caseVN/10/}{pictures/caseVN/20/}{pictures/caseVN/01/}
{pictures/caseVN/02/}{pictures/caseVN/11/}}
\begin{document}

\title[New exact  solutions  of the NVN  nonlinear equation via  $\bar{\partial}$-dressing method]{New exact
solutions with constant asymptotic values at infinity  of the  NVN integrable   nonlinear evolution equation  via $\bar{\partial}$-dressing method}

\author{M.Yu. Basalaev, V.G. Dubrovsky and A.V. Topovsky}
\address{Novosibirsk State Technical University, Karl Marx prosp. 20, Novosibirsk 630092, Russia}
\eads{\mailto{dubrovsky@academ.org}}

\begin{abstract}
The classes of exact multi line soliton, periodic solutions and solutions with functional parameters, with constant asymptotic values at infinity $u|_{\xi^2+\eta^2\rightarrow \infty}\rightarrow -\epsilon$,  for the hyperbolic and elliptic versions of  the Nizhnik-Veselov-Novikov (NVN) equation via $\bar{\partial}$-dressing method of Zakharov and
Manakov were constructed.

At fixed time these solutions are exactly solvable potentials correspondingly for  one-dimensional perturbed telegraph and  two-dimensional stationary
Schr\"{o}dinger equations. Physical meaning of stationary states of quantum particle in exact one line and two line soliton potential valleys was discussed.

In the limit $\epsilon\rightarrow 0$ exact special solutions $u^{(1)}$, $u^{(2)}$ (line solitons and periodic solutions) were found which sum $u^{(1)}+u^{(2)}$ (linear superposition) is also exact solution of NVN equation.
\end{abstract}


\section{Introduction}
\label{Section_1}
\setcounter{equation}{0}

Exact solutions of differential equations of mathematical physics, linear and nonlinear, are very
important for the understanding
of various physical phenomena. In the last three decades the Inverse Scattering Transform (IST)
method has been generalized and successfully applied to several two-dimensional nonlinear
evolution equations such as Kadomtsev-Petviashvili, Davey-Stewartson, Nizhnik-Veselov-Novikov,
Zakharov-Manakov system, Ishimori, two-dimensional integrable Sin-Gordon and others (see books~
\cite{NovZakh_BookSoliton}-\cite{KonopBook_93}
and references therein).

The extension  of nonlocal Riemann-Hilbert problem by Zakharov and Manakov~\cite{Manakov_81} and   $\bar{\partial}$-problem approach~\cite{BealsCoifman}  led to the discovery of more general $\bar{\partial}$-dressing method~\cite{ZakhManak_85}-\cite{FokasZakh_92} which became very powerful method for solving two-dimensional integrable nonlinear evolution equations.
In the present paper the $\bar{\partial}$-dressing method of Zakharov and Manakov was used for
the construction of the classes of exact
 multisoliton and periodic solutions of the famous (2+1)-dimensional
Nizhnik-Veselov-Novikov (NVN) integrable equation 
%
%
\begin{equation}\label{NVN}
 \fl u_{t} + \kappa_1 u_{\xi\xi\xi} + \kappa_2 u_{\eta\eta\eta} +
  3\kappa_1(u\partial^{-1}_{\eta}u_{\xi})_{\xi} +
  3\kappa_2(u\partial^{-1}_{\xi}u_{\eta})_{\eta}=0,
\end{equation}
where $u(\xi,\eta,t)$ is scalar function, $\kappa_1,\,\kappa_2$ are arbitrary constants,
$\xi=x+\sigma y,\,\eta=x-\sigma y$, and $\sigma^{2}=\pm 1$;
$\partial_{\xi}\equiv\frac{\partial}{\partial
\xi},\,\partial_{\eta}\equiv\frac{\partial}{\partial\eta}$ and
$\partial^{-1}_{\xi}$,  $\partial^{-1}_{\eta}$ are operators inverse to
$\partial_{\xi}$ and $\partial_{\eta}$:
$\partial^{-1}_{\eta}\partial_{\eta}=\partial^{-1}_{\xi}\partial_{\xi}=1$.
Equation (\ref{NVN}) was first introduced and studied by Nizhnik~ \cite{Nizhnik_80} for hyperbolic version (NVN-II equation) with $\sigma=1$ and independently by Veselov and Novikov~\cite{VeselovNovikov_84} for elliptic version (NVN-I equation) with $\sigma=i$,
$\kappa_1=\overline\kappa_2=\kappa$. The NVN equation is integrable by the IST due to representation of it as the compatibility condition for two linear auxiliary problems \cite{Nizhnik_80},\cite{VeselovNovikov_84}:
\begin{equation}\label{NVN dressing auxiliary problems}
\fl L_{1}\psi = \big(\partial_{\xi\eta}^{2}+u\big)\psi=0,
\end{equation}
\begin{equation}\label{NVN dressing auxiliary problems2}
\fl      L_{2}\psi = \big(\partial_{t}+\kappa_{1}\partial_{\xi}^{3}+\kappa_{2}\partial_{\eta}^{3}+
      3\kappa_{1}\big(\partial_{\eta}^{-1}u_{\xi}\big)+
      3\kappa_{2}\big(\partial_{\xi}^{-1}u_{\eta}\big)\big)\psi=0
   \end{equation}
in the form of the Manakov's triad
\begin{equation}\label{NVN dressing triada Manakova}
 \fl [L_{1}, L_{2}] = BL_{1},\quad
  B =
  3\big(\kappa_{1}\partial_{\eta}^{-1}u_{\xi\xi}+
  \kappa_{2}\partial_{\xi}^{-1}u_{\eta\eta}\big).
\end{equation}

The present paper is the  continuation of Dubrovsky et al   work  and
follows the notations, review of the subject and general considerations presented in  the previous papers
\cite{DubrovskyKonopelchenko_93}-\cite{DubrForm_03}.
We apply the $\bar{\partial}$-dressing method of Zakharov and Manakov for the construction of
classes of exact solutions with non-zero constant asymptotic values at infinity:
\begin{equation}\label{NVN dressing asymptotic value of u}
  u(\xi,\eta,t)=\tilde{u}(\xi,\eta,t)+u_\infty=\tilde{u}(\xi,\eta,t)-\epsilon,
\end{equation}
where $\tilde{u}(\xi,\eta,t)\rightarrow 0$ as $\xi^{2}+\eta^{2}\rightarrow\infty$. In this case
the first linear auxiliary problem in (\ref{NVN dressing auxiliary problems}) has the form:
\begin{equation}\label{The_first_aux_probl}
  \big(\partial_{\xi\eta}^{2}+\tilde{u}\big)\psi=\epsilon \psi.
\end{equation}
For  $\sigma=1$ with real space variables $\xi\Rightarrow t-x ,\,\eta \Rightarrow t+y$
equation (\ref{The_first_aux_probl}) can be interpreted as perturbed telegraph
equation with potential $u=\tilde u-\epsilon$ or perturbed string equation for $\epsilon=0$.
For $\sigma=i$ with complex space variables $\xi\Rightarrow x+iy=z
,\,\eta \Rightarrow x-iy=\overline{z}$ equation (\ref{The_first_aux_probl}) coincides with the
famous two-dimensional stationary Schr\"{o}dinger equation
\begin{equation}\label{2DSchr}
\big(-2\partial_{z\bar{z}}^{2}+V_{Schr}\big)\psi=E\psi
\end{equation}
with $V_{Schr}=-2\tilde{u}$ and $E=-2\epsilon$.
For this reason the construction via $\bar{\partial}$-dressing method of exact solutions of the NVN equations with constant
asymptotic values at infinity means simultaneous  calculation of exact eigenfunctions (wave functions) $\psi$ and exactly solvable potentials $u=\tilde u-\epsilon$ and  $V_{Schr}=-2\tilde{u}$ for above mentioned famous linear equations.

The
inverse scattering transform for the first auxiliary linear problem (\ref{The_first_aux_probl}) (or in particular for 2D Schr\"{o}dinger equation (\ref{2DSchr})) has been developed in a number of papers. Detailed review one can find in the book of Konopelchenko \cite{KonopBook_92}. On the basis of developed for (\ref{The_first_aux_probl}) IST using time evolution given by second auxiliary problem (\ref{NVN dressing auxiliary problems2})
several classes of exact solutions of NVN equation were constructed  \cite{KonopBook_92},
\cite{KonopBook_93},\cite{Nizhnik_80}-\cite{AthorneNimmo_91}.
Some exact solutions of NVN-II equation with $\sigma=1$   were obtained in the work \cite{Nizhnik_80} via the transformation operators.
Veselov et al   constructed finite zone solutions of NVN equation \cite{VeselovNovikov_84}. The classes of rational localized solutions of so called $NVN-I_\pm$-equation (with $E>0$ and $E<0$ for (\ref{2DSchr})) corresponding to the case of simple poles of wave function $\psi$ were presented in the works \cite{Grinev_86}-\cite{GrinevNovikov_88}. Special care requires the case of $E=0$ for (\ref{2DSchr}), i. e. the case of $NVN-I_0$ equation \cite{BoitiPempinelli}. The use of Darbu transformations for the construction of exact solutions of NVN equation was demonstrated by Matveev et al  \cite{MatveevSalle_Book}.  The class of dromion-like solutions of NVN equation  via Mottard transformations was constructed by Athorne et al \cite{AthorneNimmo_91}. We have already constructed classes of exact potentials for
perturbed telegraph
equation (\ref{The_first_aux_probl}) with potential $u=\tilde u-\epsilon$ and perturbed string equation with $u=\tilde u, \epsilon=0$   via $\overline{\partial}$-dressing method in the paper \cite{DubrovskyKonopelchenko_93} and obtained some rationally localized solutions of NVN-equation  with simple and multiple pole wave functions $\psi$ via $\overline{\partial}$-dressing method \cite{DubrForm_01},\cite{DubrForm_03}.

Present  work is concentrated on  further use of  $\overline{\partial}$-dressing method for the construction of exact solutions of two-dimensional integrable nonlinear evolution equations, exact potentials and wave functions of famous linear auxiliary problems (\ref{The_first_aux_probl}) or  (\ref{2DSchr}) and the study of their possible applications. While many studies of this subject were performed the question of physical interpretation and exploitation of results obtained via $\overline{\partial}$-dressing are still of great interest.

The paper is organized as described further.
Basic ingredients of the $\bar{\partial}$-dressing
method for the NVN equation (\ref{NVN}) in brief
are presented in sections 2,3 and general determinant formula for  multi line soliton solutions and useful formulas for the conditions of reality and potentiality of $u$ are obtained. In sections 4 and 5 the classes of exact multi line soliton
solutions  for hyperbolic version with $\sigma=1$ and for elliptic version with  $\sigma=i$ of the NVN equation respectively
are constructed.  The classes of periodic solutions for both versions of NVN equation are constructed in section 6. The classes of solutions with functional parameters are constructed in section 7. The simplest examples of exact one, two line soliton solutions with corresponding exact wave functions of auxiliary linear problems, periodic solutions and solutions with functional parameters  are presented in sections 3,4 and 5,6,7 of the paper.
%
%
%
%
\section{Basic ingredients of the $\bar{\partial}$-dressing method and general determinant formulas for
exact solutions}\label{Section_2}
As a matter of convenience here we briefly reviewed  the basic ingredients of the
$\bar{\partial}$-dressing
method \cite{ZakhManak_85}-\cite{FokasZakh_92} for the NVN equation (\ref{NVN}) in the case of
$u(\xi,\eta,t)$ with generically non-zero asymptotic value at infinity (\ref{The_first_aux_probl}).
We followed the treatment of the papers \cite{DubrForm_01},\cite{DubrForm_03} without repetition of theirs detailed calculations.

At first one postulates the non-local $\bar{\partial}$-problem:
\begin{equation}\label{dibar_problem}
\fl  \frac{\partial \chi (\lambda ,\bar {\lambda })}{\partial \bar {\lambda }} =
  (\chi * R)(\lambda ,\bar {\lambda }) = {\int\int}_C \chi (\mu ,\overline \mu )R(\mu
  ,\overline \mu ;\lambda,\bar {\lambda })d\mu \wedge d\overline \mu
\end{equation}
where in our case $\chi$ and $R$ are the scalar complex-valued functions and $\chi$ has
canonical
normalization:
$\chi\rightarrow1$ as $\lambda\rightarrow\infty$. It should be assumed that the problem
(\ref{dibar_problem})
is unique solvable. Then one introduces the dependence of kernel $R$ of the
$\bar{\partial}$-problem (\ref{dibar_problem})
on the space and time variables $\xi$,$\eta$,$t$:
\begin{eqnarray}\label{dependence R from xi, eta, t}
\fl  \frac{\partial R}{\partial \xi} &=& i\mu
  R(\mu,\overline{\mu};\lambda,\overline{\lambda};\xi,\eta,t)-
  R(\mu,\overline{\mu};\lambda,\overline{\lambda};\xi,\eta,t)i\lambda, \nonumber\\
\fl  \frac{\partial R}{\partial \eta} &=&
  -i\frac{\epsilon}{\mu}R(\mu,\overline{\mu};\lambda,\overline{\lambda};\xi,\eta,t)+
  R(\mu,\overline{\mu};\lambda,\overline{\lambda};\xi,\eta,t)i\frac{\epsilon}{\lambda},\\
\fl  \frac{\partial R}{\partial t} &=&  i(\kappa_1\mu^3-\kappa_2\frac{\epsilon^3}{\mu^3})
  R(\mu,\overline{\mu};\lambda,\overline{\lambda};\xi,\eta,t)-
  R(\mu,\overline{\mu};\lambda,\overline{\lambda};\xi,\eta,t)
  i(\kappa_1\lambda^3-\kappa_2\frac{\epsilon^3}{\lambda^3}). \nonumber
\end{eqnarray}
Integrating (\ref{dependence R from xi, eta, t}) one obtains
\begin{equation}
\fl  R(\mu ,\overline \mu ;\lambda,\bar {\lambda };\xi,\eta,t)
  =R_0 (\mu ,\overline \mu ;\lambda,\bar {\lambda })
  e^{F(\mu; \xi,\eta,t)-F(\lambda; \xi,\eta,t)}
\end{equation}
where
\begin{equation}\label{F_formula}
  F(\lambda;\xi,\eta,t)=i\big[\lambda\xi-\frac{\epsilon}{\lambda}\eta+\big(\kappa_{1}\lambda^{3}-\kappa_{2}
  \frac{\epsilon^{3}}{\lambda^{3}}\big)t\big].
\end{equation}

By the use of "long"~derivatives
\begin{equation}\label{NVN dressing long derivatives}
  D_{1} = \partial_{\xi}+i\lambda, \quad D_{2} = \partial_{\eta}-i\frac{\epsilon}{\lambda},\quad
  D_{3} =
  \partial_{t}+i\Big(\kappa_{1}\lambda^{3}-\kappa_{2}\frac{\epsilon^{3}}{\lambda^{3}}\Big)
\end{equation}
expressing the dependence (\ref{dependence R from xi, eta, t}) of kernel $R$ of the
$\bar{\partial}$-problem (\ref{dibar_problem})
on the space and time variables $\xi$,$\eta$,$t$ in the following equivalent form
\begin{equation}\label{dependence R from xi, eta, t_EquivForm}
  [D_{1},R] = 0 , \quad [D_{2},R] = 0,\quad [D_{3},R] = 0
\end{equation}
one  can  construct the operators of auxiliary linear problems
\begin{equation}
  \tilde{L} = \sum\limits_{l,m,n}u_{lmn}(\xi,\eta,t)D_{1}^{l}D_{2}^{m}D_{3}^{n}.
\end{equation}
These operators must satisfy to the conditions
\begin{equation}\label{conditions on aux problem}
  \Big[\frac{\partial}{\partial\overline{\lambda}},\tilde{L}\Big]\chi = 0, \quad
  \tilde{L}\chi(\lambda,\overline{\lambda})|_{\lambda\rightarrow\infty}\rightarrow0
\end{equation}
of absence singularities at the points $\lambda=0$ and $\lambda=\infty$ of the complex plane of
spectral
variable $\lambda$. For such operators $\tilde{L}$ the function $\tilde{L}\chi$ obeys the same
$\bar{\partial}$-equation
as the function $\chi$. There are may be several operators $\tilde{L}_i$ of this type, by virtue
of
the unique solvability of (\ref{dibar_problem}) one has $\tilde{L}_i\chi=0$ for each of them. In
considered
case one  constructs two such operators:
\begin{eqnarray}
\fl \tilde L_1  = D_{1}D_{2}+u_{1}D_{1}+u_{2}D_{2}+u, \label{Aux operators in long derivatives}\\
\fl \tilde L_2  = D_{3}+\kappa_{1}D_{1}^{3}+\kappa_{2}D_{2}^{3}+V_{1}D_{1}^{2}+V_{2}D_{2}^{2}+
  V_{3}D_{1}+V_{4}D_{2}+V.\label{Aux operators in long derivatives2}
\end{eqnarray}
Using the conditions (\ref{conditions on aux problem}) and series expansions of wave functions
$\chi$
near the points $\lambda=0$ and $\lambda=\infty$
\begin{equation}\label{series of chi}
\fl \chi =\chi_{0}+\chi_{1}\lambda+\chi_{2}\lambda^{2}+\ldots, \quad
  \chi =\chi_{\infty}+\frac{\chi_{-1}}{\lambda}+\frac{\chi_{-2}}{\lambda^{2}}+\ldots,
\end{equation}
one obtains  the reconstruction
formulas for the field variables $u_1$,$u_2$ and $V_1$,$V_2$,$V_3$,$V_4$ through the
coefficients
$\chi_0$ and $\chi_\infty$
of expansions (\ref{series of chi}) (for calculation details see  papers \cite{DubrForm_01},\cite{DubrForm_03}):
\begin{equation}\label{V_1_W_1_reconstructFormulae}
  u_1=-\frac{\chi_{\infty\eta}}{\chi_{\infty}}, \quad
  V_1=-3\kappa_1\frac{\chi_{\infty\xi}}{\chi_{\infty}};
\end{equation}
\begin{equation}\label{V_2_W_2_reconstructFormulae}
  u_2=-\frac{\chi_{0\xi}}{\chi_{0}}, \quad V_2=-3\kappa_2\frac{\chi_{0\eta}}{\chi_{0}};
\end{equation}
\begin{equation}
  V_3=3i\kappa_2\epsilon\chi_{1\eta}, \quad V_4=-3i\kappa_1\chi_{-1\xi}.
\end{equation}

According to well known terminology the operator $\tilde{L}_1$ in (\ref{Aux operators in long
derivatives}) is pure potential operator when its first derivatives
 are absent. Due to canonical normalization of wave
function $\chi|_{\lambda\rightarrow\infty}\rightarrow1$ ($\chi_\infty=1$):
\begin{equation}\label{V_1_u_1_reconstructFormulaeSimpl}
  u_1=-\frac{\chi_{\infty\eta}}{\chi_{\infty}}=0, \quad
  V_1=-3\kappa_1\frac{\chi_{\infty\xi}}{\chi_{\infty}}=0.
\end{equation}
For zero value of the term $u_2\partial_\eta$ in $\tilde{L}_1$ one must to require $\chi_0=const$,
without restriction we can choose $\chi_0=1$, and then due to
(\ref{V_2_W_2_reconstructFormulae})
\begin{equation}\label{V_2_u_2_reconstructFormulaeSimpl}
  u_2=-\frac{\chi_{0\xi}}{\chi_{0}}=0, \quad V_2=-3\kappa_2\frac{\chi_{0\eta}}{\chi_{0}}=0.
\end{equation}
Using (\ref{conditions on aux problem}),(\ref{V_1_W_1_reconstructFormulae}) -
(\ref{V_2_u_2_reconstructFormulaeSimpl})(for calculation details see also
\cite{DubrForm_01},\cite{DubrForm_03}) one obtains the following expressions for $V_3$,$V_4$
and $u$:
\begin{equation}\label{V_3_V_4_u_reconstructFormulaeSimpl}
  V_3 = 3i\kappa_2\epsilon\chi_{1\eta} = 3\kappa_2\partial_{\xi}^{-1}u_{\eta},\quad V_4 =
  -3i\kappa_1\chi_{-1\xi} = 3\kappa_1\partial_{\eta}^{-1}u_{\xi},
\end{equation}
\begin{equation}\label{u_reconstructFormulae}
  u=-\epsilon-i\chi_{-1\eta} = -\epsilon+i\epsilon\chi_{1\xi}.
\end{equation}
The field variable $V$ in (\ref{Aux operators in long derivatives2}) due to gauge freedom \cite{DubrGramol} in the present paper is chosen  to be equal to zero.
In terms of the wave function
\begin{equation}\label{psi_hi}
  \psi: = \chi e^{F(\lambda;\xi,\eta,t)} = \chi e^{i\big[\lambda\xi - \frac{\epsilon}{\lambda}\eta +
  \big(\kappa_{1}\lambda^{3} - \kappa_{2}\frac{\epsilon^{3}}{\lambda^{3}}\big)t\big]},
\end{equation}
under the reduction $u_1=0$ and $u_2=0$ (the condition of potentiality $\tilde{L}_1$), one obtains from
(\ref{Aux operators in long derivatives}),(\ref{Aux operators in long derivatives2})
due to (\ref{conditions on aux problem}) and (\ref{V_1_u_1_reconstructFormulaeSimpl})-(\ref{V_3_V_4_u_reconstructFormulaeSimpl})
the linear auxiliary system (\ref{NVN dressing auxiliary problems}),(\ref{NVN dressing auxiliary problems2}) and NVN integrable nonlinear equation (\ref{NVN}) as compatibility condition (\ref{NVN dressing triada Manakova}) of  linear auxiliary problems in (\ref{NVN dressing auxiliary problems}), (\ref{NVN dressing auxiliary problems2}).

The solution of the $\bar{\partial}$-problem (\ref{dibar_problem}) with constant normalization
$\chi_\infty=1$
is equivalent to the solution of the following singular integral equation:
\begin{eqnarray}\label{di_problem1}
\fl  \chi (\lambda) = 1 + \int\int\limits_C {\frac{d{\lambda }'\wedge
  d{\bar {\lambda }}'}{2\pi i(\lambda'-\lambda)}}
  \int\int\limits_C  \chi(\mu,\bar{\mu})
  R(\mu ,\overline \mu ;\lambda,\bar {\lambda }){d\mu \wedge d\bar{\mu }}.
\end{eqnarray}
From (\ref{di_problem1}) one obtains for the coefficients $\chi_0$ and $\chi_{-1}$ of the series
expansions
(\ref{series of chi}) of $\chi$ the following expressions:
\begin{eqnarray}\label{di_problem_chi_0}
\fl  \chi_0  = 1 + \int\int\limits_C{\frac{d{\lambda }\wedge d{\bar {\lambda }}}{2\pi i\lambda}}
  \int\int\limits_C  \chi(\mu,\bar{\mu})
  R_0(\mu ,\overline \mu ;\lambda
  ,\bar {\lambda })e^{F(\mu) -
  F(\lambda)}{d\mu \wedge d\bar{\mu }}
\end{eqnarray}
and
\begin{eqnarray}\label{di_problem_chi_-1}
\fl  \chi_{-1}  = - \int\int\limits_C {\frac{d{\lambda }\wedge
  d{\bar {\lambda }}}{2\pi i}}
  \int\int\limits_C \chi(\mu,\bar{\mu})
  R_0(\mu ,\overline \mu ;\lambda
  ,\bar {\lambda })e^{F(\mu) -
  F(\lambda)}{d\mu \wedge d\bar{\mu }}
\end{eqnarray}
where $F(\lambda)$ is short notation for $F(\lambda;\xi,\eta,t)$ given by the formula (\ref{F_formula}).
The conditions of reality $u$ and of potentiality of the operator $\tilde{L}_1$ give some restrictions
for the kernel $R_0$ of the $\bar{\partial}$-problem (\ref{dibar_problem}). In the Nizhnik case
($\sigma=1,\bar{\kappa}_1=\kappa_1,\bar{\kappa}_2=\kappa_2$) of the NVN equations (\ref{NVN})
with real
 space variables $\xi=x+y$, $\eta=x-y$ the condition of reality of $u$ leads from
 (\ref{u_reconstructFormulae}) and
(\ref{di_problem_chi_-1}) in the limit of "weak" fields
($\chi=1$ in (\ref{di_problem_chi_-1})) to the following restriction for the kernel
$R_0$
of the $\bar{\partial}$ - problem:
\begin{equation}\label{real_condition_NVN}
  R_0(\mu,\overline{\mu};\lambda,\overline{\lambda})
  =\overline{R_0(-\overline{\mu},-\mu; -\overline{\lambda},-\lambda)}.
\end{equation}
For the Veselov-Novikov case ($\sigma=i,\kappa_1=\kappa_2=\kappa=\bar{\kappa}$) of the NVN
equations (\ref{NVN}) with complex space variables
$\xi=z=x+iy$, $\eta=\bar{z}=x-iy$  the condition of reality of $u$ leads from
(\ref{u_reconstructFormulae}) and
(\ref{di_problem_chi_-1}) in the limit of "weak" fields to another restriction on the kernel $R_0$
of the $\bar{\partial}$ - problem:
\begin{equation}\label{real_condition_VN}
  R_0(\mu,\overline{\mu};\lambda,\overline{\lambda})
  =\frac{\epsilon^3}{{|\mu|}^2{|\lambda|}^2
  \overline{\mu}\overline{\lambda}}\;
  \overline{R_0(-\frac{\epsilon}{\overline{\lambda}},
  -\frac{\epsilon}{\lambda}, -\frac{\epsilon}{\overline{\mu}}
  -\frac{\epsilon}{\mu})}.
\end{equation}
The potentiality condition for the operator $\tilde{L}_1$ in (\ref{Aux operators in long derivatives}) for
the choice $\chi_0=1$ due to (\ref{di_problem_chi_0}) has the following form:
\begin{eqnarray}\label{potencial_condition}
\fl  \chi_0-1=\int\int\limits_C \frac{d\lambda\wedge
  d\overline{\lambda}}{2\pi i \lambda}\; \int\int\limits_C \;\chi(\mu,\overline{\mu})\,
  R_0(\mu,\overline{\mu};\lambda,\overline{\lambda})
  e^{F(\mu)-F(\lambda)}d\mu\wedge
  d\overline{\mu}=0 .
\end{eqnarray}

Here we obtained general formulas for multisoliton solutions
corresponding
to the degenerate delta-kernel $R_0$:
\begin{equation}\label{delta_kernel_sec2}
  R_0(\mu,\bar{\mu};\lambda,\bar{\lambda})=\pi\sum\limits_k A_k
  \delta(\mu-M_k)\delta(\lambda-\Lambda_k).
\end{equation}
In this case the wave function $\chi(\lambda)$ due to (\ref{di_problem1}) has the form:
\begin{equation}\label{chi_sec2}
  \chi(\lambda)=1+2i\sum\limits_k\frac{A_k}{\Lambda_k-\lambda}\chi(M_k)e^{F(M_k)-F(\Lambda_k)}.
\end{equation}
%
The coefficient $\chi_{-1}$ due to (\ref{di_problem_chi_-1}) and (\ref{delta_kernel_sec2}) has
the form:
\begin{equation}\label{chi{-1}_sec2}
  \chi_{-1}=-2i\sum\limits_k A_k\chi(M_k)e^{F(M_k)-F(\Lambda_k)}.
\end{equation}
For the wave functions $\chi(M_k)$ from (\ref{chi_sec2}) one obtains the following system of
equations:
\begin{equation}\label{system_sec2}
 \sum\limits_l \tilde{A}_{kl}\chi(M_l)=1,\quad
\tilde{A}_{lk}=\delta_{lk}+\frac{2iA_k}{M_l-\Lambda_k}e^{F(M_k)-F(\Lambda_k)}.
\end{equation}
%
Instead of matrix $\tilde{A}$ in (\ref{system_sec2}) it is convenient  to introduce   matrix $A$
given by expression
\begin{equation}\label{matrix_soliton_FP}
  A_{lk}=\delta_{lk}+ \frac{2iA_{k}}{M_{l}-\Lambda_{k}}e^{F(M_{l})-F(\Lambda_{k})}.
\end{equation}
Both these matrices $\tilde A$ in (\ref{system_sec2}) and $A$ (\ref{matrix_soliton_FP}) are connected by the relation
\begin{equation}\label{A_tilde_A}
  {A}_{lk}=e^{F(M_l)}\tilde{A}_{lk}e^{-F(M_k)}.
\end{equation}
From (\ref{system_sec2}) due to (\ref{A_tilde_A}) one derives the expression
for the wave function $\chi$ at discrete values of  spectral variable:
\begin{equation}\label{chi_through_lambda}
  \chi(M_l)=\sum \limits_k\tilde{A}_{lk}^{-1}=\sum\limits_k e^{F(M_k)-F(M_l)}A_{lk}^{-1}.
\end{equation}
As a matter of  convenience hereafter we described some useful formulas for wave functions satisfying to linear
auxiliary problems (\ref{NVN dressing auxiliary problems}),(\ref{NVN dressing auxiliary problems2}). From (\ref{psi_hi}) and
(\ref{chi_through_lambda}) one obtains the wave function $\psi(M_l,\xi,\eta,t) =
\chi(M_l)e^{F(M_l)}$ at discrete points $\lambda = M_l$ in the space of spectral variables:
\begin{equation}\label{BI psi(M_l)}
  \psi(M_l,\xi,\eta,t) = \chi(M_l)e^{F(M_l)} = \sum\limits_k e^{F(M_k)}A_{lk}^{-1}.
\end{equation}
For the wave function (\ref{psi_hi}) at arbitrary point $\lambda$ from (\ref{chi_sec2}) -
(\ref{chi_through_lambda}) follows the expression:
\begin{eqnarray}\label{Psi-Chi}
 \fl   \psi(\lambda,\xi,\eta,t) = \chi(\lambda)e^{F(\lambda)} =  \Big[1 + 2i\sum\limits_k\frac{A_k}{\Lambda_k - \lambda}e^{F(M_k) -
    F(\Lambda_k)}\chi(M_k)\Big]e^{F(\lambda)}= \nonumber \\
\fl  \Big[1 + 2i\sum\limits_{k,l}\frac{A_k}{\Lambda_k -
    \lambda}e^{-F(\Lambda_k)}A_{kl}^{-1}e^{F(M_l)}\Big]e^{F(\lambda)}. \label{BI psi(lambda)}
\end{eqnarray}
Inserting (\ref{chi_through_lambda}) into (\ref{chi{-1}_sec2}) one obtains  for the coefficient $\chi_{-1}$
\begin{eqnarray}
 \fl     \chi_{-1} = -2i\sum\limits_{k,l}A_{k}e^{F(M_k) - F(\Lambda_k)}e^{F(M_l) - F(M_k)}A_{kl}^{-1}= \nonumber \\
\fl    = -2i\sum\limits_{k,l}A_{k}e^{F(M_l) - F(\Lambda_k)}A_{kl}^{-1} = i\,tr\Big(\frac{\partial
    A}{\partial \xi}A^{-1}\Big). \label{chi_-1_sec2}
 \end{eqnarray}
and due to reconstruction formula $u=-\epsilon-i\chi_{-1\eta}$ the convenient determinant
formula for the solution $u$ of NVN equation (\ref{NVN}):
\begin{equation}\label{Solution_general_formula_sec2}
  u=-\epsilon+\frac{\partial}{\partial\eta}tr\Big(\frac{\partial A}{\partial \xi}A^{-1}\Big)=
  -\epsilon+\frac{\partial^2}{\partial\xi\partial\eta}\ln(\det A).
\end{equation}
Here and below useful determinant identities
\begin{equation}\label{useful_identities_FP}
  Tr(\frac{\partial A}{\partial \xi}A^{-1})=\frac{\partial}{\partial \xi} \ln (\det A),\quad
  1+trD=det(1+D)
\end{equation}
are used; the matrix $D$ from last identity of (\ref{useful_identities_FP}) is degenerate with
rank 1.

Potentiality condition (\ref{potencial_condition}) by the use of
(\ref{delta_kernel_sec2})-(\ref{chi_through_lambda}) can be transformed to the form:
\begin{equation}\label{potent_cond_FP}
  \chi_0-1=-\frac{1}{2\epsilon}\sum\limits_{k,l=1}^{N} A_{kl}^{-1} B_{lk}=0
\end{equation}
where degenerate matrix $B$ with rank 1 is defined by the formula
\begin{equation}\label{matrixB_soliton_FP}
  B_{lk}=-\frac{4i\epsilon}{\Lambda_k}A_k e^{F(M_l)-F(\Lambda_k)}.
\end{equation}
Due to
(\ref{useful_identities_FP})-(\ref{matrixB_soliton_FP})
potentiality condition (\ref{potencial_condition}) takes the form:
\begin{equation}\label{potent_cond1_FP}
  0=\sum\limits_{k,m=1}^{N} A_{km}^{-1} B_{mk}=tr(A^{-1}B)=det(B A^{-1}+1)-1,
\end{equation}
here matrix $BA^{-1}$ is degenerate of rank 1 and in deriving the last equality in
(\ref{potent_cond1_FP}) the second  matrix identity of (\ref{useful_identities_FP})
is used. Equivalently due to (\ref{potent_cond1_FP}) the potentiality condition takes the form
\begin{equation}\label{potent_cond2_FP}
  \det(A+B)=\det{A}.
\end{equation}

\section{Fulfilment of potentiality condition. General formulas for one line and two line solitons}\label{Section_3}
Formula (\ref{Solution_general_formula_sec2}) for exact solutions $u(\xi,\eta,t)$ of NVN
equations (\ref{NVN})
is effective if the reality $\bar{u}=u$ conditions (\ref{real_condition_NVN}),(\ref{real_condition_VN})  and potentiality condition
(\ref{potencial_condition}) of operator $L_1$ are satisfied. This is the major and the most difficult part of all constructions.
Here we demonstrated  how one can to fulfil the condition of potentiality
(\ref{potencial_condition}) by delta-kernel with two terms:
\begin{equation}\label{delta_kernel}
 \fl R_0(\mu,\bar{\mu};\lambda,\bar{\lambda})=\pi\Big(A\delta(\mu-\mu_1)\delta(\lambda-\lambda_1)+
  B\delta(\mu-\mu_2)\delta(\lambda-\lambda_2)\Big).
\end{equation}
Inserting (\ref{delta_kernel}) into (\ref{potencial_condition}) one obtains in the limit
of weak fields ($\chi=1$ in (\ref{potencial_condition})):
\begin{eqnarray}
\fl     \chi_0 - 1 = \int\int\limits_C\frac{1}{2
    i\lambda}\Big(A\delta(\mu-\mu_1)\delta(\lambda-\lambda_1)+
    B\delta(\mu-\mu_2)\delta(\lambda-\lambda_2)\Big)\times \nonumber \\
\fl   \times e^{F(\mu)-F(\lambda)}d\mu \wedge d\overline{\mu}\,d\lambda \wedge
    d\overline{\lambda} = 2i\bigl (  \frac{A}{\lambda_1}e^{F(\mu_1)-F(\lambda_1)}+
  \frac{B}{\lambda_2}e^{F(\mu_2)-F(\lambda_2)}\bigr)=0\label{potential_cond_weak_field}.
\end{eqnarray}
The equality (\ref{potential_cond_weak_field}) is valid if
\begin{equation}\label{1eq_from_potenWF}
  F(\mu_1)-F(\lambda_1)=F(\mu_2)-F(\lambda_2),\quad
  \frac{A}{\lambda_1}=-\frac{B}{\lambda_2}.
\end{equation}
Due to the definition of $F(\lambda) = i\big[\lambda\xi - \frac{\epsilon}{\lambda}\eta +
\big(\kappa_{1}\lambda^{3} - \kappa_{2}\frac{\epsilon^{3}}{\lambda^{3}}\big)t\big]$ from
space-dependent part of (\ref{1eq_from_potenWF}) the system of equations follows:
\begin{equation}\label{relations_from_potentiality}
        \mu_1-\lambda_1= \mu_2-\lambda_2,
      \quad
      \frac{\epsilon}{\mu_1}-\frac{\epsilon}{\lambda_1}=
      \frac{\epsilon}{\mu_2}-\frac{\epsilon}{\lambda_2}.
    \end{equation}
One can show that time-dependent part of (\ref{1eq_from_potenWF}) doesn't lead to new equation
and satisfies due to the system (\ref{relations_from_potentiality}).
The system (\ref{relations_from_potentiality}) has the following solutions:
\begin{equation}\label{solutions_system_potent}
  1)\,\mu_1=\lambda_1,\;\mu_2=\lambda_2;\qquad 2)\,\mu_1=-\lambda_2,\;\mu_2=-\lambda_1.
\end{equation}
The solution $\mu_1=\lambda_1,\;\mu_2=\lambda_2$ corresponds to lump solution and will not be
considered
here, (for more information about lump solutions see [20], [21]). For the
second
solution
$\mu_1=-\lambda_2,\;\mu_2=-\lambda_1$ taking into account second relation from (\ref{1eq_from_potenWF}) one obtains:
\begin{equation}\label{relations_from_potent2}
  \frac{A}{\lambda_1}=-\frac{B}{\lambda_2}=\frac{B}{\mu_1}=a,
\end{equation}
where $a$ is some arbitrary complex constant.
It is evident that to the potentiality condition (\ref{potencial_condition})  the
kernel $R_0$ (which is the sum of expressions of the type (\ref{delta_kernel})
with parameters defined by (\ref{relations_from_potentiality})-(\ref{relations_from_potent2}))
\begin{eqnarray}\label{sum delta_kernel_satisfied_potent}
\fl  R_0(\mu,\bar{\mu};\lambda,\bar{\lambda})=\pi\sum\limits_{k=1}^{N}\Big[a_k\lambda_k\delta(\mu-\mu_k)
  \delta(\lambda-\lambda_k) + a_k\mu_k\delta(\mu+\lambda_k)\delta(\lambda+\mu_k)\Big]=\nonumber \\
  = \pi\sum\limits_{k=1}^{2N}A_k\delta(M-M_k)  \delta(\Lambda-\Lambda_k)
\end{eqnarray}
with the sets of amplitudes $A_k$ and spectral parameters $M_k,\Lambda_k$
\begin{eqnarray}\label{ParamOfAgenPosit}
(A_1,.., A_{2N}):=(a_1\lambda_1,...,a_N\lambda_N;a_1\mu_1,...,a_N\mu_N);\nonumber \\
(M_1,...,M_{2N}):=(\mu_1,...,\mu_N;-\lambda_1,...,-\lambda_N),\nonumber \\ (\Lambda_1,...,\Lambda_{2N}):=(\lambda_1,...,\lambda_N;-\mu_1,...,-\mu_N)
\end{eqnarray}
satisfies.

In order to avoid  repetition of  similar calculations in the following sections
we prepared some useful formulas in general position for calculating one- and two- line soliton
solutions and corresponding wave functions. The determinants of matrix $A$ (\ref{matrix_soliton_FP}) with parameters  (\ref{ParamOfAgenPosit}) corresponding to the
simplest kernels (\ref{sum delta_kernel_satisfied_potent}) with $N=1$ and $N=2$ have the forms:
\begin{eqnarray}
\label{BI det A, N=1} 1.\quad N=1: \det A =
 \Big(1+p_1e^{\Delta F(\mu_1,\lambda_1)}\Big)^2; \\
 \label{BI det A, N=2}\fl 2.\quad N=2:\det A = \Big(1+p_1e^{\Delta F(\mu_1,\lambda_1)}+
 p_2e^{\Delta F(\mu_2,\lambda_2)}+qe^{\Delta F(\mu_1,\lambda_1)+\Delta F(\mu_2,\lambda_2)}\Big)^2
\end{eqnarray}
 here $p_k$, $\Delta F(\mu_k,\lambda_k)$ $(k=1,2)$ and $q$ are given by the expressions
\begin{eqnarray}\label{p_kDeltaFAndq}
 p_k :=ia_k\frac{\mu_k + \lambda_k}{\mu_k - \lambda_k}; \quad \Delta F(\mu_k,\lambda_k):=F(\mu_k) -  F(\lambda_k), \\
 q:=-p_1 p_2 \cdot \frac{(\lambda_{1}-\lambda_{2})(\lambda_{2}+\mu_{1})(\mu_{1}-\mu_{2})(\lambda_{1}+\mu_{2})}
{(\lambda_{1}+\lambda_{2})(\lambda_{2}-\mu_{1})(\mu_{1}+\mu_{2})(\lambda_{1}-\mu_{2})}.\label{q_GF}
\end{eqnarray}
The formula for one line soliton solution due to (\ref{Solution_general_formula_sec2}),(\ref{BI det A, N=1}) is:
\begin{equation}\label{uN=1Gen}
u(\xi,\eta,t)=-\epsilon-\epsilon{\frac{2p_1(\mu_1-\lambda_1)^2}{\mu_1\lambda_1}}
{\frac{e^{\Delta F(\mu_1,\lambda_1)}}
{(1+p_1e^{\Delta F(\mu_1,\lambda_1)})^2}}.
\end{equation}
By using the equations   (\ref{chi_sec2}),(\ref{matrix_soliton_FP}) and (\ref{chi_through_lambda})  corresponding to
one line soliton solution (\ref{uN=1Gen}) wave functions one calculates:
\begin{equation}\label{WFN=1Gen}
\tilde{\chi_1}:=\chi_1(\mu_1) = \chi_1(-\lambda_1) = \frac{1}{1 + p_1 e^{\Delta F(\mu_1,\lambda_1)}};
\end{equation}
\begin{equation}\label{WFN=1ChiLambda}
  \chi_1(\lambda) = 1 - \Big(\frac{\lambda_1}{\lambda - \lambda_1} + \frac{\mu_1}{\lambda +
  \mu_1}\Big)\frac{2 ia_{1} e^{\Delta F(\mu_1,\lambda_1)}}{1 + p_1e^{\Delta F(\mu_1,\lambda_1)}}.
\end{equation}
 Considering   (\ref{WFN=1Gen}), (\ref{WFN=1ChiLambda}) wave
functions $\psi_1(\mu_1)=\chi_1(\mu_1)e^{F(\mu_1)}$, $\psi_1(-\lambda_1) =
\chi_1(-\lambda_1)e^{F(-\lambda_1)}$ and $\psi_1(\lambda) = \chi_1(\lambda)e^{F(\lambda)}$
satisfy  to linear auxiliary problems (\ref{NVN dressing auxiliary problems}), (\ref{NVN dressing auxiliary problems2})
and at the same time  to famous linear equations (\ref{The_first_aux_probl}), (\ref{2DSchr})
and  have the following forms:
\begin{equation}\label{PsiWFNVNN=1,1}
\fl \psi_1(\mu_1) = \frac{e^{F(\mu_1)}}{1 + p_1 e^{\Delta F(\mu_1,\lambda_1)}},\quad
  \psi_1(-\lambda_1)= \frac{e^{-F(\lambda_1)}}{1 + p_1e^{\Delta F(\mu_1,\lambda_1)}};
\end{equation}
\begin{equation}\label{PsiWFNVNN=1,2}
\fl  \psi_1(\lambda) = e^{F(\lambda)} - \Big(\frac{\lambda_1}{\lambda - \lambda_1} + \frac{\mu_1}{\lambda +
  \mu_1}\Big)\frac{2 ia_{1} e^{\Delta F(\mu_1,\lambda_1)}e^{F(\lambda)}}{1 + p_1e^{\Delta F(\mu_1,\lambda_1)}}.
\end{equation}
For two line soliton solution one obtains via (\ref{Solution_general_formula_sec2}),(\ref{BI det A, N=2})  after simple calculations the expression:
\begin{equation}\label{uN=2Gen}
  u(\xi,\eta,t) =- \epsilon-2\epsilon\frac{N(\xi,\eta,t)}{D(\xi,\eta,t)},
\end{equation}
where the nominator $N$ and denominator $D$ are given by the expressions
\begin{eqnarray}\label{NTwo-solitGen}
\fl    N(\xi,\eta,t) = {\frac{(\lambda_{1}-\mu_{1})^2}{\lambda_{1}\mu_{1}}}e^{\Delta F(\mu_1,\lambda_1)}(q p_2 e^{2\Delta F(\mu_2,\lambda_2)}+p_1)+ \nonumber \\
  +  {\frac{(\lambda_{2}-\mu_{2})^2}{\lambda_{2}\mu_{2}}} e^{\Delta F(\mu_2,\lambda_2)}(q p_1e^{2\Delta F(\mu_1,\lambda_1)}+p_2) + \nonumber \\
 \fl    +p_1 p_2 (\lambda_{1}-\mu_{1}-\lambda_{2}+\mu_{2})\Big({\frac{\lambda_{1}-\mu_{1}}{\lambda_{1}\mu_{1}}}
   -{\frac{\lambda_{2}-\mu_{2}}{\lambda_{2}\mu_{2}}}\Big)e^{\Delta F(\mu_1,\lambda_1)+\Delta F(\mu_2,\lambda_2)}+
    \nonumber \\
 \fl   +q (\lambda_{1}-\mu_{1}+\lambda_{2}-\mu_{2})\Big({\frac{\lambda_{1}-\mu_{1}}{\lambda_{1}\mu_{1}}}+
 {\frac{\lambda_{2}-\mu_{2}}{\lambda_{2}\mu_{2}}}\Big)
e^{\Delta F(\mu_1,\lambda_1)+\Delta F(\mu_2,\lambda_2)},
\end{eqnarray}
\begin{equation}\label{DTwo-solitGen}
\fl  D(\xi,\eta,t)=(1+p_1e^{\Delta F(\mu_1,\lambda_1)}+p_2e^{\Delta F(\mu_2,\lambda_2)}+q e^{\Delta F(\mu_1,\lambda_1) + \Delta F(\mu_2,\lambda_2)} )^2.
\end{equation}
It is remarkable that for the choice $q=p_1p_2$, i. e. under the condition
\begin{equation}
\frac{(\lambda_{1}-\lambda_{2})(\lambda_{2}+\mu_{1})(\mu_{1}-\mu_{2})(\lambda_{1}+\mu_{2})}
{(\lambda_{1}+\lambda_{2})(\lambda_{2}-\mu_{1})(\mu_{1}+\mu_{2})(\lambda_{1}-\mu_{2})}=-1
\end{equation}
or for equivalent
\begin{equation}\label{SeparatCondition}
(\lambda_1\mu_1+\lambda_2\mu_2)(\lambda_1-\mu_1)(\lambda_2-\mu_2)=0
\end{equation}
the formula for two line soliton solution (\ref{uN=2Gen}) with $N$, $D$ given by (\ref{NTwo-solitGen}),(\ref{DTwo-solitGen}) reduces to very simple expression:
\begin{eqnarray}\label{uTwoSolGen}
\fl u(\xi,\eta,t) = -\epsilon-\epsilon{\frac{2p_1(\mu_1-\lambda_1)^2}{\mu_1\lambda_1}}
{\frac{e^{\Delta F(\mu_1,\lambda_1)}}
{(1+p_1e^{\Delta F(\mu_1,\lambda_1)})^2}}- \nonumber \\
-\epsilon{\frac{2p_2(\mu_2-\lambda_2)^2}{\mu_2\lambda_2}}
{\frac{e^{\Delta F(\mu_2,\lambda_2)}}
{(1+p_2e^{\Delta F(\mu_2,\lambda_2)})^2}}.
\end{eqnarray}
It should be emphasized  that in the present paper multi line soliton solutions are considered, for such solutions
by construction $\mu_k\neq\lambda_k, (k=1,2)$. Considering this  due to (\ref{SeparatCondition}) the condition
$q=p_1p_2$ satisfies if
\begin{equation}\label{EquivSeparatCondit}
\lambda_1\mu_1+\lambda_2\mu_2=0.
\end{equation}

The corresponding to two line soliton solution (\ref{uTwoSolGen}) wave functions calculated in described case by the formulas (\ref{chi_sec2}),(\ref{chi_through_lambda}),
 under condition $p_1p_2=q$, have the following simple forms:
\begin{eqnarray}\label{NVNWaveFunctN=2Gen}
\fl  \chi_2(\mu_1) &=& \tilde{\chi_1}\tilde{\chi_2}\Big[1 - i\,a_2e^{\Delta F(\mu_2,\lambda_2)}\frac{(\lambda_1 + \lambda_2)(\lambda_2 + \mu_1)(\lambda_2 + \mu_2)}{(\lambda_1 - \lambda_2)(\lambda_2 - \mu_1)(\lambda_2 - \mu_2)}\Big], \\
\fl  \chi_2(-\lambda_1) &=& \tilde{\chi_1}\tilde{\chi_2}\Big[1 - i\,a_2e^{\Delta F(\mu_2,\lambda_2)}\frac{(\lambda_1 - \lambda_2)(\lambda_2 -\mu_1)(\lambda_2 + \mu_2)}{(\lambda_1 + \lambda_2)(\lambda_2 + \mu_1)(\lambda_2 - \mu_2)}\Big], \\
\fl  \chi_2(\mu_2) &=& \tilde{\chi_1}\tilde{\chi_2}\Big[1 + i\,a_1e^{\Delta F(\mu_1,\lambda_1)}\frac{(\lambda_1 + \lambda_2)(\lambda_2 - \mu_1)(\lambda_1 + \mu_1)}{(\lambda_1 - \lambda_2)(\lambda_2 + \mu_1)(\lambda_1  -\mu_1)}\Big], \\
\fl  \chi_2(-\lambda_2) &=& \tilde{\chi_1}\tilde{\chi_2}\Big[1 + i\,a_1e^{\Delta F(\mu_1,\lambda_1)}\frac{(\lambda_1 - \lambda_2)(\lambda_2 + \mu_1)(\lambda_1 + \mu_1)}{(\lambda_1 + \lambda_2)(\lambda_2 - \mu_1)(\lambda_1 - \mu_1)}\Big],
\end{eqnarray}
\begin{eqnarray}\label{NVNWaveFunctN=2Gen1}
\fl    \chi_2(\lambda) = 1 + 2i\Big(\frac{\lambda_1 a_1}{\lambda_1 - \lambda}\chi_2(\mu_1)e^{\Delta F(\mu_1,\lambda_1)} + \frac{\mu_1 a_1}{-\mu_1 - \lambda}\chi_2(-\lambda_1)e^{\Delta F(\mu_1,\lambda_1)} + \nonumber \\
\fl    +\frac{\lambda_2 a_2}{\lambda_2 -\lambda}\chi_2(\mu_2)e^{\Delta F(\mu_2,\lambda_2)} +
\frac{\mu_2 a_2}{-\mu_2-\lambda}\chi_2(-\lambda_2)e^{\Delta F(\mu_2,\lambda_2)}\Big),
\end{eqnarray}
where $\tilde{\chi_1}$ è $\tilde{\chi_2}$ are the  wave functions (see (\ref{WFN=1Gen}))
\begin{eqnarray}\label{OneSolWFGen}
\tilde{\chi_1}= \chi_1(\mu_1) = \chi_1(-\lambda_1) = \frac{1}{1 + p_1 e^{\Delta F(\mu_1,\lambda_1)}},\nonumber \\
\tilde{\chi_2}= \chi_1(\mu_2) = \chi_1(-\lambda_2) = \frac{1}{1 + p_2 e^{\Delta F(\mu_2,\lambda_2)}}
\end{eqnarray}
corresponding to one line soliton solutions.
%
Two soliton $\psi_2$ wave functions  (\ref{BI psi(M_l)}), (\ref{Psi-Chi}) satisfying to linear auxiliary problems (\ref{NVN dressing auxiliary problems}), (\ref{NVN dressing auxiliary problems2}) and
at the same time  to famous linear equations (\ref{The_first_aux_probl}), (\ref{2DSchr})  due to (\ref{NVNWaveFunctN=2Gen})-(\ref{NVNWaveFunctN=2Gen1}) have following forms:
\begin{equation}\label{NVNPsiDWaveFunctN=2Gen1}
\fl  \psi_2(\mu_1) = {\frac{e^{F(\mu_1)}}{1 + p_1 e^{\Delta F(\mu_1,\lambda_1)}}}{\frac{1 +p_2e^{\Delta F(\mu_2,\lambda_2)}\frac{(\lambda_1 + \lambda_2)(\lambda_2 + \mu_1)}{(\lambda_1 - \lambda_2)(\lambda_2 - \mu_1)}}{1 + p_2 e^{\Delta F(\mu_2,\lambda_2)}}},
\end{equation}
\begin{equation}\label{NVNPsiDWaveFunctN=2Gen2}
\fl  \psi_2(-\lambda_1) = {\frac{e^{F(-\lambda_1)}}{1 + p_1 e^{\Delta F(\mu_1,\lambda_1)}}}{\frac{1 +p_2e^{\Delta F(\mu_2,\lambda_2)}\frac{(\lambda_1 - \lambda_2)(\lambda_2 -\mu_1)}{(\lambda_1 + \lambda_2)(\lambda_2 + \mu_1)}}{1 + p_2 e^{\Delta F(\mu_2,\lambda_2)}}},
\end{equation}
\begin{equation}\label{NVNPsiDWaveFunctN=2Gen3}
\fl  \psi_2(\mu_2) = {\frac{e^{F(\mu_2)}}{1 + p_2 e^{\Delta F(\mu_2,\lambda_2)}}}{\frac{1 - p_1e^{\Delta F(\mu_1,\lambda_1)}\frac{(\lambda_1 + \lambda_2)(\lambda_2 - \mu_1)}{(\lambda_1 - \lambda_2)(\lambda_2 + \mu_1)}}{1 + p_1 e^{\Delta F(\mu_1,\lambda_1)}}},
\end{equation}
\begin{equation}\label{NVNPsiDWaveFunctN=2Gen4}
\fl  \psi_2(-\lambda_2) = {\frac{e^{-F(\lambda_2)}}{1 + p_2 e^{\Delta F(\mu_2,\lambda_2)}}}{\frac{1 -p_1e^{\Delta F(\mu_1,\lambda_1)}\frac{(\lambda_1 - \lambda_2)(\lambda_2 + \mu_1)}{(\lambda_1 + \lambda_2)(\lambda_2 - \mu_1)}}{1 + p_1 e^{\Delta F(\mu_1,\lambda_1)}}},
\end{equation}
\begin{eqnarray}\label{NVNPsiWaveFunctN=2Gen1}
\fl    \psi_2(\lambda) = e^{F(\lambda)} + 2i\Big(\frac{\lambda_1 a_1}{\lambda_1 - \lambda}\psi_2(\mu_1)e^{- F(\lambda_1)} + \frac{\mu_1 a_1}{-\mu_1 - \lambda}\psi_2(-\lambda_1)e^{ F(\mu_1)} + \nonumber \\
\fl    +\frac{\lambda_2 a_2}{\lambda_2 -\lambda}\psi_2(\mu_2)e^{- F(\lambda_2)} +
\frac{\mu_2 a_2}{-\mu_2-\lambda}\psi_2(-\lambda_2)e^{ F(\mu_2)}\Big)e^{F(\lambda)}.
\end{eqnarray}

All formulas (\ref{BI det A, N=1})-(\ref{NVNPsiWaveFunctN=2Gen1}) derived in the present section  will be effective if
 the reality  conditions (\ref{real_condition_NVN}), (\ref{real_condition_VN}) are satisfied.
 The reality condition $u=\bar{u}$ imposes additional restrictions
  on the parameters $a_k$, $\lambda_k$, $\mu_k$  (\ref{ParamOfAgenPosit}) of the kernel (\ref{sum
delta_kernel_satisfied_potent}). These restrictions
and the calculations of exact multi line soliton solutions $u$ with corresponding wave functions
are suitable  for hyperbolic and elliptic
versions of NVN equation (\ref{NVN}) separately.
%
%
%
%
%
%

\section{Exact multi line soliton solutions of NVN-II equation}\label{Section_4}

 In the present
section the hyperbolic version of NVN equation (\ref{NVN}) or NVN-II equation, i. e.
the case $\sigma^2 = 1$ with real space variables $\xi = x + y$ and $\eta = x - y$, will be
covered. In order to satisfy the reality condition (\ref{real_condition_NVN})
let us require for each term in the sum (\ref{sum delta_kernel_satisfied_potent}):
\begin{eqnarray}\label{N real condition}
\fl    a_k\lambda_k\delta(\mu-\mu_k)\delta(\lambda-\lambda_k) +
    a_k\mu_k\delta(\mu+\lambda_k)\delta(\lambda+\mu_k) = \nonumber\\
\fl    = \overline{a}_k\overline{\lambda}_k\delta(\mu+\overline{\mu}_k)\delta(\lambda+
    \overline{\lambda}_k)+\overline{a}_k\overline{\mu}_k\delta(\mu-
    \overline{\lambda}_k)\delta(\lambda-\overline{\mu}_k).
\end{eqnarray}

From (\ref{N real condition}) two
possibilities follow:
\begin{equation}\label{N relation points and amplitudes possibility 1}
 \fl 1. \: a_k\lambda_k = \overline{a}_k\overline{\lambda}_k,  a_k\mu_k =
  \overline{a}_k\overline{\mu}_k,  \mu_k =
  -\overline{\mu}_k,  \lambda_k = -\overline{\lambda}_k;\quad 2.\: a_k\lambda_k = \overline{a}_k\overline{\mu}_k,  \mu_k = \overline{\lambda}_k.
\end{equation}
In the first case in (\ref{N relation points and amplitudes possibility 1}) one obtains
that the spectral points $\mu_k$, $\lambda_k$ and amplitudes $a_k$ are pure imaginary:
\begin{equation}\label{N points and amplitudes possibility 1}
  \mu_k = -\overline{\mu}_k := i\mu_{k0}, \:\: \lambda_k = -\overline{\lambda}_k :=
  i\lambda_{k0}, \:\: a_k = -\overline{a}_k := -ia_{k0}.
\end{equation}
For the second case in (\ref{N relation points and amplitudes possibility 1}) it is appropiate to introduce
 the following notations for amplitudes and spectral points
\begin{equation}\label{N points and amplitudes possibility 2}
a_k=\overline a_k:=a'_{k0}; \quad \lambda'_k,\quad \mu'_k:=\overline\lambda'_k.
\end{equation}
So the kernel (\ref{delta_kernel_sec2}), (\ref{sum delta_kernel_satisfied_potent}) satisfying to
potentiality (\ref{potencial_condition}) and reality (\ref{real_condition_NVN}) conditions in
considered two cases (\ref{N relation points and amplitudes possibility 1}) due to  (\ref{N points and
amplitudes possibility 1}),  (\ref{N points and
amplitudes possibility 2}) can be chosen in the following form
\begin{equation}\label{N kernel possibility 1}
\fl    R_0(\mu,\overline{\mu},\lambda,\overline{\lambda}) = \pi\sum\limits_{k = 1}^{2(L+N)}A_k
    \delta(\mu - M_k)\delta(\lambda - \Lambda_k)
\end{equation}
of $L$ pairs of the type
$\pi\big(a_{l0}\lambda_{l0}\delta(\mu - i\mu_{l0})\delta(\lambda - i\lambda_{l0})+a_{l0}\mu_{l0}\delta(\mu + i\lambda_{l0})\delta(\lambda + i\mu_{l0})\big), \quad (l=1,...,L)$ and $N$ pairs of the type
$\pi\big(a'_{n0}\lambda'_n\delta(\mu - \mu'_n)\delta(\lambda - \lambda'_n)+a'_{n0}\overline\lambda'_n\delta(\mu + \lambda'_n)\delta(\lambda + \mu'_n)\big)$, \quad with $\mu'_n=\bar\lambda'_n \quad (n=1,...,N)$ of corresponding items. In (\ref{N kernel possibility 1})
 for application of general determinant formulas (\ref{matrix_soliton_FP}),
(\ref{Solution_general_formula_sec2}) and (\ref{potent_cond2_FP})  due to (\ref{N points and
amplitudes possibility 1})-(\ref{N kernel possibility 1})  the following
sets of amplitudes $A_k$ and spectral parameters $M_k$, $\Lambda_k$
\begin{eqnarray}\label{N_parameters_1}
\fl    (A_1,.., A_{2(L+N)}) = \nonumber \\
\fl = (a_{10}\lambda_{10},.., a_{L0}\lambda_{L0}; a_{10}\mu_{10},..
    a_{L0}\mu_{L0};a'_{10}\lambda'_1,.., a'_{N0}\lambda'_N; a'_{10}\mu'_1,.. a'_{N0}\mu'_N ),\nonumber \\
\fl    (M_1,.., M_{2(L+N)}) = (i\mu_{10},.., i\mu_{L0}; -i\lambda_{10},.., -i\lambda_{L0};\mu'_1,.., \mu'_N; -\lambda'_1,..,
    -\lambda'_N),\nonumber \\
\fl    (\Lambda_1,.., \Lambda_{2(L+N)}) = (i\lambda_{10},.., i\lambda_{L0}; -i\mu_{10},.., -i\mu_{L0};\lambda'_1,.., \lambda'_N; -\mu'_1,..,
   -\mu'_N).
\end{eqnarray}
are introduced.

General determinant formula (\ref{Solution_general_formula_sec2}) with matrix $A$ from (\ref{matrix_soliton_FP}) with corresponding parameters (\ref{N_parameters_1}) of kernel $R_0$ (\ref{N kernel possibility 1})  of
$\overline{\partial}$-problem (\ref{dibar_problem}) gives exact multi line soliton
solutions $u(\xi,\eta,t)$ with constant asymptotic value $-\epsilon$ at infinity of
hyperbolic version of NVN equation. At the same time an application of general scheme of
$\overline{\partial}$-dressing method gives exact potentials
$u$ and corresponding wave functions $\chi^{[L,N]}(M_l)$, $\psi^{[L,N]}(M_l) = \chi^{[L,N]}(M_l)e^{F(M_l)}$ at discrete
spectral parameters $M_l$ and $\chi^{[L,N]}(\lambda)$, $\psi^{[L,N]}(\lambda) = \chi^{[L,N]}(\lambda)e^{F(\lambda)}$ at continuous spectral parameter $\lambda$ of linear auxiliary problems
(\ref{NVN dressing auxiliary problems}),(\ref{NVN dressing auxiliary problems2})
and one-dimensional perturbed telegraph equation (\ref{The_first_aux_probl}). For the convenience here and henceforth the symbols $\chi^{[L,N]}, \psi^{[L,N]}$ denote the wave functions of multi line soliton exact solution  corresponding to the general kernel (\ref{N kernel possibility 1}) with  $L+N$ pairs of items.

The rest
of the present section is devoted to the presentation for considered   case (\ref{N relation points and amplitudes possibility 1}) of the explicit forms of some one line of types $[1,0],[0,1]$ and two line  of types $[2,0],[0,2],[1,1]$ soliton solutions of hyperbolic version of NVN equation and exact potentials with corresponding wave functions of one-dimensional perturbed telegraph equation (\ref{The_first_aux_probl}).

{\bf {4.1 \quad $[1,0]$ and $[2,0]$ line solitons}}

The kernels of  type $R_0$
(\ref{N kernel possibility 1}) with values $L=1,2;\quad N=0$ (i. e. $a_{l0}\neq 0,l=1,2; a'_{n0}=0, n=1,...,N$) in (\ref{N_parameters_1})  are correspond to  $[1,0]$, $[2,0]$ solitons.
For nonsingular one line $[1,0]$ and two line $[2,0]$ soliton solutions of hyperbolic version of NVN equation
parameters $\mu_k,\lambda_k, a_k$ in general formulas (\ref{BI det A, N=1})-(\ref{NVNPsiWaveFunctN=2Gen1}) of Section 3
must be identified due to (\ref{N_parameters_1}) by the following way:
\begin{equation}\label{[1,2,0]Solitons}
\fl \mu_k = -\overline{\mu}_k := i\mu_{k0}, \:\: \lambda_k = -\overline{\lambda}_k :=
  i\lambda_{k0}, \:\: a_k = -\overline{a}_k := -ia_{k0}, \quad (k=1,2)
\end{equation}
and real parameters $p_k$ (\ref{p_kDeltaFAndq})
\begin{equation}\label{p_kVHyperb1}
p_k= a_{k0}\frac{\mu_{k0}+\lambda_{k0}}{\mu_{k0}-\lambda_{k0}} = e^{\phi_{0k}} > 0, \quad (k=1,2)
\end{equation}
since positive constants must be chosen. The real phases $\Delta F(\mu_k,\lambda_k)=F(\mu_k)-F(\lambda_k):=\varphi_k, (k=1,2)$ (\ref{BI det A, N=1})-(\ref{NVNPsiWaveFunctN=2Gen1}) are given in considered case by the expressions:
\begin{equation}\label{N two-solution 1 varphi1}
\fl  \varphi_k(\xi,\eta,t) = (\lambda_{k0}-\mu_{k0})\xi+\Big(\frac{\epsilon}{\lambda_{k0}}-
  \frac{\epsilon}{\mu_{k0}}\Big)\eta-\kappa_1\big(\lambda_{k0}^3-\mu_{k0}^3\big)t-
  \kappa_2\Big(\frac{\epsilon^3}{\lambda_{k0}^3}-\frac{\epsilon^3}{\mu_{k0}^3}\Big)t.
\end{equation}
One line soliton $[1,0]$ solution  generating by simplest kernel $R_0$ of the type (\ref{N kernel
possibility 1}) with $L = 1, N=0$ and parameters (\ref{N_parameters_1}) due to (\ref{uN=1Gen})
and (\ref{p_kVHyperb1}), (\ref{N two-solution 1 varphi1}) is nonsingular line soliton:
\begin{equation}\label{N solution 1 u}
  u = -\epsilon - \frac{\epsilon(\lambda_{10} - \mu_{10})^2}{2\lambda_{10}\mu_{10}}
  \frac{1}{\cosh^2\frac{\varphi + \phi_{01}}{2}}.
\end{equation}


\begin{figure}[h]
\begin{center}
\includegraphics[width=0.50\textwidth,keepaspectratio]{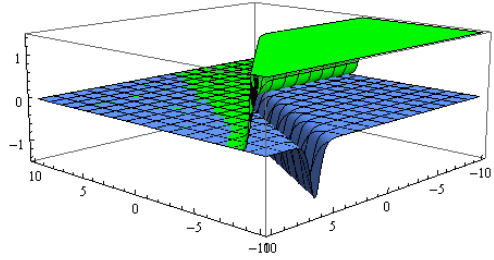}
\fl\parbox[t]{1\textwidth}{\caption{One line soliton $[1,0]$ solution $\tilde{u}(x,y,t=0)=u(x,y,t=0)+\epsilon$ (\ref{N solution 1 u}) (blue) and squared absolute value of corresponding wave function $|\psi^{[1,0]}(i\mu_{10})|^{2}$ (green) (\ref{N solution 1 chi(-i lambda10)}) with parameters $a_{10}=-1,\epsilon=1,\lambda_{10}=1,\mu_{10}=4$.}\label{graphNVNOne+WF}}
\end{center}
\end{figure}
Wave functions $\psi^{[1,0]}(i\mu_{10})$, $\psi^{[1,0]}(-i\lambda_{10})$ and $\psi^{[1,0]}(\lambda)$ due to
formulas (\ref{PsiWFNVNN=1,1}), (\ref{PsiWFNVNN=1,2}) and
(\ref{[1,2,0]Solitons})-(\ref{N two-solution 1 varphi1}) have the following  forms:
\begin{equation}\label{N solution 1 chi(-i lambda10)}
\fl  \psi^{[1,0]}(i\mu_{10}) =  \frac{e^{F(i\mu_{10})}}{1 + e^{\varphi_1+\phi_0}}, \quad
  \psi^{[1,0]}(-i\lambda_{10}) =  \frac{e^{-F(i\lambda_{10})}}{1 + e^{\varphi_1+\phi_{01}}};
\end{equation}
\begin{equation}\label{N solution 1 chi(lambda)}
\fl  \psi^{[1,0]}(\lambda) = e^{F(\lambda)} - \Big(\frac{i\lambda_{10}}{\lambda - i\lambda_{10}} +
  \frac{i\mu_{10}}{\lambda + i\mu_{10}}\Big)
  {\frac {2a_{10}e^{\varphi_1+F(\lambda)}}{1 + e^{\varphi_1+\phi_{01}}}}.
\end{equation}
Graphs of one line $[1,0]$ soliton (\ref{N solution 1 u}) and the squared absolute value of wave function $\psi^{[1,0]}(i\mu_{10})$ (\ref{N solution 1 chi(-i lambda10)}) for certain values of parameters are presented in Fig.\ref{graphNVNOne+WF}.
Graph of the squared absolute value of another wave function - $\psi^{[1,0]}(-i\lambda_{10})$ has the similar form but with localization along another one half of potential valley

Two line soliton $[2,0]$ solution  in considered case of kernel $R_0$ (\ref{N kernel possibility 1})
with parameters (\ref{q_GF}),(\ref{N_parameters_1})-(\ref{p_kVHyperb1})
is given by the formula  (\ref{uN=2Gen}).
%
%
It is remarkable that under the condition $q=p_1 p_2$ (see (\ref{SeparatCondition})) which is equivalent to the relation:
\begin{equation}\label{N two-solution 1 condtion}
  (\lambda_{10}\mu_{10}+\lambda_{20}\mu_{20})(\lambda_{10}
  -\mu_{10})(\lambda_{20}-\mu_{20})=0,
\end{equation}
i. e. to relation  $\lambda_{10}\mu_{10}+\lambda_{20}\mu_{20}=0$ (due to  $\lambda_{n0}\neq\mu_{n0}$,
we do not consider in the present paper lumps!), the solution (\ref{uN=2Gen}) radically simplifies and due to (\ref{uTwoSolGen}) takes the form:
\begin{equation}\label{N two-solution 1 u simple}
\fl  u(\xi,\eta,t) = -\epsilon - \frac{\epsilon(\lambda_{10}-\mu_{10})^2}{2\lambda_{10}\mu_{10}}
    \frac{1}{\cosh^2{\frac{\varphi_1(\xi,\eta,t)+\phi_{01}}{2}}}- \frac{\epsilon(\lambda_{20}-\mu_{20})^2}{2\lambda_{20}\mu_{20}}
    \frac{1}{\cosh^2{\frac{\varphi_2(\xi,\eta,t)+\phi_{02}}{2}}}.
\end{equation}
%
%
%
\begin{figure}[h]
\begin{center}
a)\includegraphics[width=0.45\textwidth,keepaspectratio]{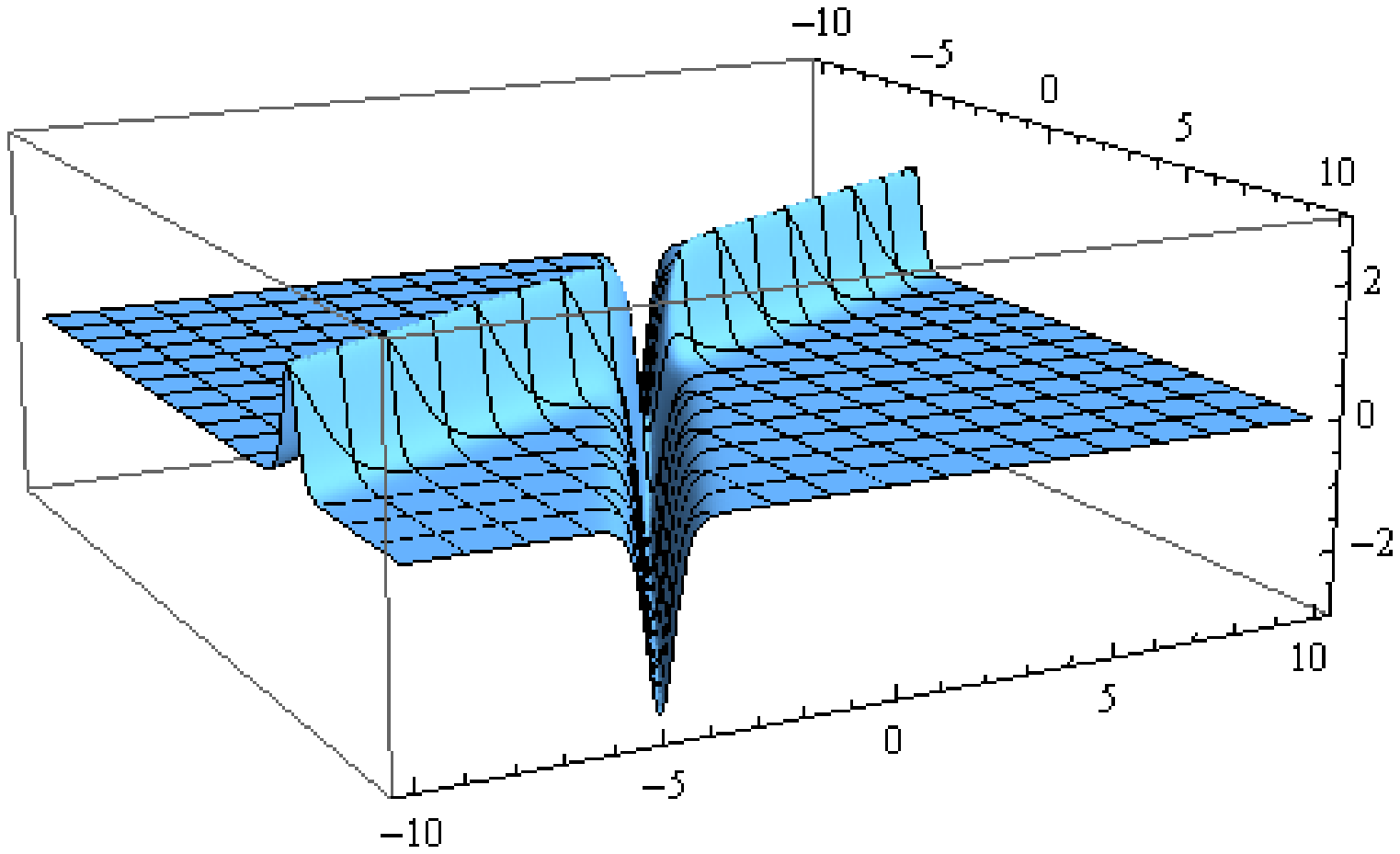}\hfill
b)\includegraphics[width=0.50\textwidth,keepaspectratio]{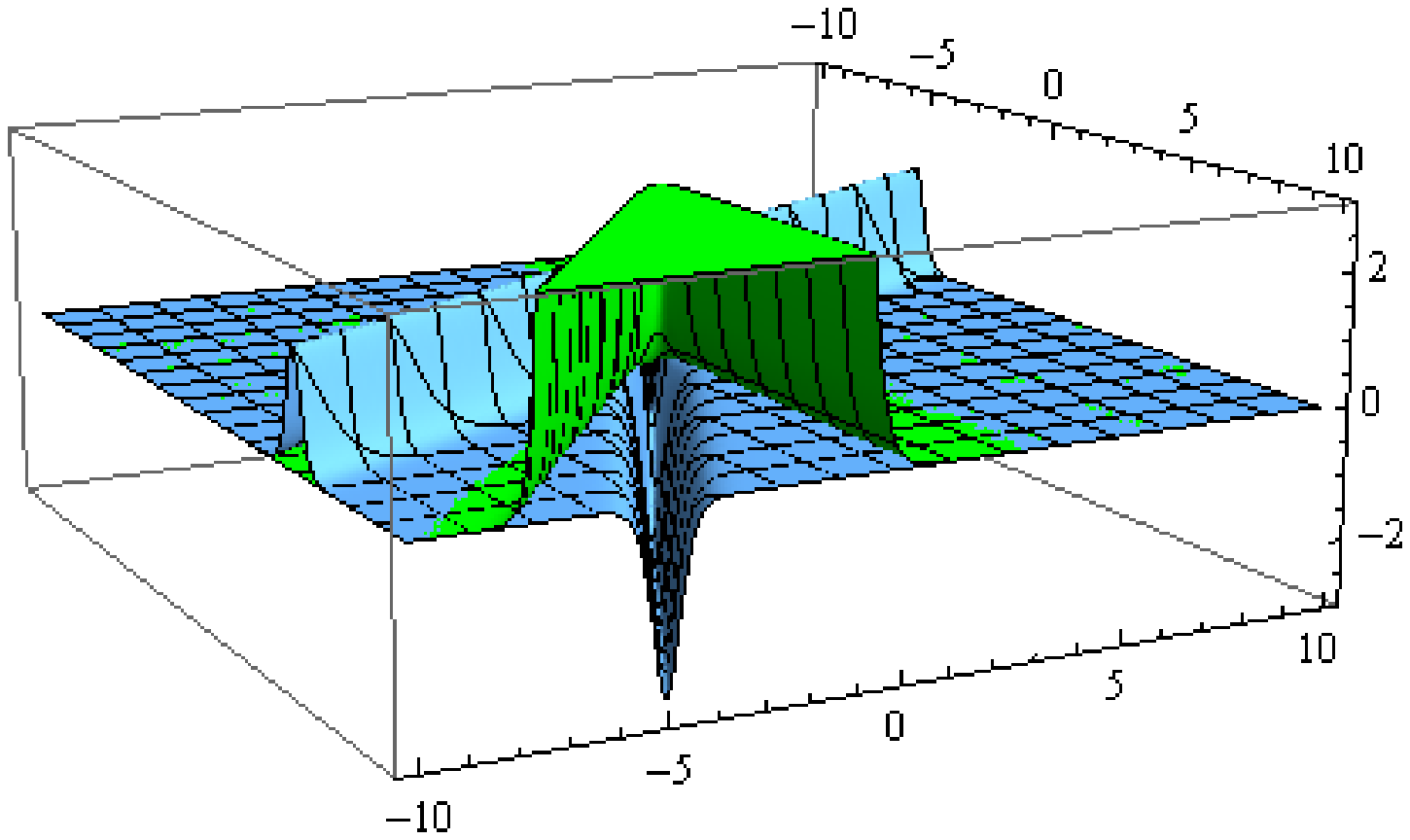}
\fl\parbox[t]{1\textwidth}{\caption{Two line soliton $[2,0]$ solution ${\tilde{u}(x,y,{t=0})=u(x,y,{t=0})+\epsilon}$ (\ref{N two-solution 1 u simple}) (a) and squared absolute value of corresponding wave function $|\psi^{[2,0]}(i\mu_{10})|^2$ (green) (b), with parameters $a_{10}=1,\lambda_{10}=1,\mu_{10}=-3;a_{20}=-1,\lambda_{20}=4,\epsilon=-1$.}\label{graphNVNTwo}}
\end{center}
\end{figure}

The corresponding  wave functions $\chi^{[2,0]},\psi^{[2,0]}$ calculated in considered case of kernel $R_0$ (\ref{N kernel possibility 1}) with parameters (\ref{N_parameters_1})-(\ref{p_kVHyperb1})   by the formulas (\ref{chi_sec2})-(\ref{Psi-Chi}),
 under condition $p_1p_2=q$, i. e. under $\lambda_{10}\mu_{10}+\lambda_{20}\mu_{20}=0$,
 are given by the  simple formulas (\ref{NVNWaveFunctN=2Gen})-(\ref{NVNPsiWaveFunctN=2Gen1}).
Graphs of two line $[2,0]$ soliton (\ref{N two-solution 1 u simple}) and the squared absolute value of wave function - $\psi^{[2,0]}(i\mu_{10})$ (\ref{NVNPsiDWaveFunctN=2Gen1}) for certain values of parameters are presented in  Fig.\ref{graphNVNTwo} (the squared absolute values of other wave functions (\ref{NVNPsiDWaveFunctN=2Gen2}-\ref{NVNPsiDWaveFunctN=2Gen4}) have 
similar forms but with localization along another three possible halves of two potential
valleys).

${\overline{\partial}}$-dressing in  present paper is carried out for the fixed nonzero value of parameter $\epsilon$. Nevertheless one can correctly set $\epsilon=c_{k}\mu_{k0}, (k=1,2)$ ($c_{k}$-arbitrary complex constant) and consider the limit $\epsilon \rightarrow 0$ in all derived formulas and obtain some interesting results also for the case of $\epsilon=0$. Limiting procedure  $\epsilon=c_{k}\mu_{k0}\rightarrow 0, (k=1,2)$  can be correctly performed by the following settings in all required formulas:  $\epsilon \rightarrow 0$ and $\mu_{k0}\rightarrow 0$ in cases when uncertainty is absent, but $\frac{\mu_{20}}{\mu_{10}}=-\frac{\lambda_{10}}{\lambda_{20}}\rightarrow \frac{c_{1}}{c_{2}}$ in accordance with the relations $\epsilon=c_{k}\mu_{k0}$ and $\mu_{10}\lambda_{10}+\mu_{20}\lambda_{20}=0$; the last relation is assumed to be valid in considered limit.
The two line soliton solution  (\ref{N two-solution 1 u simple}) in the limit $\epsilon \rightarrow 0$ takes the form:
\begin{equation}\label{NVNE0Limit}
u=-\frac{c_{1}\lambda_{10}}{2\cosh^2{\frac{\varphi_{1}(\xi,\eta,t) + \phi_{01}}{2}}} -\frac{c_{2}\lambda_{20}}{2\cosh^2{\frac{\varphi_{2}(\xi,\eta,t) + \phi_{02}}{2}}},
\end{equation}
where the phases  $\varphi_{k}(\xi,\eta,t)$ and $\phi_{0k}$ due to (\ref{p_kVHyperb1}), (\ref{N two-solution 1 varphi1}) have in considered limit the forms:
\begin{equation}\label{NVNPhazesE0Limit}
\fl\varphi_k(\xi,\eta,t) =\lambda_{k0}\xi - c_k\eta -
\kappa_1 \lambda_{k0}^{3}t+\kappa_2 c_{k}^{3}t,\quad
\phi_{0k}=\ln(-a_{k0}).
\end{equation}
One can check by direct substitution that NVN-II equation (\ref{NVN})
with $\sigma=1$ satisfies by $u$
given by (\ref{NVNE0Limit}), it satisfies also by each item
\begin{equation}\label{uE0LimitSolsNVN}
u^{(k)}=-\frac{c_{k}\lambda_{k0}}{2\cosh^2{\frac{\varphi_{k}(\xi,\eta,t) + \phi_{0k}}{2}}} , \quad (k=1,2)
\end{equation}
of the sum (\ref{NVNE0Limit}). So in considered case the linear principle
of superposition $u=u^{(1)}+u^{(2)}$ for such special solutions $u^{(1)},u^{(2)}$  (\ref{uE0LimitSolsNVN}) is valid.

{\bf {4.2 \quad $[0,1]$ and $[0,2]$ line solitons}}

To  $[0,1]$, $[0,2]$ solitons  the kernels of  type $R_0$
(\ref{N kernel possibility 1}) with values $L=0;\quad N=1,2$ (i. e. $a_{l0}= 0,l=1,...,L; a'_{n0}\neq 0, n=1,2$)
in (\ref{N_parameters_1})  are correspond.
For nonsingular one line $[0,1]$ and two line $[0,2]$ soliton solutions of hyperbolic version of NVN equation
parameters $\mu_k,\lambda_k, a_k$ in general formulas (\ref{BI det A, N=1})-(\ref{NVNPsiWaveFunctN=2Gen1}) of Section 3
must be identified due to (\ref{N_parameters_1}) by the following way:
\begin{equation}\label{[1,2,0]Solitons1}
\fl \mu_k = \overline\lambda_k, \:\: a_k = \overline{a}_k := a_{k0}, \quad (k=1,2).
\end{equation}
The parameters $p_k$, $(k=1,2)$, $q$  in (\ref{BI det A, N=1})-(\ref{NVNPsiWaveFunctN=2Gen1}) due to (\ref{[1,2,0]Solitons1})
 are given by the expressions:
\begin{equation}\label{N two-solution 2 p_1}
\fl  p_k = -a_{k0}\frac{\lambda_{kR}}{\lambda_{kI}}:=e^{\phi_{0k}} > 0,\quad q = p_1 p_2 \cdot \Bigg|\frac{(\lambda_1 - \lambda_2)(\lambda_1 + \bar{\lambda}_2)}
{(\lambda_1 + \lambda_2)(\lambda_1 - \bar{\lambda}_2)}\Bigg|^2,
\end{equation}
where the parameters $p_k:=e^{\phi_{0k}}>0$ are chosen as positive constants.

The real phases $\Delta F(\mu_k,\lambda_k)=F(\mu_k)-F(\lambda_k):=\varphi_k, (k=1,2)$ in (\ref{BI det A, N=1})-(\ref{NVNPsiWaveFunctN=2Gen1}) are given
due to (\ref{F_formula}) in considered case by the expressions:
\begin{equation}\label{N two-solution 1 varphi2}
\fl  \varphi_k(\xi,\eta,t) = i\Big[(\overline{\lambda}_k - \lambda_k)\xi -
  \epsilon\Big(\frac{1}{\overline{\lambda}_k} - \frac{1}{\lambda_k}\Big)\eta +
  \kappa_1(\overline{\lambda}_k^3 - \lambda_k^3)t -
  \kappa_2\epsilon^3\Big(\frac{1}{\overline{\lambda}_k^3} - \frac{1}{\lambda_k^3}\Big)t\Big].
\end{equation}
\begin{figure}[h]
\begin{center}
\includegraphics[width=0.50\textwidth,keepaspectratio]{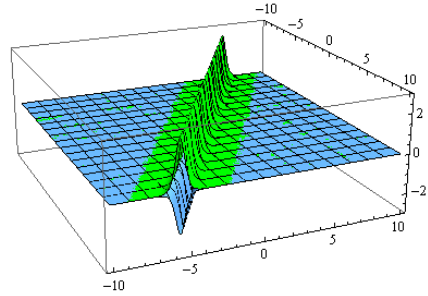}
\fl\parbox[t]{1\textwidth}{\caption{One line soliton $[0,1]$ solution $\tilde{u}(x,y,t=0)=u(x,y,t=0)+\epsilon$ (\ref{N solution 2 u}) (blue) and squared absolute value of corresponding wave functions $|\psi^{[0,1]}(\overline{\lambda}_1)|^2=|\psi^{[0,1]}(-\lambda_{1})|^2$ (green) (\ref{N solution 2 chi(bar lambda1)}) with parameters $a_{10}=-1,\lambda_{1R}=0.2,\lambda_{1I}=2,\epsilon=-1$.}\label{graphNVN3One}}
\end{center}
\end{figure}
One line soliton $[0,1]$ solution generated by simplest kernel $R_0$ of the type (\ref{N kernel
possibility 1}) with $L = 0, N=1$ and parameters (\ref{N_parameters_1}) due to (\ref{uN=1Gen})
and (\ref{[1,2,0]Solitons1})-(\ref{N two-solution 1 varphi2}) is nonsingular line soliton:
\begin{equation}\label{N solution 2 u}
  u = -\epsilon + \frac{2\epsilon\lambda_{1I}^2}{|\lambda_1|^2}
  \frac{1}{\cosh^2\frac{\varphi_1(\xi,\eta,t) + \phi_{01}}{2}}.
\end{equation}
\begin{figure}[h]
\begin{center}
\includegraphics[width=0.50\textwidth,keepaspectratio]{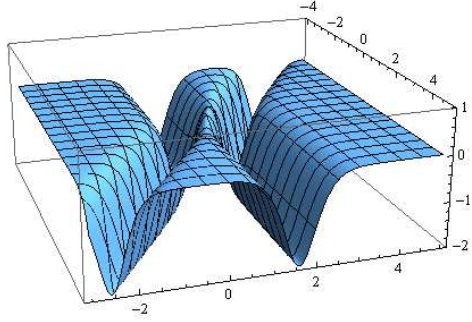}
\fl\parbox[t]{1\textwidth}{\caption{Two line soliton $[0,2]$ solution $\tilde{u}(x,y,t=0)=u(x,y,t=0)+\epsilon$ with parameters $a_{10}=-1,\lambda_{1R}=0.2,\lambda_{1I}=2;a_{20}=-1,\lambda_{2R}=0.1,\lambda_{2I}=1,\epsilon=-2$.}\label{graphNVN[0,2]}}
\end{center}
\end{figure}
The corresponding  wave functions $\psi^{[0,1]}(\mu_1)=\chi^{[0,1]}(\mu_1)e^{F(\mu_1)}$, $\psi^{[0,1]}(-\lambda_1) = \chi^{[0,1]}(-\lambda_1)e^{F(-\lambda_1)}$ and $\psi^{[0,1]}(\lambda) = \chi^{[0,1]}(\lambda)e^{F(\lambda)}$
of linear auxiliary problems (\ref{NVN dressing auxiliary problems}),(\ref{NVN dressing auxiliary problems2})
and exact potential $u=\tilde u-\epsilon$ of one-dimensional perturbed telegraph equation (\ref{The_first_aux_probl}) due to
(\ref{PsiWFNVNN=1,1})-(\ref{PsiWFNVNN=1,2}) and (\ref{[1,2,0]Solitons1})-(\ref{N two-solution 1 varphi2})
 have the forms:
\begin{equation}\label{N solution 2 chi(bar lambda1)}
\fl  \psi^{[0,1]}(\overline{\lambda}_1) = {\frac{e^{F(\overline{\lambda}_1)}}{1+ e^{\varphi_1+\phi_{01}}}},\quad
\psi^{[0,1]}(-\lambda_1)={\frac{e^{-F(\lambda_1)}}{1 +e^{\varphi_1+\phi_{01}}}};
\end{equation}
\begin{equation}\label{N solution 2 chi(lambda)}
\fl  \psi^{[0,1]}(\lambda) = e^{F(\lambda)} - \Big(\frac{\lambda_1}{\lambda - \lambda_1} +
  \frac{\overline{\lambda}_1}{\lambda + \overline{\lambda}_1}\Big)\frac{2ia_{10}e^{\varphi_1+F(\lambda)}}{1 +
 e^{\varphi_1+\phi_{01}}}.
\end{equation}
Graphs of  one line $[0,1]$ soliton (\ref{N solution 2 u}) and the squared absolute values of wave functions (\ref{N solution 2 chi(bar lambda1)})
for certain values of parameters are shown in Fig.\ref{graphNVN3One}.

Two line soliton solution  in considered case of kernel (\ref{N kernel
possibility 1}) with $L = 0, N=2$ and parameters (\ref{N_parameters_1}),(\ref{N two-solution 2 p_1})
is given by the formula  (\ref{uN=2Gen}).
%
%
%
%
%
%
It is interesting to note that the condition $q=p_1p_2$ in the considered case
of kernel $R_0$ of the type (\ref{N kernel
possibility 1}) with $L = 0, N=2$ and parameters (\ref{N_parameters_1}),(\ref{[1,2,0]Solitons1}), (\ref{N two-solution 2 p_1}) due to (\ref{EquivSeparatCondit})
takes the form  $\lambda_1\mu_1+\lambda_2\mu_2=|\lambda_1|^2+|\lambda_2|^2=0$
and can not be satisfied for $\lambda_k\neq0$, by this reason  splitting of two line soliton solution
(\ref{uN=2Gen})-(\ref{DTwo-solitGen}) into the  simple form (\ref{uTwoSolGen})
in the present  case is impossible. Graph of  two line $[0,2]$ soliton given by (\ref{uN=2Gen})-(\ref{DTwo-solitGen})
for certain values of corresponding parameters is shown in Fig.\ref{graphNVN[0,2]}.

{\bf {4.3 \quad $[1,1]$ line soliton}}

To $[1,1]$ soliton  corresponds the kernel of  type $R_0$
(\ref{N kernel possibility 1}) with values $L=1;\quad N=1$ (i. e. $a_{10}\neq 0, a'_{10}\neq 0$)
in (\ref{N_parameters_1}). For nonsingular two line $[1,1]$ soliton solution of hyperbolic version of NVN equation
parameters $\mu_k,\lambda_k, a_k$ in general formulas (\ref{BI det A, N=1})-(\ref{NVNPsiWaveFunctN=2Gen1}) of Section 3
must be identified due to (\ref{N_parameters_1}) by the following way:
\begin{eqnarray}\label{[1,1]Soliton}
\fl \mu_1 = -\overline{\mu}_1 := i\mu_{10}, \:\: \lambda_1 = -\overline{\lambda}_1:=
i\lambda_{10}, \:\: a_1 = -\overline{a}_1 := -ia_{10},\nonumber\\
\fl \mu_2=\mu'_1,\:\:\lambda_2 =\lambda'_1=\overline{\mu'}_1, \:\: a_2=a'_1 = \overline{a'}_1 := a'_{10},
\end{eqnarray}
$a_2$, $\lambda_2$, $\mu_2$ in formulas (\ref{uN=2Gen})-(\ref{NVNPsiWaveFunctN=2Gen1}) due (\ref{[1,1]Soliton}) must be identified with $a'_1$, $\lambda'_1$, $\mu'_1$ in (\ref{N kernel possibility 1}).
The parameters $p_k$, $(k=1,2)$, $q$  in (\ref{BI det A, N=1})-(\ref{NVNPsiWaveFunctN=2Gen1}) due to (\ref{p_kDeltaFAndq}) and (\ref{[1,1]Soliton}) are given by expressions:
\begin{equation}\label{p_kVHyperb[1,1]}
\fl p_1= a_{10}\frac{\mu_{10}+\lambda_{10}}{\mu_{10}-\lambda_{10}}:= e^{\phi_{01}} > 0,\quad
p_2= -a_{20}\frac{\lambda_{2R}}{\lambda_{2I}}:=e^{\phi_{02}} > 0.
\end{equation}
Two line soliton $[1,1]$ solution in considered case with parameters (\ref{q_GF}), (\ref{[1,1]Soliton})
is given by the formula  (\ref{uN=2Gen}). It is remarkable that under the condition $q=p_1 p_2$ (see (\ref{SeparatCondition})) which is equivalent to the relation:
\begin{equation}\label{N two-solution [1,1]condtion}
  (-\lambda_{10}\mu_{10}+|\lambda_{2}|^2)(i\lambda_{10}
  -i\mu_{10})(\lambda_{2}-\overline{\lambda}_{2})=0,
\end{equation}
i. e. to relation  $-\lambda_{10}\mu_{10}+|\lambda_{2}|^2=0$ (due to $\lambda_{n0}\neq\mu_{n0}$,
we do not consider in the present paper lumps!), the solution (\ref{NTwo-solitGen}) radically simplifies and due to (\ref{uTwoSolGen}) takes the form:
\begin{equation}\label{N solution [1,1] u simple}
\fl  u(\xi,\eta,t) = -\epsilon - \frac{\epsilon(\lambda_{10}-\mu_{10})^2}{2\lambda_{10}\mu_{10}}
    \frac{1}{\cosh^2{\frac{\varphi_1(\xi,\eta,t)+\phi_{01}}{2}}}+
    \frac{2\epsilon\lambda^2_{2I}}{|\lambda_2|^2}
  \frac{1}{\cosh^2\frac{\varphi_2(\xi,\eta,t) + \phi_{02}}{2}},
\end{equation}
where phases $\varphi_1(\xi,\eta,t)$, $\varphi_2(\xi,\eta,t)$ are given by the formulas (\ref{N two-solution 1 varphi1}),(\ref{N two-solution 1 varphi2}).
\begin{figure}[h]
\begin{center}
\includegraphics[width=0.50\textwidth,keepaspectratio]{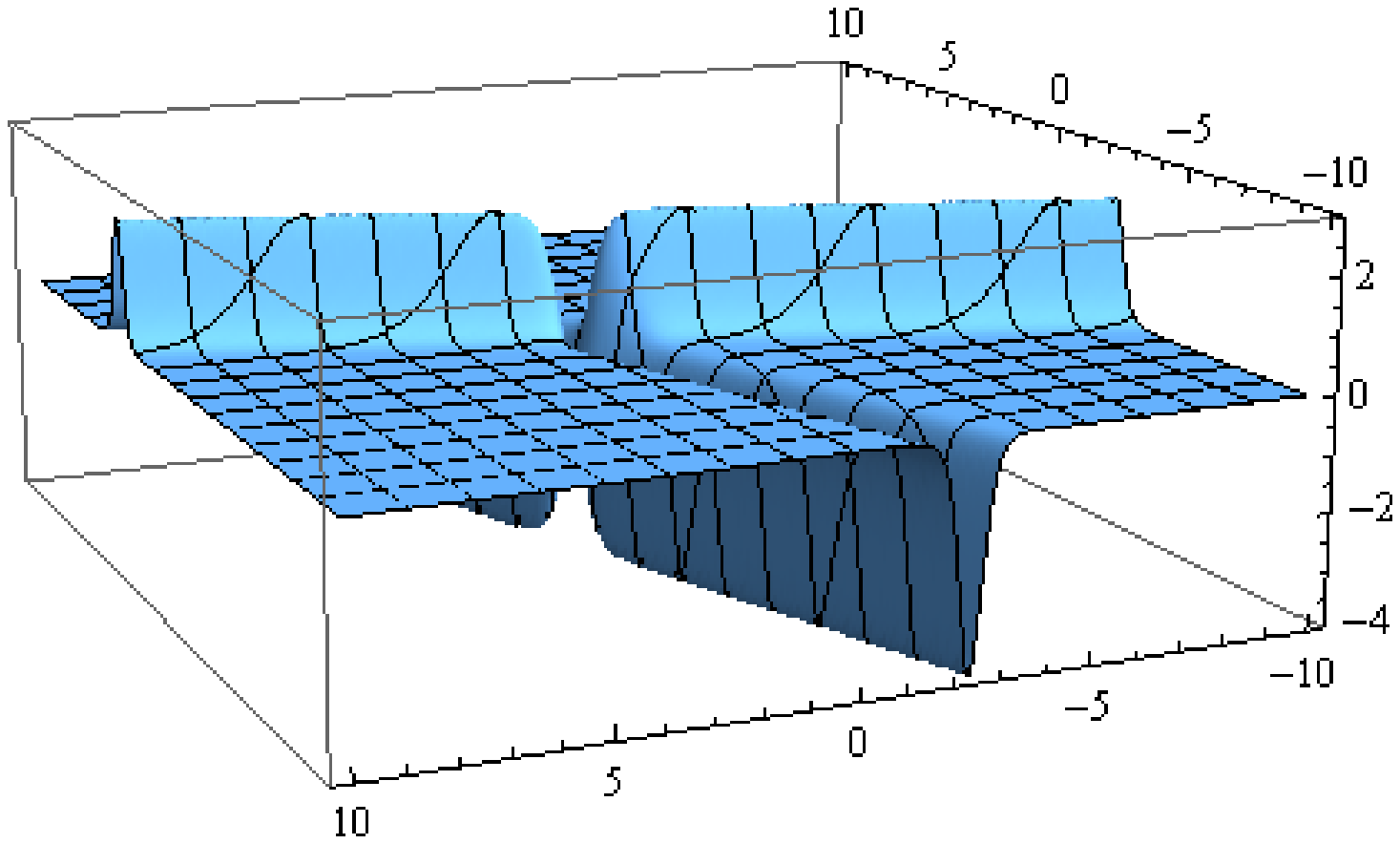}
\fl\parbox[t]{1\textwidth}{\caption{Two line soliton $[1,1]$ solution $\tilde{u}(x,y,t=0)=u(x,y,t=0)+\epsilon$ (\ref{N solution [1,1] u simple}) with parameters $a_{10}=-0.1,\lambda_{10}=2,\epsilon=-2;a_{20}=-0.1,\lambda_{2R}=0.1,\lambda_{2I}=1$.}\label{NVNSoliton[1,1]}}
\end{center}
\end{figure}

\begin{figure}[!h]
a\includegraphics[width=0.55\textwidth,keepaspectratio]{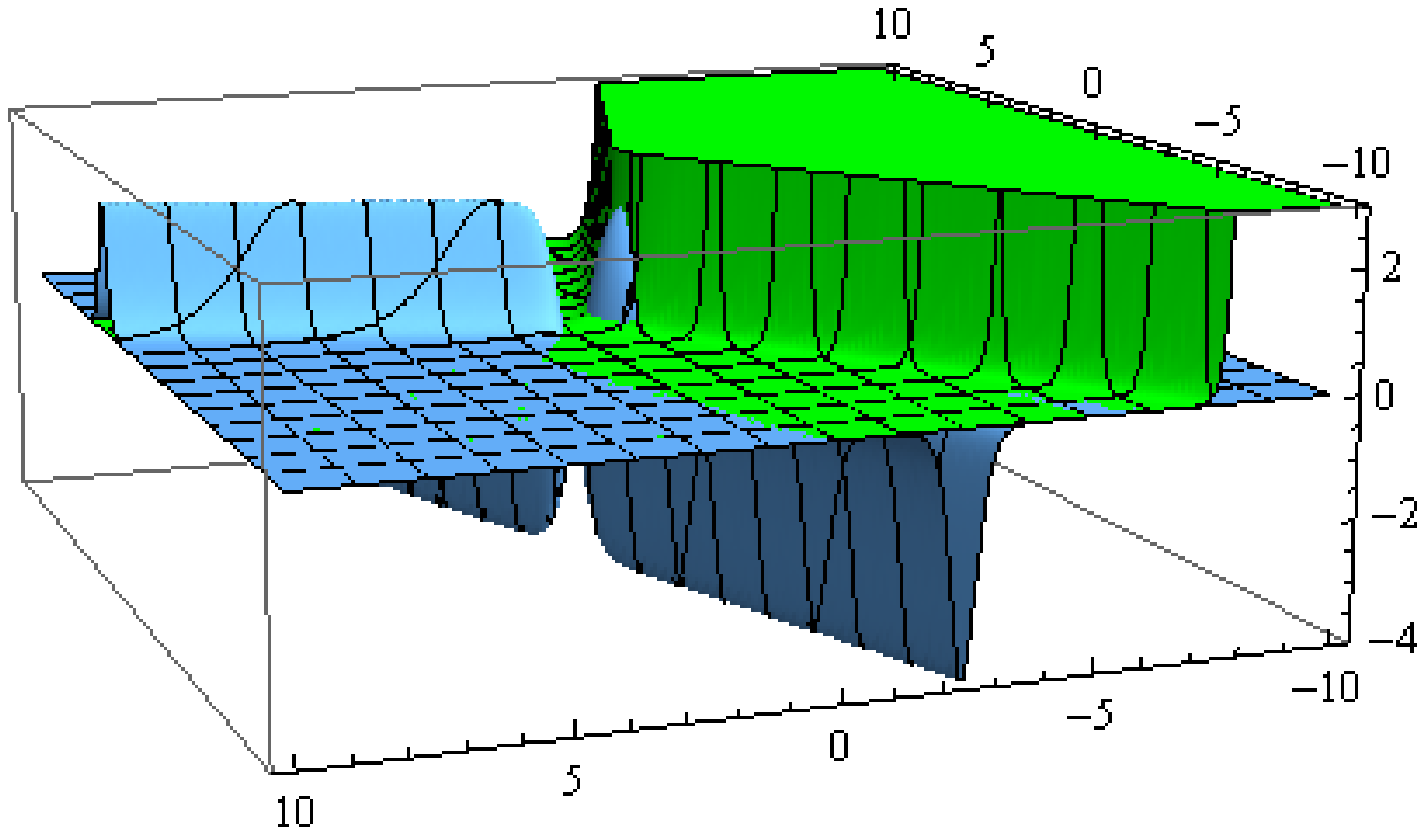}\hfill
b\includegraphics[width=0.50\textwidth,keepaspectratio]{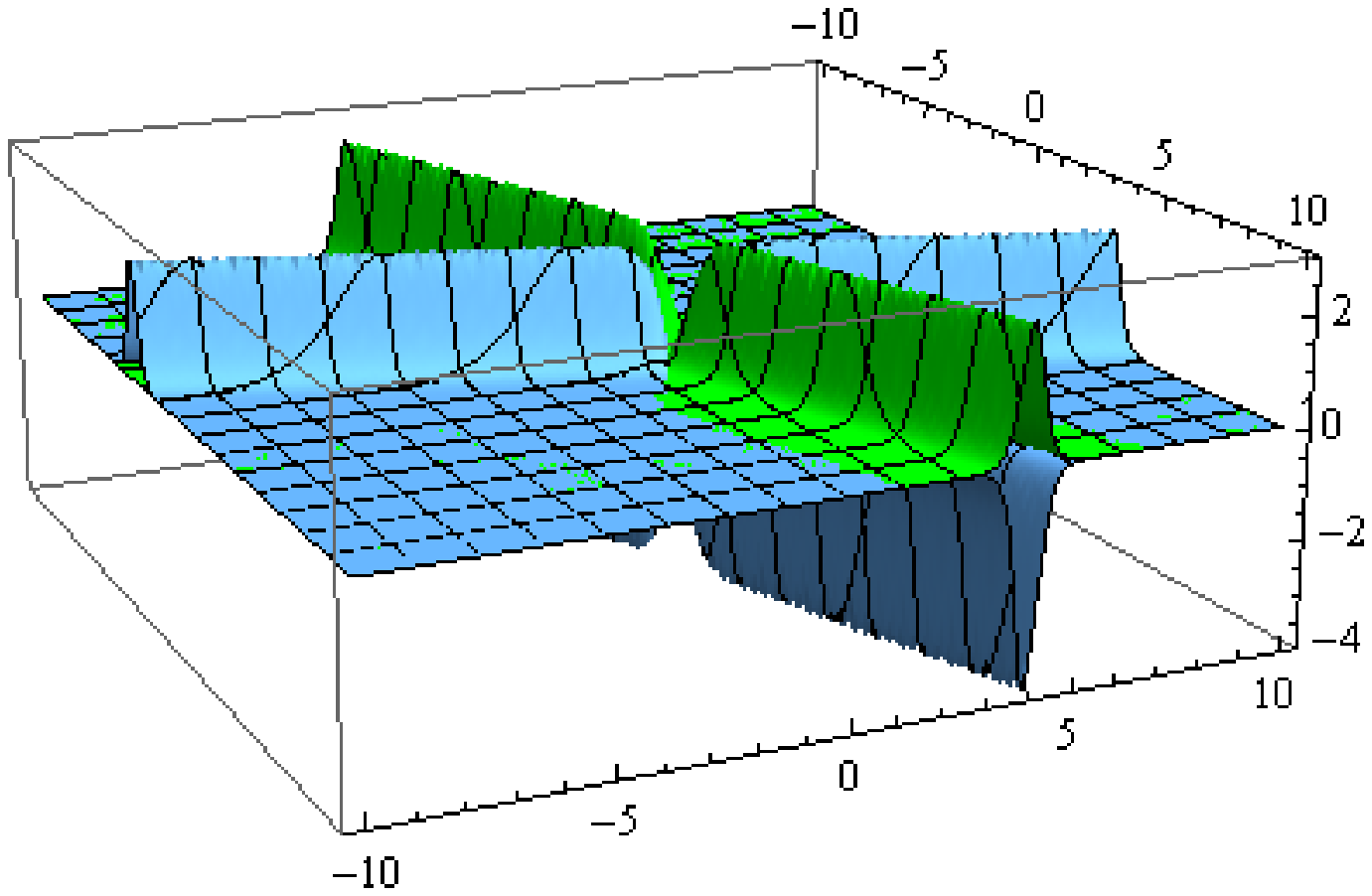}
\parbox[t]{1\textwidth}{\caption{Nonbounded $|\psi^{[1,1]}(i\mu_{10})|^2$(a) and  bounded $|\psi^{[1,1]}(\overline{\lambda}_1)|^2=|\psi^{[1,1]}(-\lambda_{1})|^2$(b) squared absolute values of wave functions (green) corresponding to solution in the Fig.\ref{NVNSoliton[1,1]}.}\label{NVNSoliton[1,1]WF}}
\end{figure}

The corresponding  wave functions $\chi^{[1,1]},\psi^{[1,1]}$ calculated in considered case of kernel $R_0$ (\ref{N kernel possibility 1}) with parameters (\ref{N_parameters_1}),(\ref{[1,1]Soliton}) and  by the formulas (\ref{chi_sec2})-(\ref{Psi-Chi}),
under condition $p_1p_2=q$, i. e. under $-\lambda_{10}\mu_{10}+|\lambda_{2}|^2=0$,
are given by the  simple formulas (\ref{NVNWaveFunctN=2Gen})-(\ref{NVNPsiWaveFunctN=2Gen1}). Graphs of  two line $[1,1]$ soliton (\ref{N solution [1,1] u simple}) and squared absolute values of some wave functions given by (\ref{NVNPsiDWaveFunctN=2Gen1})-(\ref{NVNPsiDWaveFunctN=2Gen3})
for certain values of parameters are shown in Fig.\ref{NVNSoliton[1,1]}-Fig.\ref{NVNSoliton[1,1]WF} (graphs
of $|\psi^{[1,1]}(i\mu_{10})|^2$ and $|\psi^{[1,1]}(-i\lambda_{10})|^2$ are similar to each other but with localization
along two different halves of corresponding potential valley).

In all considered cases for NVN-II equation (hyperbolic version) multi line solitons are finite but corresponding wave functions can take infinite values in some areas of the plane $(x,y)$, (Fig.\ref{graphNVNOne+WF},\,\ref{graphNVNTwo},\,\ref{NVNSoliton[1,1]WF}a). Only in two considered cases, for soliton $[0,1]$ and soliton $[1,1]$ the squared absolute value of corresponding wave functions $|\psi^{[0,1]}(\overline{\lambda}_1)|^2=|\psi^{[0,1]}(-\lambda_{1})|^2$ (Fig.\ref{graphNVN3One}) and
$|\psi^{[1,1]}(\overline{\lambda}_1)|^2=|\psi^{[1,1]}(-\lambda_{1})|^2$ (Fig.\ref{NVNSoliton[1,1]WF}b) are finite.

 We have to mention that exact potentials (of types [0,1] and [1,0]) of (\ref{The_first_aux_probl}) with corresponding wave functions (\ref{N solution 1 chi(-i lambda10)}), (\ref{N solution 2 chi(bar lambda1)}) in the paper
\cite{DubrovskyKonopelchenko_93} have been calculated and used for the construction of exact solutions of  two-dimensional
generalized integrable sine-Gordon equation (2DGSG). In the present paper time evolution (\ref{dependence R from xi, eta, t}) is taken into account and corresponding multi line soliton solutions of NVN-II equation are calculated.

%
%
%
\section{Exact multi line soliton solutions of NVN-I equation}\label{Section_5}
For elliptic version of NVN equation (\ref{NVN}), or NVN-I equation, with
 $\sigma^2 = -1$ and complex space variables $\xi :=z= x + iy$, $\eta :=\bar z= x - iy$
 an application of reality condition
(\ref{real_condition_VN}) to each term of the sum (\ref{sum delta_kernel_satisfied_potent}) for
$R_0$ gives the following relation:
\begin{eqnarray}
\fl    a_k\lambda_k\delta(\mu - \mu_k)\delta(\lambda - \lambda_k) + a_k\mu_k\delta(\mu +
    \lambda_k)\delta(\lambda + \mu_k) = \nonumber \\
\fl    = \frac{\epsilon^3}{|\lambda|^2|\mu|^2\overline{\lambda}\overline{\mu}}\Big[\overline{a}_k
    \overline{\lambda}_k\delta\Big(-\frac{\epsilon}{\overline{\lambda}} - \mu_k\Big)
    \delta\Big(-\frac{\epsilon}{\overline{\mu}}-\lambda_k\Big) + \overline{a}_k\overline{\mu}_k
    \delta\Big(-\frac{\epsilon}{\overline{\lambda}} + \lambda_k\Big)
    \delta\Big(-\frac{\epsilon}{\overline{\mu}} + \mu_k\Big)\Big] = \nonumber \\
\fl    = \frac{\epsilon\overline{a}_k}{\overline{\mu}_k}\delta\Big(\lambda +
    \frac{\epsilon}{\overline{\mu}_k}\Big)\delta\Big(\mu +
    \frac{\epsilon}{\overline{\lambda}_k}\Big) +
    \frac{\epsilon\overline{a}_k}{\overline{\lambda}_k}\delta\Big(\lambda -
    \frac{\epsilon}{\overline{\lambda}_k}\Big)\delta\Big(\mu -
    \frac{\epsilon}{\overline{\mu}_k}\Big).\label{VN real condition}
\end{eqnarray}
We should  underline that in the present paper complex delta functions (with complex arguments)
are used. The last equality in (\ref{VN real condition}) by the well known  property of complex
delta functions
$
  \delta(\varphi(z)) = \sum\limits_k\delta(z - z_k)/|\varphi'(z_k)|^2
$
is obtained; $z_k$ in last formula are simple roots of equation $\varphi(z_k) =
0$.

 From (\ref{VN real condition}) two possibilities are follow:
%
\begin{equation}\label{VN relation points and amplitudes possibility 1}
\fl  1. \quad a_k\lambda_k = \frac{\epsilon\overline{a}_k}{\overline{\mu}_k}, \:\: \lambda_k =
  -\frac{\epsilon}{\overline{\mu}_k},\:\: \mu_k =
  -\frac{\epsilon}{\overline{\lambda}_k}; \quad 2.\quad a_k\lambda_k = \frac{\epsilon\overline{a}_k}{\overline{\lambda}_k}, \:\: \lambda_k =
  \frac{\epsilon}{\overline{\lambda}_k}, \:\: \mu_k = \frac{\epsilon}{\overline{\mu}_k}.
\end{equation}
For the first case in (\ref{VN relation points and amplitudes possibility 1}) taking into account the reality of
$\epsilon$ one obtains
\begin{equation}\label{VN amplitudes and epsilon possibily 1}
\fl
  a_k = -\overline{a}_k := ia_{k0}, \:\: \epsilon = -\mu_k\overline{\lambda}_k =
  -\overline{\mu}_k\lambda_k; \quad \arg(\mu_k) = \arg(\lambda_k) + m\pi,
\end{equation}
i. e. pure imaginary amplitudes $a_k$  ($a_{k0}=\overline{a}_{k0}$) and
the relation between arguments of discrete spectral points $\mu_k$ and $\lambda_k$
%
%
with $m$ arbitrary integer.
From the second possibility in (\ref{VN relation points and amplitudes possibility 1})
for satisfying the reality condition (\ref{real_condition_VN}) the following relations
\begin{equation}\label{VN relation lambda1 and mu1}
\fl a_k = \overline{a}_k := a'_{k0}, \quad \epsilon =
|\mu'_k|^2 = |\lambda'_k|^2; \:\: \arg(\mu'_k) = \arg(\lambda'_k) + \delta_k
\end{equation}
with real amplitudes $a_k = \overline{a}_k := a'_{k0}$ and arbitrary constants $\delta_k$ are follow.
%
%

So the kernel (\ref{delta_kernel_sec2}), (\ref{sum delta_kernel_satisfied_potent}) satisfying to
potentiality (\ref{potencial_condition}) and reality (\ref{real_condition_NVN}) conditions in
considered two cases (\ref{N relation points and amplitudes possibility 1}) due to
(\ref{VN relation points and amplitudes possibility 1})-(\ref{VN relation lambda1 and mu1})
can be chosen in the following form
\begin{equation}\label{VN kernel possibility 2}
\fl    R_0(\mu,\overline{\mu},\lambda,\overline{\lambda}) = \pi\sum\limits_{k = 1}^{2(L+N)}A_k
    \delta(\mu - M_k)\delta(\lambda - \Lambda_k)
\end{equation}
%
of $L$ pairs of the type
$i\pi\big(a_{l0}\lambda_l\delta(\mu -\mu_l)\delta(\lambda - \lambda_l)+
    a_{l0}\mu_l\delta(\mu + \lambda_l)\delta(\lambda +
    \mu_l)\big)$ \quad (here $\epsilon = -\mu_l\overline{\lambda}_l =-\overline{\mu}_l\lambda_l, \quad (l=1,..,L)$); and $N$ pairs of the type
$\pi\big(a'_{n0}\lambda'_n\delta(\mu - \mu'_n)\delta(\lambda - \lambda'_n)+a'_{n0}\mu_n\delta(\mu + \lambda'_n)\delta(\lambda + \mu'_n)\big)$ \quad (here $\epsilon=|\lambda'_n|^2=|\mu'_n|^2, \quad (n=1,..,N)$) of corresponding items. Here in (\ref{VN kernel possibility 2})
 for application of general determinant formulas (\ref{matrix_soliton_FP}),
(\ref{Solution_general_formula_sec2}) and (\ref{potent_cond2_FP}) due to (\ref{VN relation points and amplitudes possibility 1})-(\ref{VN relation lambda1 and mu1})  the following
sets of amplitudes $A_k$ and spectral parameters $M_k$, $\Lambda_k$
\begin{eqnarray}\label{VN_parameters_1}
\fl    (A_1,.., A_{2(L+N)}) = \nonumber \\
\fl =(ia_{10}\lambda_1,.., ia_{L0}\lambda_L;
    ia_{10}\mu_1,..,
    ia_{L0}\mu_L;a'_{10}\lambda'_1,.., a'_{N0}\lambda'_N; a'_{10}\mu'_1,.., a'_{N0}\mu'_N ),\nonumber \\
\fl    (M_1,.., M_{2(L+N)}) = (\mu_1,..,
    \mu_L; -\lambda_1,.., -\lambda_L; \mu'_1,.., \mu'_N; -\lambda'_1,.., -\lambda'_N),\nonumber \\
\fl    (\Lambda_1,.., \Lambda_{2(L+N)}) = (\lambda_1,.., \lambda_N;
    -\mu_1,.., -\mu_N;  \lambda'_1,..,
     \lambda'_N; -\mu'_1,.., -\mu'_N)
\end{eqnarray}
are introduced.

General determinant formula (\ref{Solution_general_formula_sec2}) with matrix $A$ from (\ref{matrix_soliton_FP}) with corresponding parameters (\ref{VN_parameters_1}) of kernels $R_0$ (\ref{VN kernel possibility 2}) of $\overline{\partial}$-problem (\ref{dibar_problem}) gives exact multi line soliton
solutions $u(z,\overline{z},t)$ with constant asymptotic value $-\epsilon$ at infinity of
elliptic version of NVN equation. Simultaneously an application of general scheme of
$\overline{\partial}$-dressing method gives exact potentials
$u$ and corresponding wave functions $\chi^{[L,N]}(M_l)$, $\psi^{[L,N]}(M_l) = \chi^{[L,N]}(M_l)e^{F(M_l)}$ at discrete
spectral parameters $M_l$ and $\chi^{[L,N]}(\lambda)$, $\psi^{[L,N]}(\lambda) = \chi^{[L,N]}(\lambda)e^{F(\lambda)}$ at continuous spectral parameter $\lambda$ of linear auxiliary problems
(\ref{NVN dressing auxiliary problems}),(\ref{NVN dressing auxiliary problems2})
and two-dimensional stationary Schr\"{o}dinger equation (\ref{2DSchr}).
Here and below  the symbols $\chi^{[L,N]}, \psi^{[L,N]}$ denote the wave functions of multi line soliton exact solution  corresponding to the general kernel (\ref{VN kernel possibility 2}) with  $L+N$ pairs of items.

The rest
of the section is devoted to the presentation for considered  two cases (\ref{VN relation points and amplitudes possibility 1}) of the explicit forms of some one line of types $[1,0],[0,1]$ and two line soliton
solutions of types $[2,0],[0,2],[1,1]$ of elliptic version of NVN equation and exact potentials with corresponding wave functions of two-dimensional stationary Schr\"{o}dinger equation (\ref{2DSchr}).

{\bf{5.1 \quad  $[1,0],[2,0]$ line solitons}}

To  $[1,0]$, $[2,0]$ line solitons  the kernels of  type $R_0$
(\ref{VN kernel possibility 2}) with values $L=1,2;\quad N=0$ (i. e. $a_{l0}\neq 0,l=1,2; a'_{n0}=0, n=1,...,N$) in (\ref{VN_parameters_1})  are correspond.

For nonsingular one line $[1,0]$ and two line $[2,0]$ soliton solutions of elliptic version of NVN equation
parameters $\mu_k,\lambda_k, a_k$ in general formulas (\ref{BI det A, N=1})-(\ref{NVNPsiWaveFunctN=2Gen1}) of Section 3
must be identified due to (\ref{VN_parameters_1}) by the following way:
\begin{equation}\label{[1,2,0]VNSolitons}
\fl a_k = -\overline{a}_k := ia_{k0},\quad  \mu_k = -{\frac{\epsilon}{\overline{\lambda}_k}} \quad (k=1,2),
\end{equation}
and real parameters $p_k$ (\ref{p_kDeltaFAndq})
\begin{equation}\label{p_kVNElliptic1}
p_k= a_{k0}\frac{\lambda_{k} +\mu_k}{\lambda_{k}-\mu_k} = e^{\phi_{0k}} > 0, \quad (k=1,2)
\end{equation}
 as positive constants must be chosen. The real phases $\Delta F(\mu_k,\lambda_k)=F(\mu_k)-F(\lambda_k):=\varphi_k, (k=1,2)$ in (\ref{BI det A, N=1})-(\ref{NVNPsiWaveFunctN=2Gen1}) are given in considered case by the expressions:
\begin{equation}\label{VN two-solution 1 varphi1}
\fl  \varphi_k(z,\bar{z},t) = i\big[( \mu_k - \lambda_k)z -( \overline{\mu}_k - \overline{\lambda}_k)\overline{z} +
\kappa( \mu^3_k -\lambda_{k}^{3})t -\overline{\kappa}( \overline{\mu}_{k}^{3} - \overline{\lambda}_{k}^{3})t\big].
\end{equation}
One line soliton $[1,0]$ solution  corresponding to simplest kernel $R_0$ of the type (\ref{VN kernel possibility 2}) with parameters (\ref{VN_parameters_1}) due to (\ref{[1,2,0]VNSolitons})-(\ref{VN two-solution 1 varphi1}) is nonsingular line soliton:
\begin{equation}\label{VN solution 1 u}
\fl  u = -\epsilon - \frac{\epsilon(\lambda_1 - \mu_1)^2}{2\lambda_1\mu_1}
  \frac{1}{\cosh^2\frac{\varphi_1 + \phi_{01}}{2}}=-\epsilon + \frac{|\lambda_1 - \mu_1|^2}{2}
  \frac{1}{\cosh^2\frac{\varphi_1 + \phi_{01}}{2}};\quad \epsilon=-\lambda_1\bar\mu_1.
\end{equation}
\begin{figure}[h]
a\includegraphics[width=0.50\textwidth,keepaspectratio]{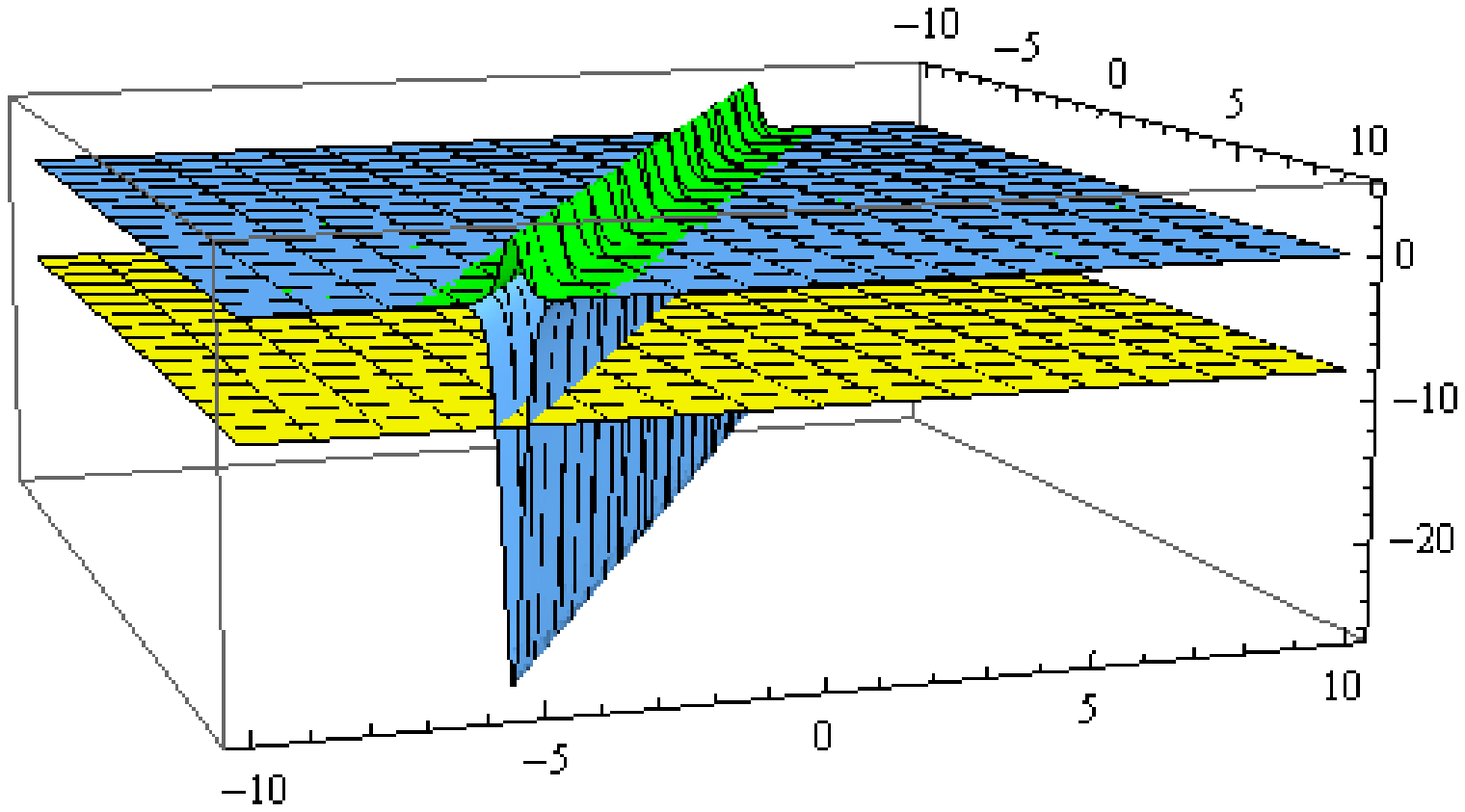}\hfill
b\includegraphics[width=0.50\textwidth,keepaspectratio]{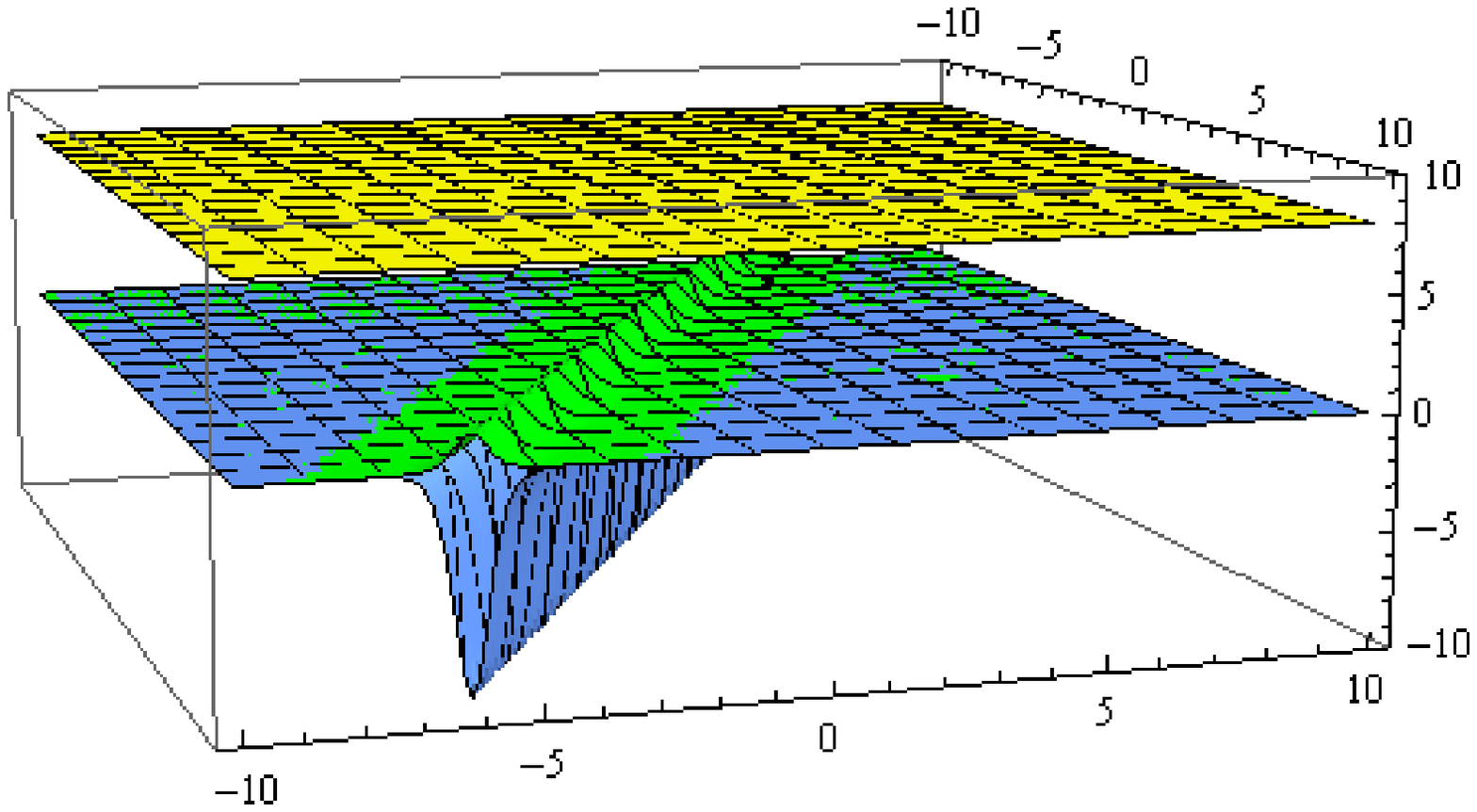}\\
\end{figure}
\begin{figure}[h]
\begin{center}
c\includegraphics[width=0.50\textwidth,keepaspectratio]{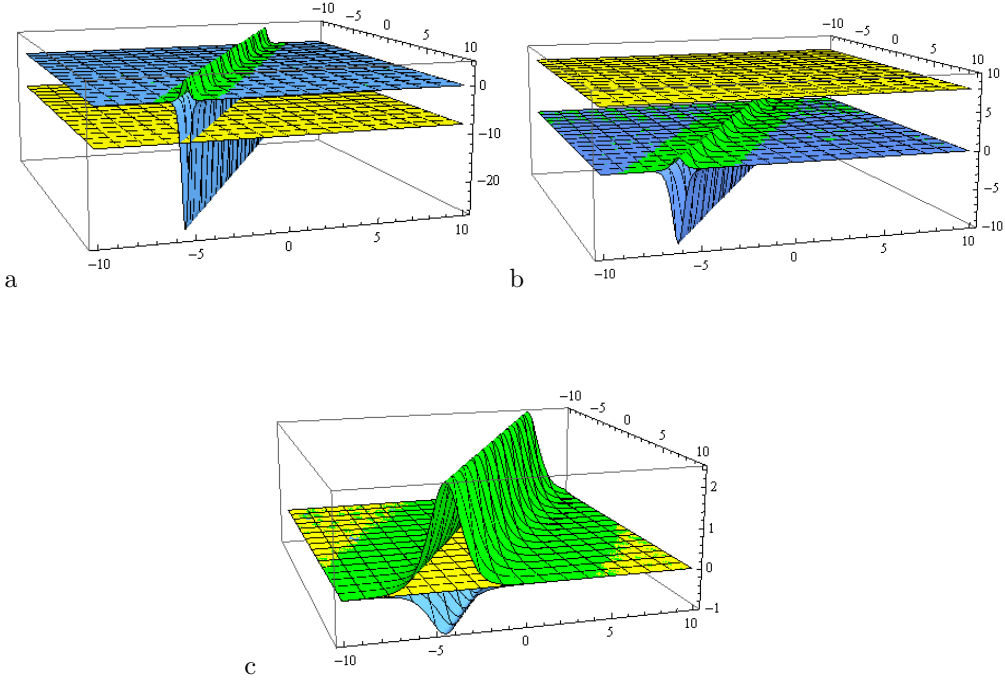}
\fl\parbox[t]{1\textwidth}{\caption{Potential $V_{Shr}$ (\ref{ExactVSchr1pot}) (blue) with the energy level $E$ (yellow) and  corresponding squared absolute values of wave functions $|\psi^{[1,0]}(\mu_1)|^2=|\psi^{[1,0]}(-\lambda_1)|^2$
(\ref{VN solution 1 chi(mu_1)}) (green) with parameters:
a) $a_{10}=-0.1,\lambda_1=e^{i\frac{\pi}{6}},\mu_1=4e^{i\frac{7\pi}{6}},E=-2\epsilon=-8$;
b) $a_{10}=-0.1,\lambda_1=e^{i\frac{\pi}{6}},\mu_1=4e^{i\frac{\pi}{6}},E=-2\epsilon=8$;
c) $a_{10}=0.1,\lambda_1=e^{i\frac{\pi}{6}},\mu_1=0,E=-2\epsilon=0$.}\label{graphVN2One+WF} }
\end{center}
\end{figure}
%
%
%
%
%
%
%
%
The corresponding  wave functions $\psi^{[1,0]}(\mu_1)=\chi^{[1,0]}(\mu_1)e^{F(\mu_1)}$, $\psi^{[1,0]}(-\lambda_1) = \chi^{[1,0]}(-\lambda_1)e^{F(-\lambda_1)}$ and $\psi^{[1,0]}(\lambda) = \chi(\lambda)e^{F(\lambda)}$
of linear auxiliary problems (\ref{NVN dressing auxiliary problems}),(\ref{NVN dressing auxiliary problems2})
and exact potential $V_{Shr}$ of 2D stationary Schr\"{o}dinger equation (\ref{2DSchr})
with energy level $E:=-2\epsilon$ due to (\ref{Psi-Chi}), (\ref{WFN=1Gen})-(\ref{PsiWFNVNN=1,2})
have the forms:
\begin{equation}\label{VN solution 1 chi(mu_1)}
\fl \psi(\mu_1) =
 \frac{e^{F(\mu_1)}}{1 + e^{\varphi_1+\phi_{01}}},\quad
 \psi(-\lambda_1) =
 \frac{e^{-F(\lambda_1)}}{1 + e^{\varphi_1+\phi_{01}}},
\end{equation}
\begin{equation}\label{VN solution 1 chi(lambda)}
\fl  \psi(\lambda) = e^{F(\lambda)} + \Big(\frac{\lambda_1}{\lambda - \lambda_1} + \frac{\mu_1}{\lambda +
  \mu_1}\Big)\frac{2 a_{10} e^{\varphi_1+F(\lambda)}}{1 + e^{\varphi_1+\phi_{01}}};
\end{equation}
\begin{equation}\label{ExactVSchr1pot}
\fl V_{Schr}=-\frac{E(\lambda_1 - \mu_1)^2}{\lambda_1\mu_1}
  \frac{1}{\cosh^2\frac{\varphi_1 + \phi_{01}}{2}}=
  -\frac{|\lambda_1 - \mu_1|^2}{\cosh^2\frac{\varphi_1 + \phi_{01}}{2}}; \quad E=-2\epsilon=2\lambda_1\bar\mu_1.
\end{equation}
\begin{figure}[h]
a\includegraphics[width=0.50\textwidth,keepaspectratio]{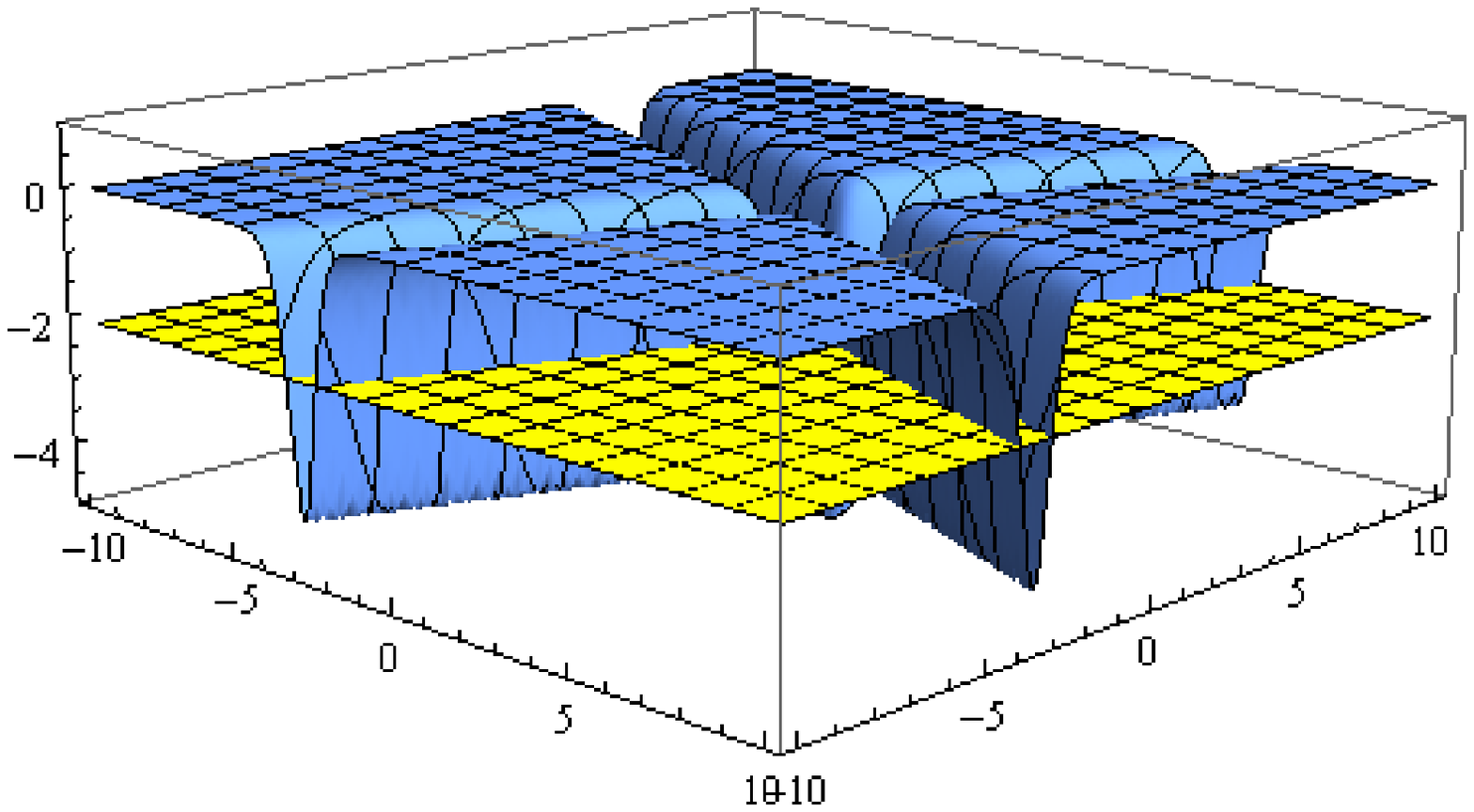}\hfill
b\includegraphics[width=0.50\textwidth,keepaspectratio]{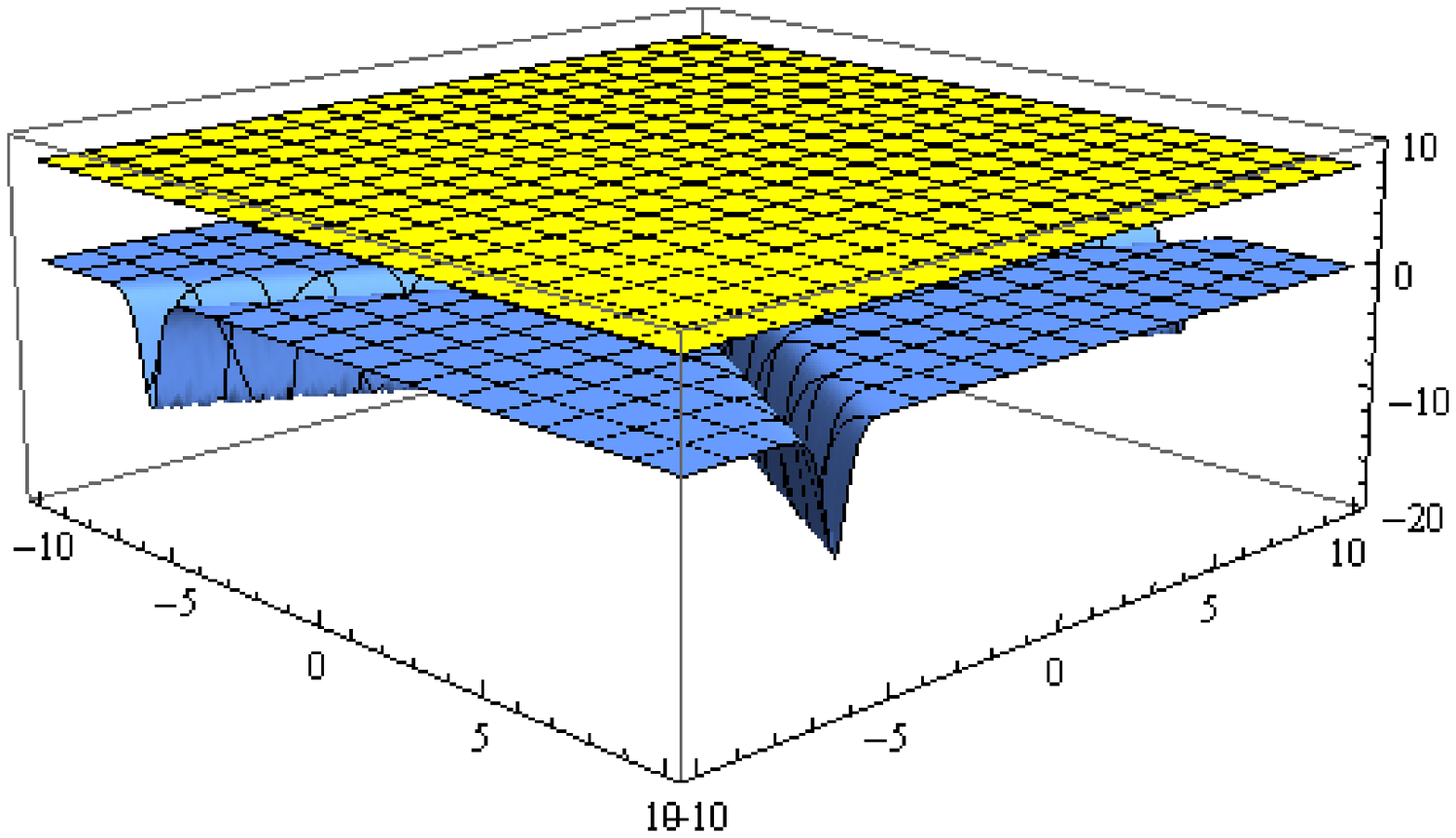}
\end{figure}
\begin{figure}[h]
\begin{center}
c\includegraphics[width=0.50\textwidth,keepaspectratio]{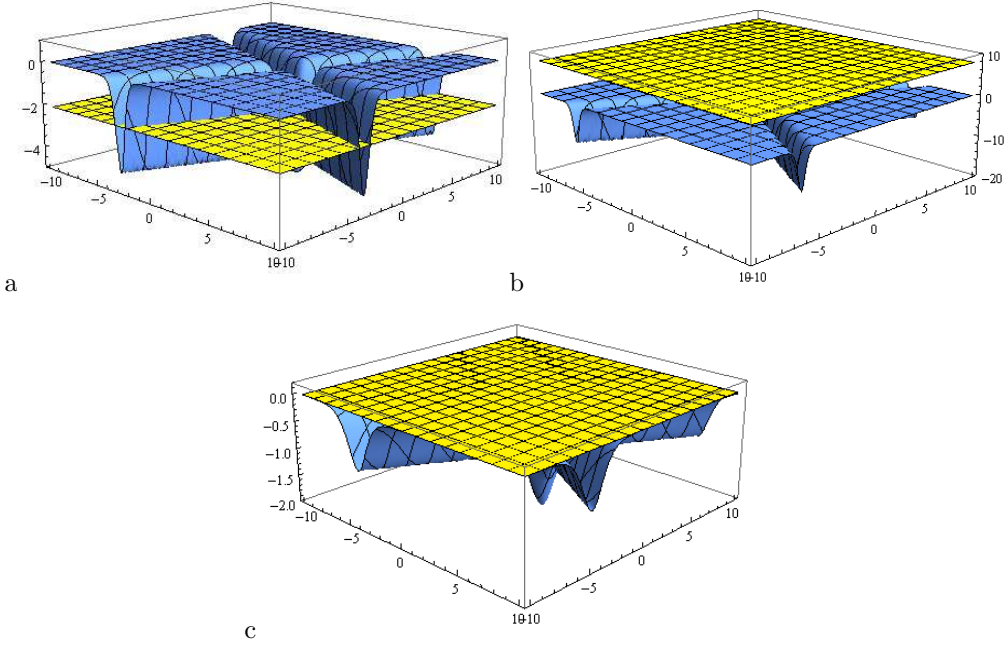}
\fl\parbox[t]{1\textwidth}{\caption{Potential $V_{Shr}$ corresponding two line soliton $[2,0]$ solution (\ref{VN solution 1 u simple}) (blue) with the energy level $E$ (yellow) with parameters:
a) $a_{10}=-1,\lambda_1=e^{i\frac{\pi}{8}},\mu_1=1.05 e^{i\frac{9\pi}{8}}; a_{20}=-1,\tau=1,E=-2\epsilon=-2.1$;
b) $a_{10}=-0.1,\lambda_1=e^{i\frac{\pi}{6}},\mu_1=4 e^{i\frac{\pi}{6}}; a_{20}=-0.1,\tau=1,E=-2\epsilon=8$;
c) $a_{10}=0.1,\lambda_1=e^{i\frac{\pi}{6}},\mu_1=0; a_{20}=0.1,\tau=1,E=-2\epsilon=0$.}\label{graphVN2}}
\end{center}
\end{figure}

Graphs of Schr\"{o}dinger potentials (\ref{ExactVSchr1pot}) (connected with one line $[1,0]$ solitons $V_{Schr}=-2\tilde u$ (\ref{VN solution 1 u})) and squared absolute values of wave functions (\ref{VN solution 1 chi(mu_1)}) for stationary states with  energies $E<0$,\quad  $E>0$ and \quad $E=0$ (equation (\ref{2DSchr}) for particle with mass $m=1$)
for certain values of corresponding parameters are shown in  Fig.\ref{graphVN2One+WF}.
One can prove that two wave functions (\ref{VN solution 1 chi(mu_1)}) for all signs of energy
correspond to  stationary states of a particle with opposite to each other
conserved projections (on direction of valley) of momentum. In all above mentioned stationary states with
 wave functions (\ref{VN solution 1 chi(mu_1)}) particle
is bounded  in transverse  direction to potential valley and moves
 freely along the direction of potential valley.

Two line soliton $[2,0]$ solution  in considered case  of kernel
$R_0$ of the type (\ref{VN kernel possibility 2}) with parameters (\ref{VN_parameters_1})
is given by the formula  (\ref{uN=2Gen}),
%
%
%
%
it is remarkable that under the condition $q=p_1 p_2$  this solution
radically simplifies. Indeed, due to (\ref{EquivSeparatCondit}) condition $q=p_1 p_2$ is satisfied if  $\lambda_1\mu_1+\lambda_2\mu_2=0$ and
in this case two line soliton solution (\ref{uN=2Gen}) takes the form (\ref{uTwoSolGen}):
\begin{eqnarray}\label{VN solution 1 u simple}
\fl  u(z,\bar{z},t)= -\epsilon - \frac{\epsilon(\lambda_1 - \mu_1)^2}{2\lambda_1\mu_1}\frac{1}{\cosh^2{\frac{\varphi_{1}(z,\bar{z},t) + \phi_{01}}{2}}} - \frac{\epsilon(\lambda_2 - \mu_2)^2}{2\lambda_2\mu_2}\frac{1}{\cosh^2{\frac{\varphi_{2}(z,\bar{z},t) + \phi_{02}}{2}}}= \nonumber\\
\fl=-\epsilon + \frac{|\lambda_1 - \mu_1|^2}{2}\frac{1}{\cosh^2{\frac{\varphi_{1}(z,\bar{z},t) + \phi_{01}}{2}}} + \frac{|\lambda_2 - \mu_2|^2}{2}\frac{1}{\cosh^2{\frac{\varphi_{2}(z,\bar{z},t) + \phi_{02}}{2}}}.
\end{eqnarray}

From the relation $\lambda_1\mu_1+\lambda_2\mu_2=0$ taking into
account the first condition (\ref{VN relation points and amplitudes possibility 1}) ($\lambda_1\overline{\mu}_1=\overline{\lambda}_1\mu_1=\lambda_2\overline\mu_2=\overline{\lambda}_2\mu_2=-\epsilon$)
follows $\overline{\mu}_2/\overline{\mu}_1=-\mu_2/\mu_1=\lambda_1/\lambda_2$ and from the last relation one obtains
\begin{equation}\label{VN solution 1 condtion4}
  \mu_2=i\tau{\mu}_1, \quad
  \lambda_2=i\tau^{-1}{\lambda}_1, \quad \tau=\bar\tau
\end{equation}
with arbitrary real constant $\tau$.

Wave functions corresponding to two line soliton $[2,0]$ solution (\ref{VN solution 1 u simple})   in  considered case of kernel $R_0$ of the type (\ref{VN kernel possibility 2}) with parameters (\ref{VN_parameters_1}) and (\ref{[1,2,0]VNSolitons})-(\ref{VN two-solution 1 varphi1}), under condition $p_1p_2=q$, are given by very simple expressions (\ref{NVNWaveFunctN=2Gen})-(\ref{NVNPsiWaveFunctN=2Gen1}).
\begin{figure}[!h]
a\includegraphics[width=0.50\textwidth,keepaspectratio]{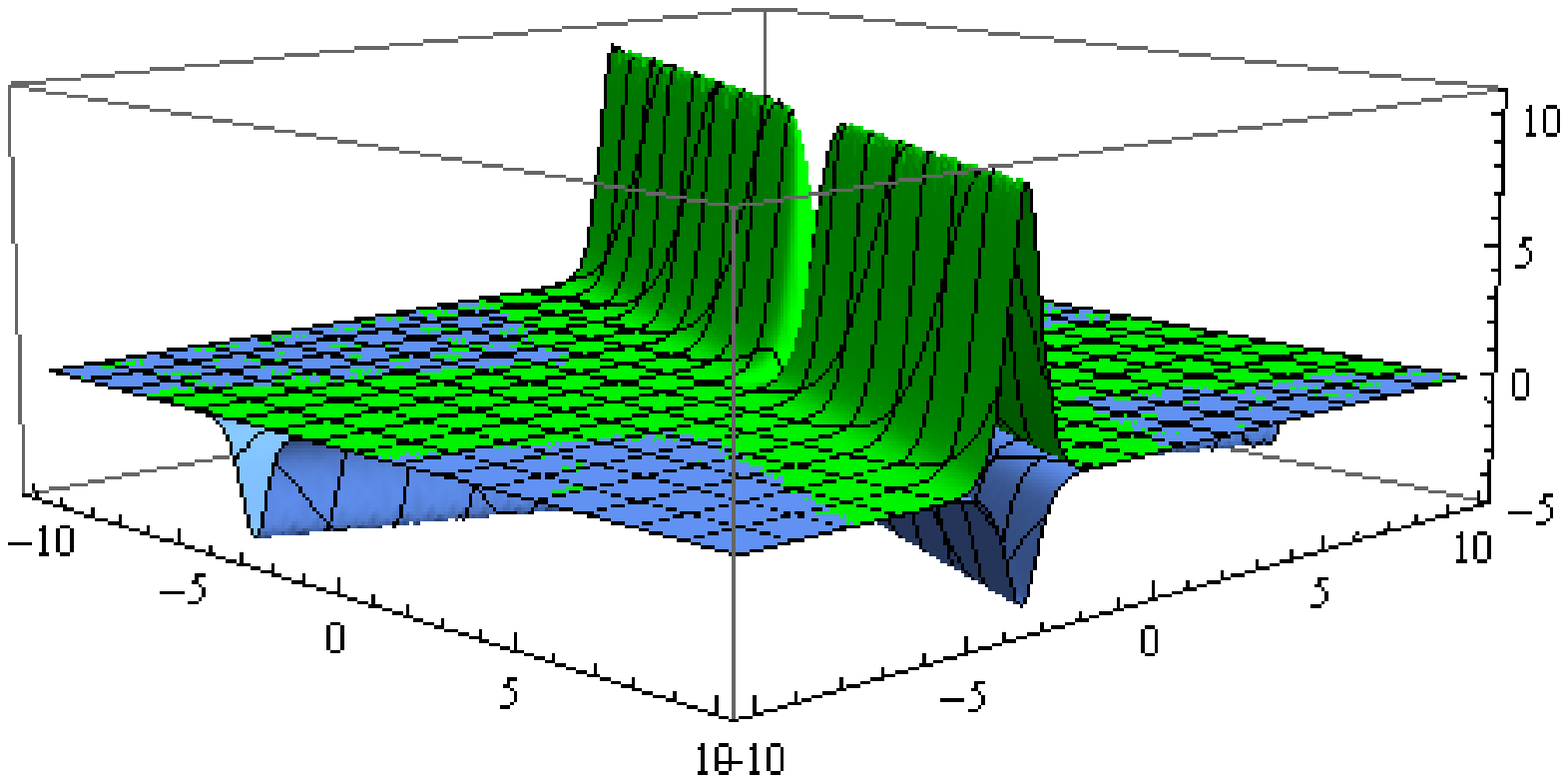}\hfill
b\includegraphics[width=0.45\textwidth,keepaspectratio]{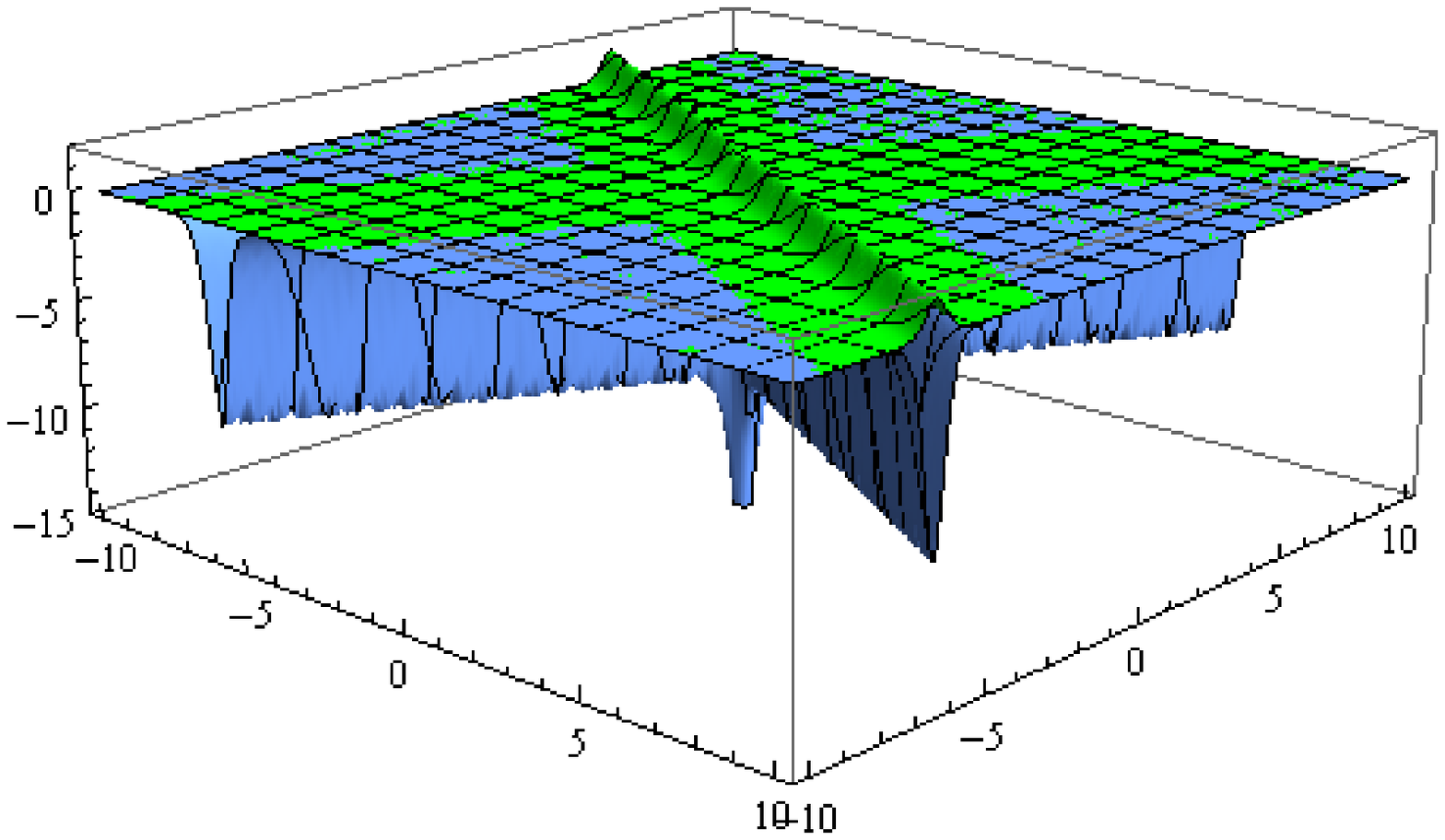}
\end{figure}
\begin{figure}[!h]
\begin{center}
c\includegraphics[width=0.50\textwidth,keepaspectratio]{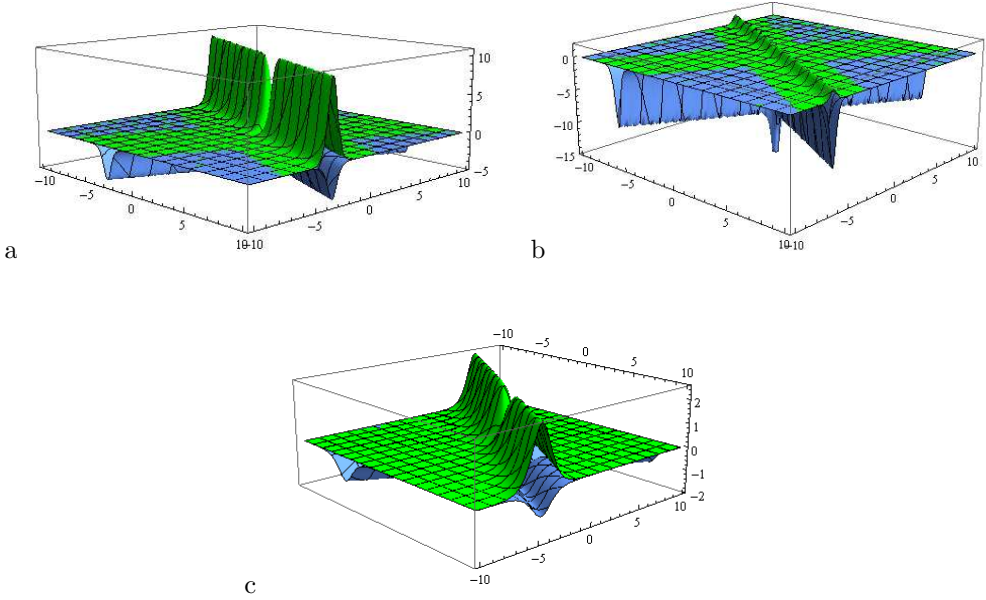}
\fl\parbox[t]{1\textwidth}{\caption{Squared absolute values of wave functions $|\psi^{[2,0]}(\mu_1)|^2=|\psi^{[2,0]}(-\lambda_1)|^2$ (green) corresponding to different values of energy $E$ in the Fig.\ref{graphVN2}(a,b,c).}\label{graphVN2+WF}}
\end{center}
\end{figure}

Graphs of Schr\"{o}dinger  potentials
(connected with two line $[2,0]$ soliton $V_{Schr}=-2 \tilde u$ solutions (\ref{VN solution 1 u simple}))
and squared absolute values $|\psi^{[2,0]}(\mu_1)|^2=|\psi^{[2,0]}(-\lambda_1)|^2$ of some wave functions from
(\ref{NVNPsiDWaveFunctN=2Gen1})-(\ref{NVNPsiDWaveFunctN=2Gen4})
for certain values of  parameters are shown in  Fig.\ref{graphVN2} and Fig.\ref{graphVN2+WF} (graphs of $|\psi^{[2,0]}(\mu_2)|^2=|\psi^{[2,0]}(-\lambda_2)|^2$ are similar to graphs of $|\psi^{[2,0]}(\mu_1)|^2=|\psi^{[2,0]}(-\lambda_1)|^2$ but with localization along another soliton valley).

Calculated via ${\overline{\partial}}$-dressing method wave functions
(\ref{NVNPsiDWaveFunctN=2Gen1})-(\ref{NVNPsiDWaveFunctN=2Gen4})
at discrete values of spectral parameters correspond to  possible physical basis
states of particle localized in the field of two potential valleys.
${\overline{\partial}}$-dressing in  present paper is carried out for the fixed nonzero value of parameter $\epsilon$ or, in  context of  present section, for nonzero energy
$E\neq  0$. Nevertheless one can correctly  consider the limit $\epsilon \rightarrow 0$ in all derived formulas and obtain some interesting results also for the case of zero energy $E=-2\epsilon=0$. Limiting procedure  $E=-2\epsilon=\mu_k\bar\lambda_k+\bar\mu_k\lambda_k \rightarrow 0, (k=1,2)$  can be correctly performed by the following settings in all required formulas:  $\epsilon \rightarrow 0$ and $\mu_k \rightarrow 0$ in cases when uncertainty is absent, but ${\frac{\epsilon}{\mu_k}}\rightarrow -\bar\lambda_k$ in accordance with the relation $\epsilon=-\mu_k\bar\lambda_k$; in addition the formula $\lambda_2=i\tau^{-1}\lambda_1$ (\ref{VN solution 1 condtion4}) (followed from the relations
$\bar\mu_k\lambda_k=\mu_k\bar\lambda_k$ and $\mu_1\lambda_1+\mu_2\lambda_2=0$) with arbitrary real constant $\tau$  is assumed to be valid. The two line soliton solution due to (\ref{VN solution 1 u simple}) in considered limit has the form:
\begin{equation}\label{VSchrE0Limit}
u=\frac{|\lambda_1|^2}{2\cosh^2{\frac{\varphi_{1}(z,\bar{z}) + \phi_{01}}{2}}} +\frac{|\lambda_2|^2}{2\cosh^2{\frac{\varphi_{2}(z,\bar{z}) + \phi_{02}}{2}}},
\end{equation}
 the phases  $\varphi_{k}(z,\bar{z})$ and $\phi_{0k}$ due to (\ref{F_formula}),(\ref{VN two-solution 1 varphi1}),(\ref{p_kVNElliptic1}) have in considered limit the forms:
\begin{equation}\label{PhazesE0Limit}
\fl\varphi_k(z,\bar{z},t) = -i\big(\lambda_kz -  \overline{\lambda}_k\overline{z} +
\kappa \lambda_{k}^{3}t - \overline{\kappa} \overline{\lambda}_{k}^{3}t\big),\quad
\phi_{0k}= \ln{a_{k0}}.
\end{equation}
One can check by direct substitution that NVN-I equation (\ref{NVN})
with $\sigma=i$ satisfies by $u=\tilde u=-V_{Schr}/2$
given by (\ref{VSchrE0Limit}), but it also satisfies by each item
\begin{equation}\label{uE0LimitSols}
u^{(k)}={\frac{|\lambda_k|^2}{2\cosh^2{\frac{\varphi_{k}(z,\bar{z}) + \phi_{0k}}{2}}}}, \quad (k=1,2)
\end{equation}
of the sum (\ref{VSchrE0Limit}). Thus, in considered case the linear principle
of superposition $u=u^{(1)}+u^{(2)}$ for such special solutions $u^{(1)},u^{(2)}$  (\ref{uE0LimitSols}) is valid. One can show using (\ref{VN solution 1 condtion4}),(\ref{PhazesE0Limit})  that line solitons $u^{(1)}$ and $u^{(2)}$ are propagate in the plane $(x,y)$ in perpendicular to each other directions. Schr\"{o}dinger potentials $V_{Schr}$ (of the types [1,0] and [2,0]) with corresponding squared absolute value wave functions of zero energy limit $E=0$ are also pictured by graphs of  Fig.\ref{graphVN2One+WF}, Fig.\ref{graphVN2} and Fig.\ref{graphVN2+WF}.

{\bf{5.2 $[0,1],[0,2]$ line solitons}}

 The kernels of  type $R_0$
(\ref{VN kernel possibility 2}) with values $L=0;\quad N=1,2$ (i. e. $a_{l0}= 0,l=1,..,L; a'_{n0}\neq 0, n=1,2$) in (\ref{VN_parameters_1})   correspond to  $[0,1]$, $[0,2]$ line solitons.
For nonsingular one line $[0,1]$ and two line $[0,2]$ soliton solutions of elliptic version of NVN equation
parameters $a_k,\mu_k,\lambda_k$ in general formulas (\ref{BI det A, N=1})-(\ref{NVNPsiWaveFunctN=2Gen1}) of Section 3
must be identified due to (\ref{VN_parameters_1}) by the following way:
\begin{equation}\label{[0,2]VNparameters}
a_k = \overline{a}_k := a_{k0}, \quad \epsilon = |\mu_k|^2=|\lambda_k|^2, \quad (k=1,2).
\end{equation}
 Real parameters $p_k$ due to (\ref{p_kDeltaFAndq}), (\ref{VN_parameters_1}) and (\ref{[0,2]VNparameters})
\begin{equation}\label{p_kVNellipt2}
\fl p_k= ia_{k0}\frac{\mu_{k} + \lambda_{k}}{\mu_{k} -\lambda_{k} } = a_{k0}\cot\frac{\delta_k}{2}:= e^{\phi_{0k}} > 0,
\quad \mu_k:=\lambda_k e^{i\delta_k}, \quad (k=1,2)
\end{equation}
appear as positive constants. The real phases $\Delta F(\mu_k,\lambda_k)=F(\mu_k)-F(\lambda_k):=\varphi_k, (k=1,2)$ are given
in considered case by the expressions:
\begin{equation}\label{VN two-solution 1 varphi2}
\fl  \varphi_k(z,\bar{z},t) = i[(\mu_k - \lambda_k)z - (\overline{\mu}_k - \overline{\lambda}_k)\overline{z} +
\kappa(\mu_{k}^{3} - \lambda_{k}^{3})t - \overline{\kappa}(\overline{\mu}_{k}^{3} - \overline{\lambda}_{k}^{3})t].
\end{equation}
One line soliton $[0,1]$ solution  corresponding to simplest kernel $R_0$ of the type (\ref{VN kernel possibility 2}) with parameters (\ref{VN_parameters_1}) due to (\ref{uN=1Gen})
and (\ref{[0,2]VNparameters})-(\ref{p_kVNellipt2}) and (\ref{VN two-solution 1 varphi2}) is nonsingular line soliton:
\begin{equation}\label{VN solution 2 u}
  u =-\epsilon + \frac{|\lambda_1 - \mu_1|^2}{2}
  \frac{1}{\cosh^2\frac{\varphi_1 + \phi_{01}}{2}} =-\epsilon +
  \frac{2\epsilon\sin^2\frac{\delta_1}{2}}{\cosh^2\frac{\varphi_1 + \phi_{01}}{2}}.
\end{equation}
%
%
%
%
%
%
%
%
The corresponding  wave functions $\psi^{[0,1]}(\mu_1)=\chi^{[0,1]}(\mu_1)e^{F(\mu_1)}$, $\psi^{[0,1]}(-\lambda_1) = \chi^{[0,1]}(-\lambda_1)e^{F(-\lambda_1)}$ and $\psi^{[0,1]}(\lambda) = \chi^{[0,1]}(\lambda)e^{F(\lambda)}$
of linear auxiliary problems (\ref{NVN dressing auxiliary problems}),(\ref{NVN dressing auxiliary problems2})
and exact potential $V_{Shr}$ of 2D stationary Schr\"{o}dinger equation (\ref{2DSchr})
with energy level $E:=-2\epsilon$ have forms:
\begin{equation}\label{VN solution 2 chi(mu_1)}
  \fl \psi(\mu_1) = \frac{e^{F(\mu_1)}}{1 + e^{\varphi_1+ \phi_{01}}},\quad
  \psi(-\lambda_1)= \frac{e^{-F(\lambda_1)}}{1 + e^{\varphi_1+ \phi_{01}}},
\end{equation}
\begin{equation}\label{VN solution 2 chi(lambda)}
 \fl  \psi(\lambda) = e^{F(\lambda)} - \Big(\frac{\lambda_1}{\lambda - \lambda_1} + \frac{\mu_1}{\lambda + \mu_1}\Big)\frac{2ia_{10} e^{\varphi_1+F(\lambda)}}{1 +
  e^{\varphi_1+ \phi_{01}}};
  \end{equation}
 \begin{equation}\label{Schr01Pot}
 \fl V_{Schr}=-\frac{|\lambda_1 - \mu_1|^2}{\cosh^2\frac{\varphi_1 + \phi_{01}}{2}} =
  -\frac{4\epsilon\sin^2(\frac{\delta_1}{2})}{\cosh^2\frac{\varphi_1 + \phi_{01}}{2}};  E=-2\epsilon=-2|\lambda_1|^2=-2|\mu_1|^2.
 \end{equation}
%
%
%
%
\begin{figure}[h]
a\includegraphics[width=0.50\textwidth,keepaspectratio]{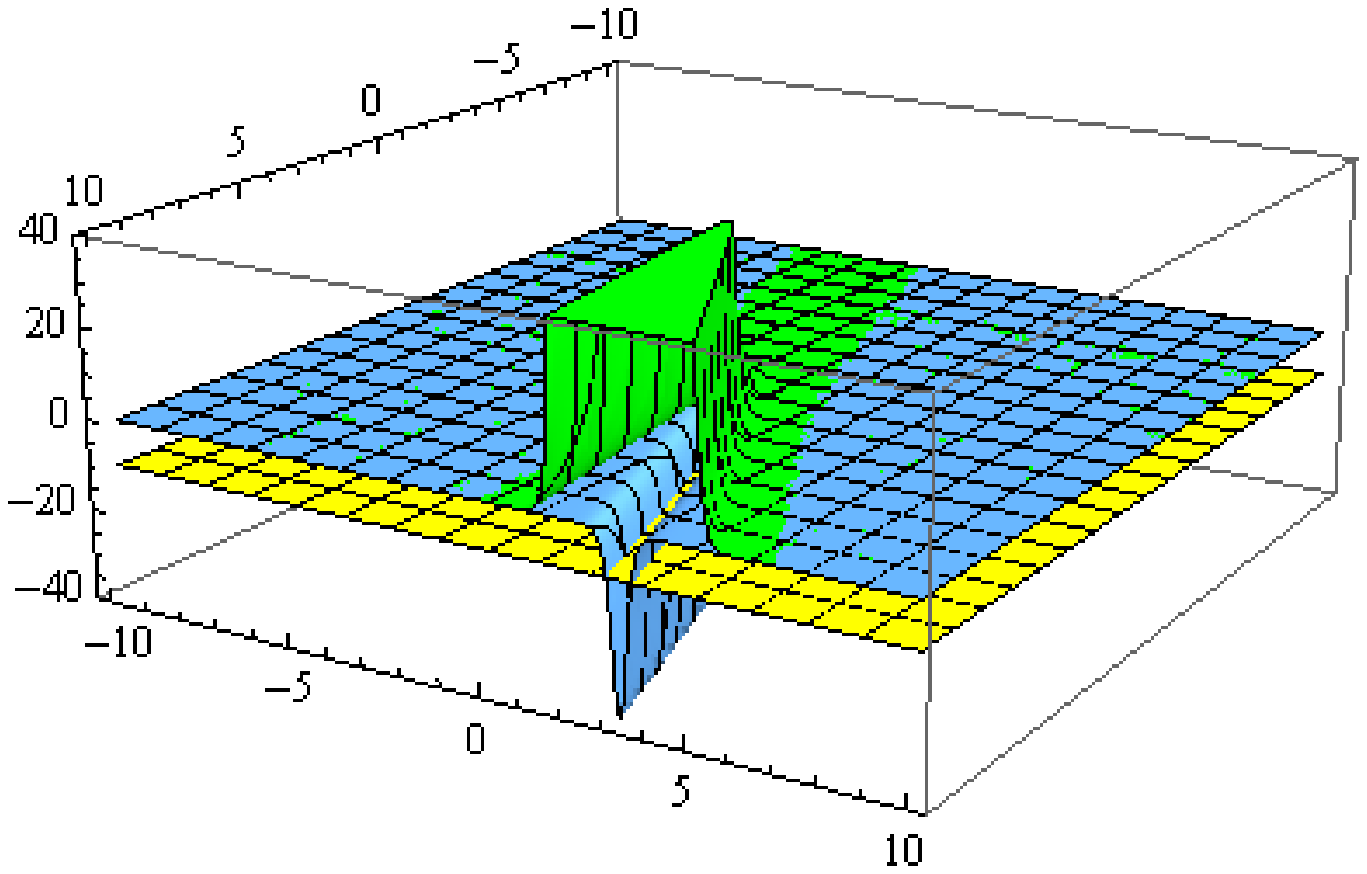}\hfill
b\includegraphics[width=0.50\textwidth,keepaspectratio]{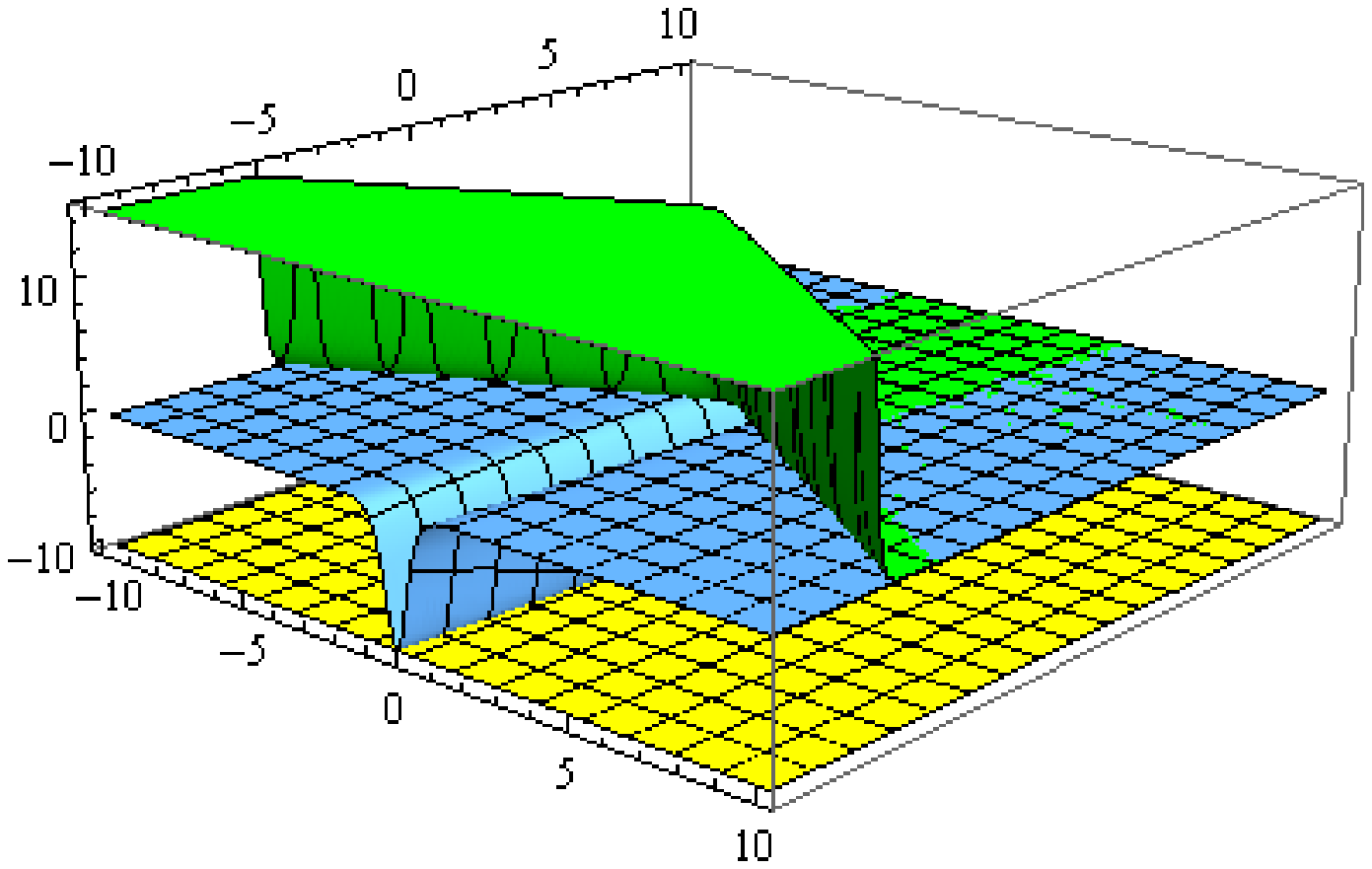}\\
\end{figure}
\begin{figure}[h]
\begin{center}\fl\parbox[t]{1\textwidth}{\caption{Potential $V_{Shr}$ (\ref{VN solution 2 u}) (blue) with the energy level $E$ (yellow) and corresponding squared absolute value of wave function $|\psi^{[0,1]}(\mu_1)|^2$
(\ref{VN solution 2 chi(mu_1)}) (green) with parameters:
a) $a_{10}=-1,\lambda=2-i,\delta=\frac{10\pi}{9},E=-2\epsilon=-10$;
b) $a_{10}=-1,\lambda=2-i,\delta=\frac{\pi}{3},E=-2\epsilon=-10$.}\label{VNSoliton01}}
\end{center}
\end{figure}
%
Graphs of Schr\"{o}dinger  potential $V_{Schr}$ (\ref{Schr01Pot})
(connected with one line $[0,1]$ soliton $V_{Schr}=-2 \tilde u$ solution (\ref{VN solution 2 u}))
and the squared absolute value of wave function $\psi^{[0,1]}(\mu_1)$ from
(\ref{VN solution 2 chi(mu_1)}) for certain values of  parameters are shown in Fig.\ref{VNSoliton01}: a) $(V_{Shr})_{min}<E<0$, b) $(V_{Shr})_{min}=E<0$ (the squared absolute value  $|\psi^{[0,1]}(-\lambda_1)|^2$ has the similar form but with localization along another one half of potential valley).

Two line soliton $[0,2]$ solution  in considered case  of kernel kernel $R_0$ of the type (\ref{VN kernel possibility 2}) with parameters (\ref{VN_parameters_1}) and (\ref{[0,2]VNparameters}),(\ref{p_kVNellipt2}) and (\ref{VN two-solution 1 varphi2}) is given by the formula  (\ref{uN=2Gen}). It is remarkable that under the condition $q=p_1 p_2$  this solution
radically simplifies. Indeed, due to (\ref{EquivSeparatCondit})
condition $q=p_1 p_2$ is satisfied if  $\lambda_1\mu_1+\lambda_2\mu_2=0$,
in this case two line soliton solution (\ref{uN=2Gen}) takes the form (\ref{uTwoSolGen}):
\begin{eqnarray}\label{VN solution 2 u simple}
\fl  u(z,\bar{z},t)= -\epsilon + {\frac{|\lambda_1 - \mu_1|^2}{2}}\frac{1}{\cosh^2{\frac{\varphi_{1}(z,\bar{z},t)+
  \phi_{01}}{2}}}+\frac{|\lambda_1 - \mu_1|^2}{2}\frac{1}{\cosh^2{\frac{\varphi_{2}(z,\bar{z},t),
  +\phi_{02}}{2}}}=\nonumber \\
 \fl = -\epsilon + \frac{2\epsilon\sin^2{\frac{\delta_{1}}{2}}}{\cosh^2{\frac{\varphi_{1}(z,\bar{z},t)+
  \phi_{01}}{2}}}+\frac{2\epsilon\sin^2{\frac{\delta_{2}}{2}}}{\cosh^2{\frac{\varphi_{2}(z,\bar{z},t)
  +\phi_{02}}{2}}}, \quad  \mu_k=\lambda_k e^{i\delta_k}, \epsilon=|\lambda_k|^2=|\mu_k|^2.
\end{eqnarray}
\begin{figure}[h!]
a\includegraphics[width=0.50\textwidth,keepaspectratio]{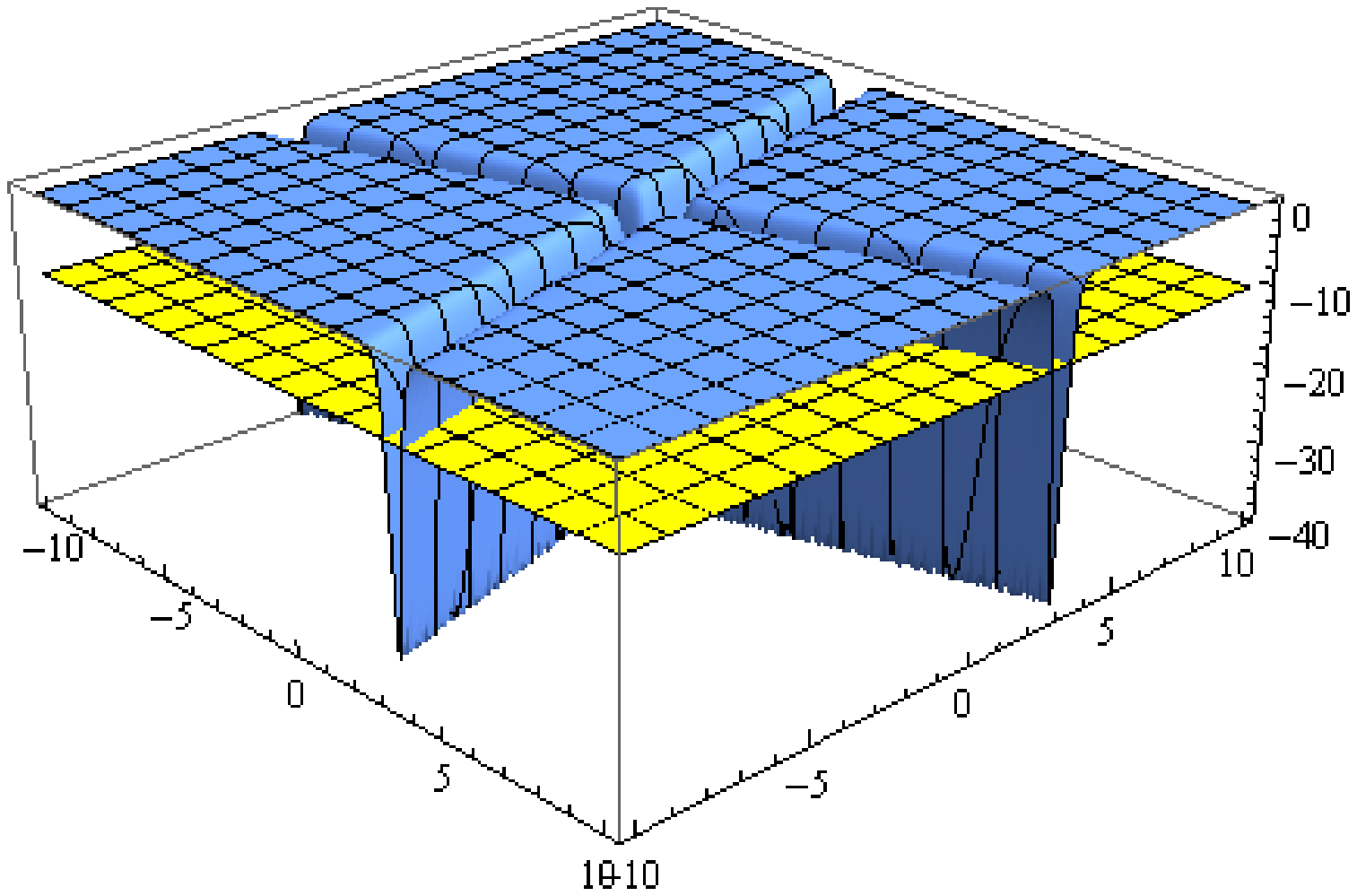}\hfill
b\includegraphics[width=0.50\textwidth,keepaspectratio]{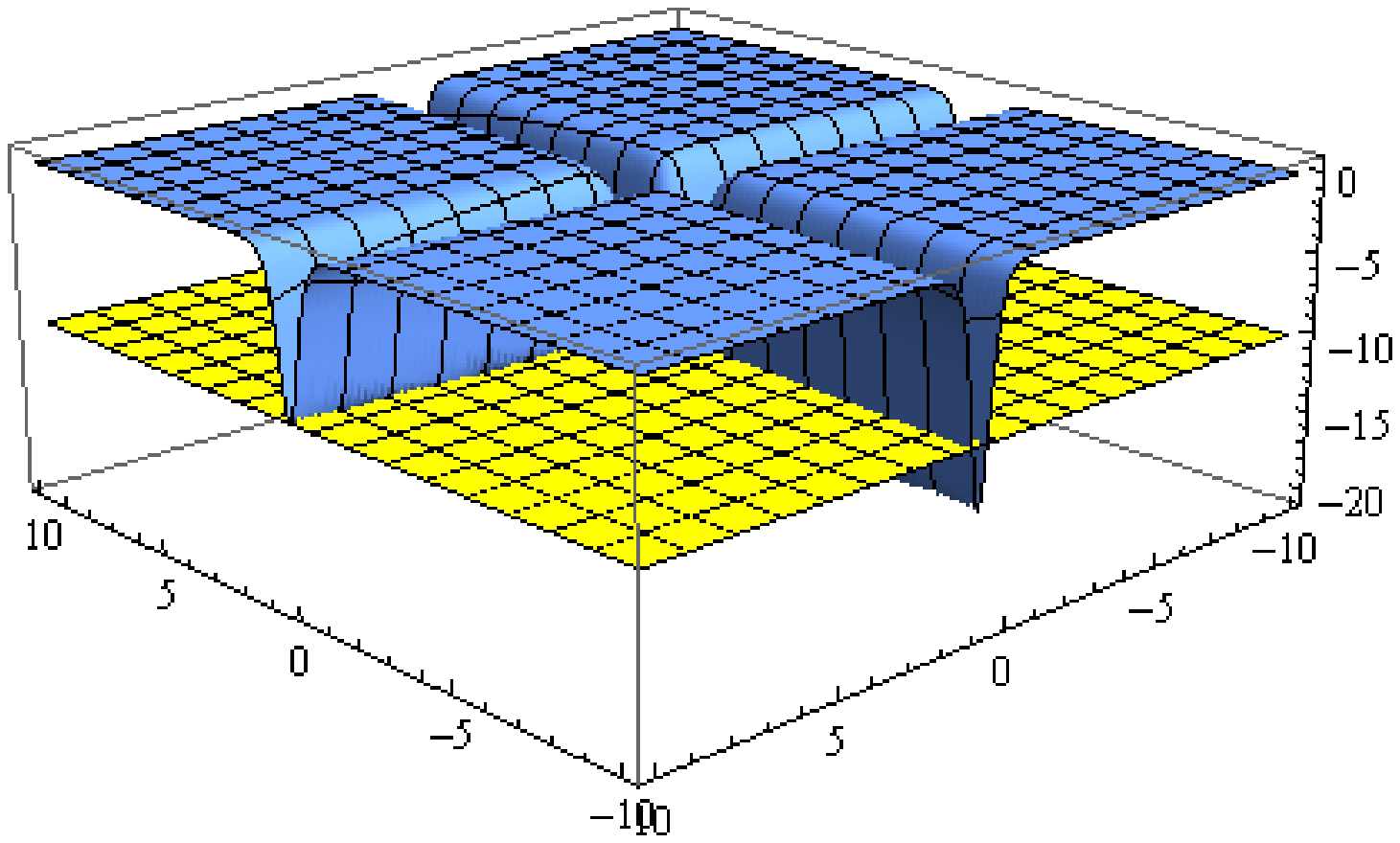}\\
\end{figure}
\begin{figure}[h!]
\begin{center}
\fl\parbox[t]{1\textwidth}{\caption{Potential $V_{Shr}$ corresponding two line soliton $[0,2]$ solution (\ref{VN solution 2 u simple})(blue) with the energy level $E$(yellow) with parameters:
(a) $a_{10}=-1,\lambda_1=2-i,\delta_1=\frac{10\pi}{9};a_{20}=1,\delta_2=\frac{3\pi}{10},E=-2\epsilon=-10$;
(b) $a_{10}=1,\lambda_1=2-i,\delta_1=\frac{\pi}{3};a_{20}=1,\delta_2=\frac{3\pi}{5},E=-2\epsilon=-10$.}\label{graphVN1}}
\end{center}
\end{figure}

The corresponding to two line soliton solution (\ref{VN solution 2 u simple}) wave functions  in  considered case of kernel $R_0$ of the type (\ref{VN kernel possibility 2}) with parameters (\ref{VN_parameters_1}) and (\ref{[0,2]VNparameters}),(\ref{p_kVNellipt2}) and (\ref{VN two-solution 1 varphi2}),
under condition $p_1p_2=q$, are given by very simple expressions (\ref{NVNWaveFunctN=2Gen})-(\ref{NVNPsiWaveFunctN=2Gen1}).
\begin{figure}[!h]
a\includegraphics[width=0.50\textwidth,keepaspectratio]{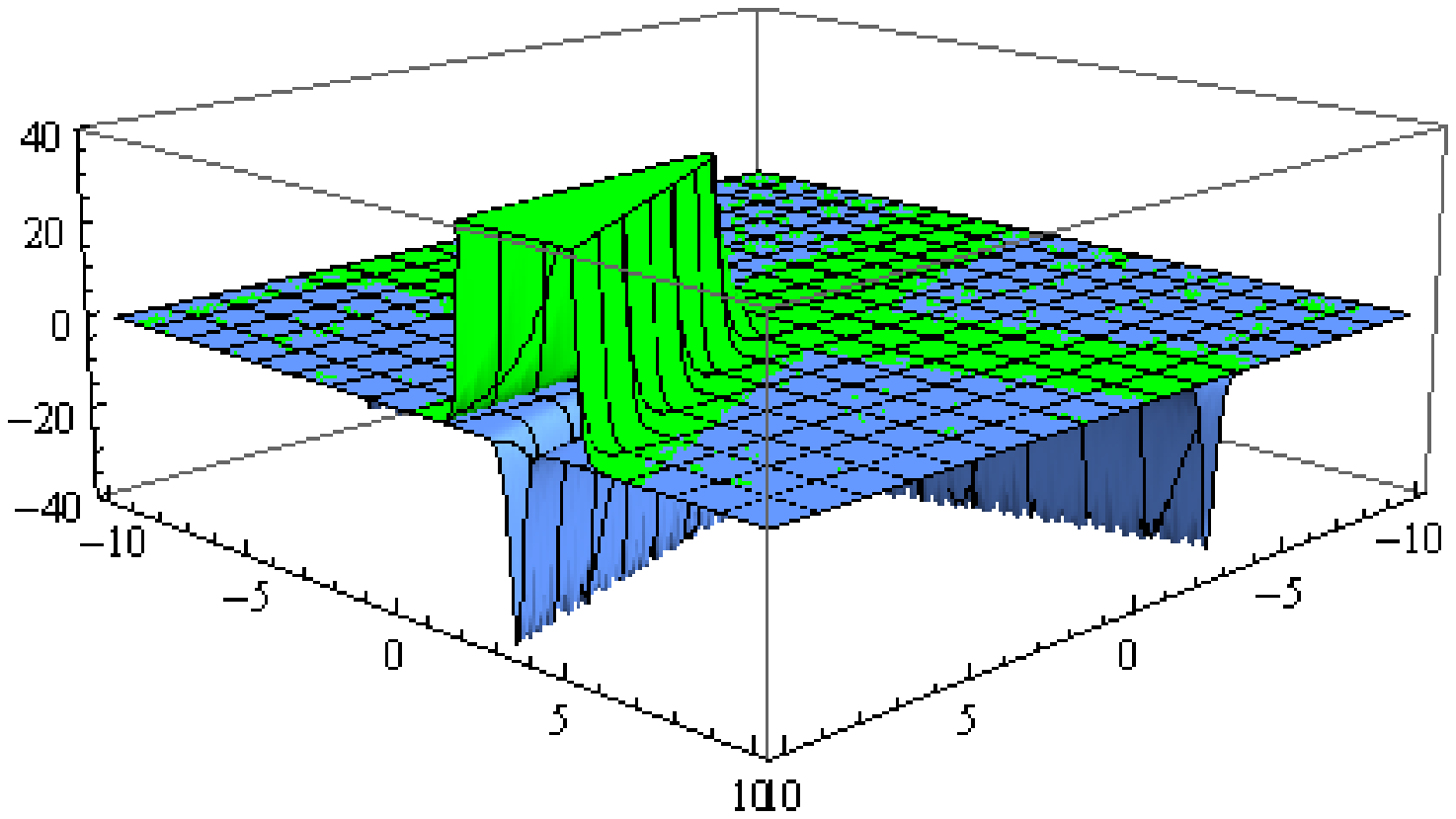}\hfill
b\includegraphics[width=0.50\textwidth,keepaspectratio]{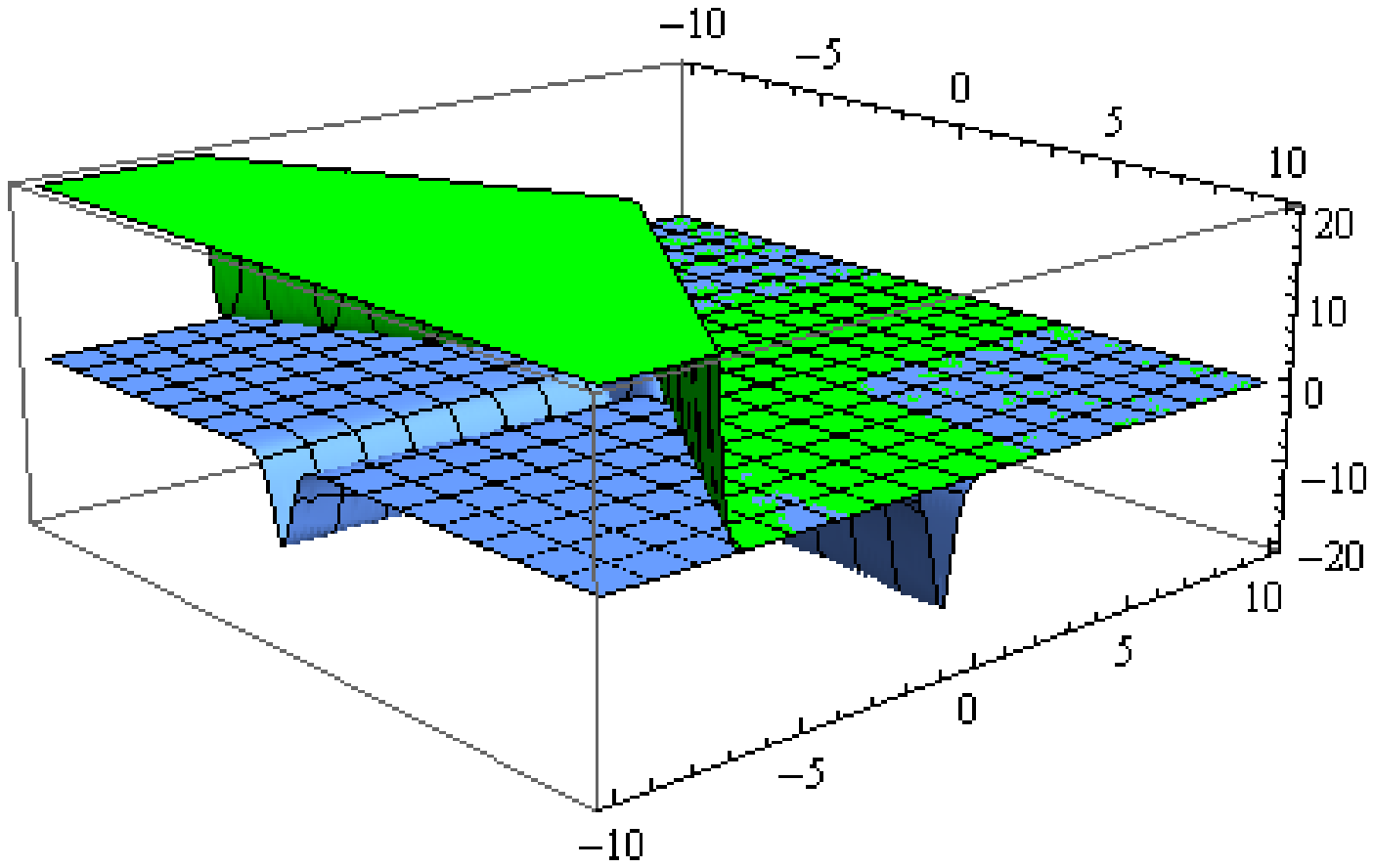}
\end{figure}
\begin{figure}[!h]
\begin{center}
\fl\parbox[t]{1\textwidth}{\caption{Squared absolute value of wave function $|\psi^{[0,2]}(\mu_1)|^2$ (green) for the different types of crossings of potentials valleys by energy planes in the Fig.\ref{graphVN1} (a,b).}\label{graphVN1+WF}}
\end{center}
\end{figure}

Graphs of Schr\"{o}dinger potentials (connected with two line $[0,2]$ solitons $V_{Schr}=-2\tilde u$ (\ref{VN solution 2 u simple})) and squared absolute value $|\psi^{[0,2]}(\mu_1)|^2$ of one wave function from four linear independent partners (\ref{NVNPsiDWaveFunctN=2Gen1})-(\ref{NVNPsiDWaveFunctN=2Gen4})
for certain values of parameters are shown in Fig.\ref{graphVN1} and Fig.\ref{graphVN1+WF} (the squared absolute values of other wave functions have the similar forms but with localization along another three possible halves of two potential valleys).
In all considered in the present section cases of one line [0,1] and two line [0,2] solitons $u=\tilde u-\epsilon$ and
Schr\"{o}dinger potentials $V_{Schr}=-2\tilde u$  corresponding wave functions (Fig.\ref{VNSoliton01}, Fig.\ref{graphVN1+WF}) are not bounded.

{\bf {5.3 \quad $[1,1]$ line soliton}}

The kernel of  type $R_0$ (\ref{VN kernel possibility 2}) with values $L=1;\quad N=1$ (i. e. $a_{10}= 1; a'_{10}= 1$) in (\ref{VN_parameters_1})   correspond to  $[1,1]$ line soliton.
For this soliton solution parameters $a_k,\mu_k,\lambda_k$ in general formulas (\ref{BI det A, N=1})-(\ref{NVNPsiWaveFunctN=2Gen1}) of Section 3
must be identified due to (\ref{VN_parameters_1}) by the following way:
\begin{eqnarray}\label{[1,1]VNparameters}
a_1 = -\overline{a}_1:= ia_{10},\quad  \epsilon = -\mu_1\overline{\lambda}_1\nonumber\\
a_2= a'_1 = \overline{a}'_1:= a'_{10},\: \mu_2=\mu'_1,\:\lambda_2=\lambda'_1,\quad \epsilon = |\mu'_1|^2=|\lambda'_1|^2 .
\end{eqnarray}
$a_2$, $\lambda_2$, $\mu_2$ in formulas (\ref{uN=2Gen})-(\ref{NVNPsiWaveFunctN=2Gen1}) due (\ref{[1,1]VNparameters}) must be identified with $a'_1$, $\lambda'_1$, $\mu'_1$ in (\ref{VN kernel possibility 2}).
Real parameters $p_1$, $p_2$ due to (\ref{p_kDeltaFAndq}), (\ref{VN_parameters_1}) and (\ref{[1,1]VNparameters})
\begin{equation}\label{p_kVNellipt[1,1]}
\fl p_1= -a_{10}\frac{\mu_1+\lambda_1}{\mu_1-\lambda_1}: = e^{\phi_{01}} > 0,\quad
p_2= ia_{20}\frac{\mu_{2} + \lambda_{2}}{\mu_{2} -\lambda_{2}} = a_{20}\cot\frac{\delta_2}{2}:= e^{\phi_{02}} > 0,
\end{equation}
appear as positive constants.

Two line soliton $[1,1]$ solution  in considered case  of kernel kernel $R_0$ of the type (\ref{VN kernel possibility 2}) with parameters (\ref{q_GF}) and (\ref{[1,1]VNparameters}),(\ref{p_kVNellipt[1,1]}) is given by the formula  (\ref{uN=2Gen}). It is remarkable that under the condition $q=p_1 p_2$  this solution
radically simplifies. Indeed, due to (\ref{EquivSeparatCondit})
condition $q=p_1 p_2$ is satisfied if  $\lambda_1\mu_1+\lambda_2\mu_2=0$,
in this case two line soliton solution (\ref{uN=2Gen}) takes the form (\ref{uTwoSolGen}:
\begin{equation}\label{VN solution[1,1] 2 u simple}
\fl  u(z,\bar{z},t)= -\epsilon +\frac{|\lambda_1 - \mu_1|^2}{2}\frac{1}{\cosh^2{\frac{\varphi_{1}(z,\bar{z},t) + \phi_{01}}{2}}}
+\frac{|\lambda_2 - \mu_2|^2}{2}\frac{1}{\cosh^2{\frac{\varphi_{2}(z,\bar{z},t)+\phi_{02}}{2}}}
\end{equation}
where $|\lambda_2|^2=|\mu_2|^2=-\mu_1\overline{\lambda}_1=-\overline{\mu}_1\lambda_1=\epsilon$ and the phases $\varphi_{1}$, $\varphi_{2}$ are given by formulas (\ref{VN two-solution 1 varphi1}),(\ref{VN two-solution 1 varphi2}).
Graphs of Schr\"{o}dinger potentials  (connected with two line $[1,1]$ solitons $V_{Schr}=-2\tilde u$ (\ref{VN solution[1,1] 2 u simple})) and squared absolute values of some wave functions from (\ref{NVNPsiDWaveFunctN=2Gen1})-(\ref{NVNPsiDWaveFunctN=2Gen4})
for certain values of parameters are shown in  Fig.\ref{VNSoliton[1,1]E-} and Fig.\ref{VNSoliton[1,1]E-WFs} (graphs of $|\psi^{[1,1]}(-\lambda_{2})|^2$ and $|\psi^{[1,1]}(\mu_{2})|^2$ are similar to each other but with localization along two different halves of corresponding potential valley).
\begin{figure}[h]
\begin{center}
\includegraphics[width=0.50\textwidth,keepaspectratio]{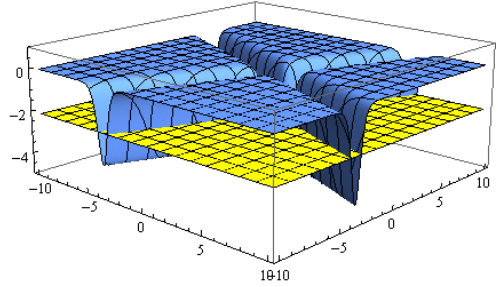}
\fl\parbox[t]{1\textwidth}{\caption{Potential $V_{Shr}$ corresponding two line soliton $[1,1]$ solution (\ref{uN=2Gen})(blue) and energy level E (yellow) with parameters $a_{1}=-1,\lambda_{1}=1e^{\frac{\pi}{8}},\mu_{1}=1.05 e^{\frac{9\pi}{8}};a_{2}=-1,\lambda_{2}=1.0247e^{\frac{\pi}{2}},\mu_{2}=1.0247 e^{\frac{7\pi}{4}},E=-2\epsilon=-2.1$.}\label{VNSoliton[1,1]E-}}
\end{center}
\end{figure}
\begin{figure}[h]
a\includegraphics[width=0.50\textwidth,keepaspectratio]{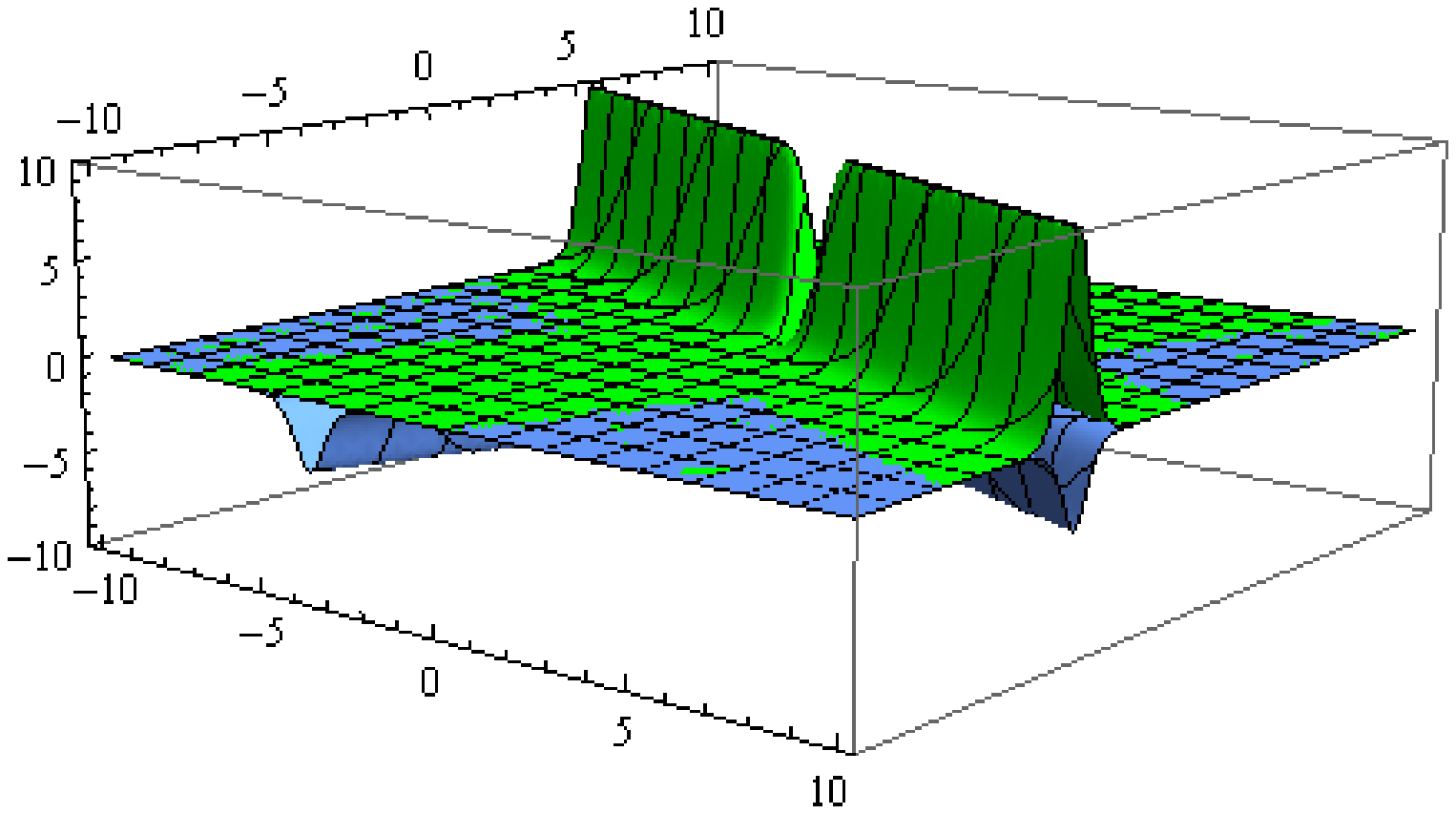}\hfill
b\includegraphics[width=0.52\textwidth,keepaspectratio]{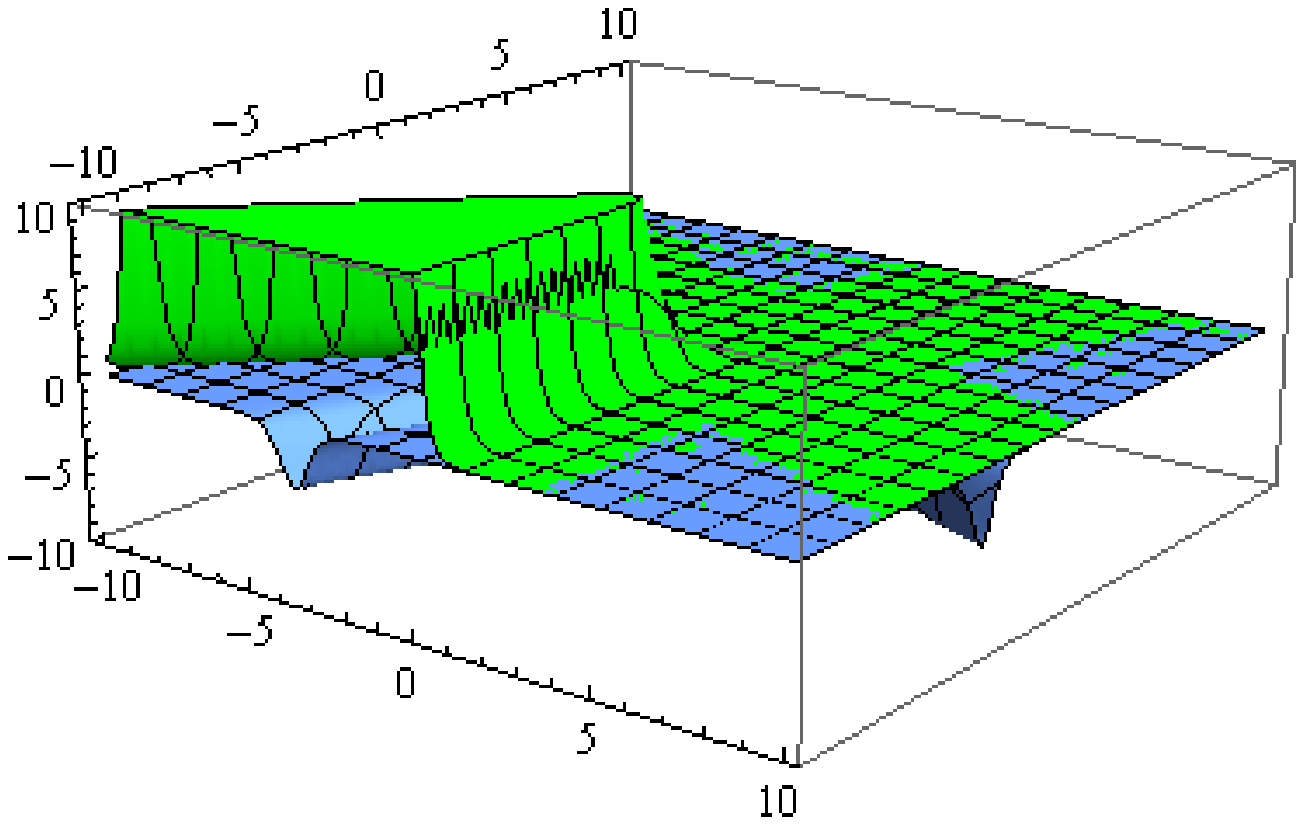}\\
\end{figure}

\begin{figure}[h]
\begin{center}
\fl\parbox[t]{1\textwidth}{\caption{Bounded $|\psi^{[1,1]}(\mu_{1})|^2=|\psi^{[1,1]}(-\lambda_{1})|^2$ (a) and nonbounded $|\psi^{[1,1]}(\mu_{2})|^2$ (b) squared absolute values of wave functions (green) given by (\ref{NVNPsiDWaveFunctN=2Gen1})-(\ref{NVNPsiDWaveFunctN=2Gen3}) corresponding to potential and energy in the Fig.\ref{VNSoliton[1,1]E-}.}\label{VNSoliton[1,1]E-WFs}}
\end{center}
\end{figure}

In considered in present section case of two line [1,1] soliton  $u=\tilde u-\epsilon$ (\ref{VN solution[1,1] 2 u simple}) with corresponding Schr\"{o}dinger potential $V_{Schr}=-2\tilde u$ squared absolute values of wave functions $|\psi^{[1,1]}(\mu_{1})|^2=|\psi^{[1,1]}(-\lambda_{1})|^2$ are bounded (Fig.\ref{VNSoliton[1,1]E-WFs}a), but the squared absolute values of other basis wave functions $|\psi^{[1,1]}(\mu_{2})|^2$ and $|\psi^{[1,1]}(-\lambda_{2})|^2$ are not bounded (Fig.\ref{VNSoliton[1,1]E-WFs}b).

In conclusion of Section 5  let us mention that all constructed in subsection 5.1 solitons and
corresponding wave functions are finite and have appropriate physical interpretation.
For example, the wave function (\ref{VN solution 1 chi(lambda)}) of continuous spectral
parameter $\lambda$  for discrete values of this parameter
$ \lambda=\mu_1$ or $\lambda=-\lambda_1$ coincides with wave functions (\ref{VN solution 1 chi(mu_1)});
for  positive values of energy $E=-2\epsilon>0$ and $\lambda\neq\mu_1, \lambda\neq-\lambda_1$, under condition
$|\lambda|^2=-\epsilon=E/2>0$, the wave function (\ref{VN solution 1 chi(lambda)}) corresponds to stationary states of nonlocalized
 on the plane $(x,y)$ particle which do not reflects from the constructed  potential  (\ref{ExactVSchr1pot}).
In considered in subsections 5.2 and 5.3 cases multi line solitons are finite but corresponding wave functions can take infinite values in some areas of the plane $(x,y)$, (Fig.\ref{VNSoliton01},\,\ref{graphVN1+WF},\,\ref{VNSoliton[1,1]E-WFs}b); only for two line soliton  $[1,1]$ squared absolute values
of  wave functions $|\psi^{[1,1]}(\mu_{1})|^2=|\psi^{[1,1]}(-\lambda_{1})|^2$ (Fig.\ref{VNSoliton[1,1]E-WFs}a) are finite.
The question of more detailed  physical interpretation and applications of exact potentials
and corresponding wave functions of 2D stationary Schr\"{o}dinger equation will be considered elsewhere.

%
%
\section{Periodic solutions of the NVN equation}\label{Section_6}
The restrictions (\ref{real_condition_NVN}) and (\ref{real_condition_VN}) on the kernel $R_0$ of
the $\bar{\partial}$-problem (\ref{dibar_problem}) which lead to real solutions $u=\bar{u}$ of the
NVN equations (\ref{NVN}) are obtained in section 2
by the use of reconstruction formula (\ref{u_reconstructFormulae})
\begin{equation}\label{exact_real_cond}
  u=-\epsilon-i\chi_{-1\eta}=-\epsilon+i\overline{\chi}_{-1\eta}
\end{equation}
in the limit of "weak" fields~, i.e. $\chi_{-1}$ in (\ref{exact_real_cond}) is calculated from its
exact  expression (\ref{di_problem_chi_-1}) with approximation $\chi\simeq1$. It is shown in section 4 and 5 that reality conditions (\ref{real_condition_NVN}) and (\ref{real_condition_VN}) work and lead to multi line soliton solutions of the NVN equation.

Such use  of reality condition was considered in all previous papers (see for example \cite{DubrovskyKonopelchenko_93}-\cite{DubrForm_03}) devoted to constructions of classes of exact solutions of integrable nonlinear evolution equations via $\overline{\partial}$-dressing method.
But there is existing possibility of non use  the limit of weak fields and imposing the reality condition $u=\overline u$
directly to exact solutions (\ref{uN=1Gen}) of NVN equation calculated in sections 2, 3 and satisfying
only to potentiality condition.

Thus one starts from the general kernel $R_0$ (\ref{sum delta_kernel_satisfied_potent}) of $\overline\partial$-dressing problem (with parameters (\ref{ParamOfAgenPosit})) which   satisfies  to potentiality condition $\chi_0-1=0$ or equivalently to (\ref{potent_cond2_FP}). All general formulas  (\ref{BI det A, N=1})-(\ref{NVNPsiWaveFunctN=2Gen1}) of section  3 are assumed to be applied here.
For simplest  kernel $R_0$ (\ref{sum delta_kernel_satisfied_potent}) with $N=1$ the requirements of reality (\ref{exact_real_cond}), i.e.
$\chi_{-1\eta}=-\overline{\chi}_{-1\eta}$, leads due to (\ref{di_problem_chi_-1}) and (\ref{BI det A, N=1})-(\ref{uN=1Gen}) to the conclusion:
\begin{equation}\label{real_condition_sec5}
\fl  \epsilon \frac{a_1(\lambda_1^2-\mu_1^2)}{\lambda_1\mu_1}\frac{1}
  {\Big[e^{-\frac{\varphi_1}{2}}-ia_1\frac{\lambda_1+\mu_1}{\lambda_1-\mu_1}e^{\frac{\varphi_1}{2}}\Big]^2}=
  -\epsilon\frac{\overline{a}_1(\overline{\lambda}_1^2-\overline{\mu}_1^2)}
  {\overline{\lambda}_1\overline{\mu}_1}
  \frac{1}
  {\Big[e^{-\frac{\overline{\varphi}_1}{2}}+i\overline{a}_1
  \frac{\overline{\lambda}_1+\overline{\mu}_1}{\overline{\lambda}_1-\overline{\mu}_1}
  e^{\frac{\overline{\varphi}_1}{2}}\Big]^2}
\end{equation}
with the phase $\varphi_1$ given due to (\ref{F_formula}) by expressions:
\begin{equation}\label{phase_N_sec5}
\fl \varphi_1(\xi,\eta,t) = F(\mu_{1}) - F(\lambda_{1}) = i\Big[(\mu_{1}-\lambda_{1})\xi-
\Big(\frac{\epsilon}{\mu_{1}}-\frac{\epsilon}{\lambda_{1}}\Big)\eta
+\kappa_1(\mu_{1}^3-\lambda_{1}^3)t
-\kappa_2\Big(\frac{\epsilon^3}{\mu_{1}^3}-\frac{\epsilon^3}{\lambda_{1}^3}\Big)t
\Big]
\end{equation}
in hyperbolic case and
\begin{equation}\label{phase_VN_sec5}
\fl  \varphi_1(z,\overline{z},t) =F(\mu_{1}) - F(\lambda_{1}) = i\Big[(\mu_1 - \lambda_1)z -
  \Big(\frac{\epsilon}{\mu_{1}}-\frac{\epsilon}{\lambda_{1}}\Big)\overline{z} + \kappa(\mu_1^3 -
  \lambda_1^3)t - \overline{\kappa}\Big(\frac{\epsilon^3}{\mu_{1}^3}-
  \frac{\epsilon^3}{\lambda_{1}^3}\Big)t\Big]
\end{equation}
in elliptic case of NVN equation (\ref{NVN}).
The condition (\ref{real_condition_sec5}) of reality can be satisfied as for real phase
$\varphi_1=\overline{\varphi_1}$ (this case leads to multi line  soliton solutions considered in sections 4,5) as long as
for imaginary phase $\varphi_1=-\overline{\varphi_1}$.
The last case leads to periodic solutions of the NVN equation.
Hereafter we described  separately the cases of the hyperbolic and elliptic NVN equations.
%
%

\emph{\bf{The hyperbolic case}}. The condition of imaginary  phase
$\varphi_1=-\overline{\varphi_1}$ due to (\ref{phase_N_sec5}) leads to relation:
\begin{eqnarray}\label{im_phase_N_sec5}
\fl  i\Big[(\mu_{1}-\lambda_{1})\xi-
  \Big(\frac{\epsilon}{\mu_{1}}-\frac{\epsilon}{\lambda_{1}}\Big)\eta
  +\kappa_1(\mu_{1}^3-\lambda_{1}^3)t
  -\kappa_2\Big(\frac{\epsilon^3}{\mu_{1}^3}-\frac{\epsilon^3}{\lambda_{1}^3}\Big)t\Big]=\nonumber\\
\fl  =i\Big[(\overline{\mu}_{1}-\overline{\lambda}_{1})\xi-
  \Big(\frac{\epsilon}{\overline{\mu}_{1}}-\frac{\epsilon}{\overline{\lambda}_{1}}\Big)\eta
  +\kappa_1(\overline{\mu}_{1}^3-\overline{\lambda}_{1}^3)t
  -\kappa_2\Big(\frac{\epsilon^3}{\overline{\mu}_{1}^3}-\frac{\epsilon^3}{\overline{\lambda}_{1}^3}\Big)t\Big].
\end{eqnarray}
From space-dependent part of (\ref{im_phase_N_sec5}) one obtains the following system of equations:
\begin{equation}\label{periodic_solution_system}
  \mu_1-\lambda_1=\overline{\mu}_1-\overline{\lambda}_1,\qquad \frac{\epsilon}{\mu_{1}}-\frac{\epsilon}{\lambda_{1}}=
  \frac{\epsilon}{\overline{\mu}_{1}}-\frac{\epsilon}{\overline{\lambda}_{1}}.
\end{equation}
Supposing that $\mu_1\neq\lambda_1$ (the solution $\mu_1=\lambda_1$ of (\ref{periodic_solution_system})
leads to lump solutions, which are not considered here, see the papers \cite{DubrForm_01}, \cite{DubrForm_03})
one obtains from (\ref{periodic_solution_system}) the equivalent system of equations
\begin{equation}\label{periodic_solution_system1}
  \mu_1-\lambda_1=\overline{\mu}_1-\overline{\lambda}_1,\qquad {\mu_{1}}{\lambda_{1}}={\overline{\mu}_{1}}{\overline{\lambda}_{1}}.
\end{equation}
The system (\ref{periodic_solution_system1}) has two solutions:
\begin{equation}\label{periodic_solution_system1_solve}
  1)\,\mu_1=-\overline{\lambda}_1,\qquad 2)\,\lambda_1=\lambda_{10},\;\mu_1=\mu_{10}
\end{equation}
where $\lambda_{10}$ and $\mu_{10}$ are real constants. One can show that time-dependent part of (\ref{im_phase_N_sec5}) doesn't lead to new equations and satisfies due to the system (\ref{periodic_solution_system1}).
For solution $\mu_1=-\overline{\lambda}_1$ of the system (\ref{periodic_solution_system1}) the phase $\varphi_1$
 given by (\ref{phase_N_sec5}) is pure imaginary and has form:
 \begin{equation}\label{phase_periodic_solution1}
\fl  \varphi_1(\xi,\eta,t) =  -i\Big[(\lambda_{1}+\overline{\lambda}_{1})\xi-
  \Big(\frac{\epsilon}{\lambda_{1}}+\frac{\epsilon}{\overline{\lambda}_{1}}\Big)\eta
  +\kappa_1(\lambda_{1}^3+\overline{\lambda}_{1}^3)t
  -\kappa_2\Big(\frac{\epsilon^3}{\lambda_{1}^3}+\frac{\epsilon^3}{\overline{\lambda}_{1}^3}\Big)t
  \Big]:=-i\tilde\varphi_1.
\end{equation}
Inserting $\mu_1=-\overline{\lambda}_1$ and (\ref{phase_periodic_solution1})
into (\ref{real_condition_sec5}) one obtains the relation:
\begin{equation}\label{Periodic_solutions_relations1}
\Big(\frac{\lambda_1}{\mu_1}-\frac{\mu_1}{\lambda_1}\Big)\big[a_1 e^{i\tilde\varphi_1} - \overline{a}_1 e^{-i\tilde\varphi_1}\big]
\Big[1 + |a_1|^2\Big(\frac{\lambda_1+\mu_1}{\lambda_1-\mu_1}\Big)^2\Big] = 0,
\end{equation}
which nontrivially satisfies under the condition:
\begin{equation}\label{Periodic_solutions_condition1}
    |a_1| = \pm i\frac{\lambda_1-\mu_1}{\lambda_1+\mu_1}=\pm\frac{\lambda_{1R}}{\lambda_{1I}}.
\end{equation}
The solution of the NVN equation (\ref{NVN}) due to (\ref{u_reconstructFormulae}) and (\ref{Periodic_solutions_condition1})
for the choice $|a_1|=\frac{\lambda_{1R}}{\lambda_{1I}}$  has the form:
\begin{equation}\label{periodic_solution_cos_1}
\fl    u=
    -\epsilon-2i\epsilon\frac{|a_1|(\lambda_1^2-\overline{\lambda}_1^2)}{|\lambda_1|^2}\frac{e^{i\arg{a_1}}}
    {\Big[e^{i\frac{\tilde\varphi_1}{2}}+e^{i\arg{a_1}}e^{-i\frac{\tilde\varphi_1}{2}}\Big]^2}=-\epsilon+2\epsilon
    \frac{\lambda_{1R}^2}{|\lambda_1|^2}\frac{1}{\cos^2({\frac{\tilde\varphi_1-\arg{a_1}}{2}})}.
\end{equation}
The solution of the NVN equation (\ref{NVN})
for $|a_1|=-\frac{\lambda_{1R}}{\lambda_{1I}}$  due
to (\ref{u_reconstructFormulae}) and (\ref{Periodic_solutions_condition1}) has the form:
\begin{equation}\label{periodic_solution_sin_1}
\fl    u=-\epsilon-2i\epsilon\frac{|a_1|(\lambda_1^2-\overline{\lambda}_1^2)}{|\lambda_1|^2}\frac{e^{i\arg{a_1}}}
    {\Big[e^{i\frac{\tilde\varphi_1}{2}}-e^{i\arg{a_1}}e^{-i\frac{\tilde\varphi_1}{2}}\Big]^2}=-\epsilon+2\epsilon
    \frac{\lambda_{1R}^2}{|\lambda_1|^2}\frac{1}{\sin^2({\frac{\tilde\varphi_1-\arg{a_1}}{2}})}.
\end{equation}
For the second solution $\lambda_1=\lambda_{10}$, $\mu_1=\mu_{10}$ of the
system (\ref{periodic_solution_system1_solve}) pure imaginary phase $\varphi_1$ given by (\ref{phase_N_sec5}) has the form:
\begin{equation}\label{phase_periodic_solution2}
\fl  \varphi_1(\xi,\eta,t) =  i\Big[(\mu_{10}-\lambda_{10})\xi-
  \Big(\frac{\epsilon}{\mu_{10}}-\frac{\epsilon}{\lambda_{10}}\Big)\eta
  +\kappa_1(\mu_{10}^3-\lambda_{10}^3)t
  -\kappa_2\Big(\frac{\epsilon^3}{\mu_{10}^3}-\frac{\epsilon^3}{\lambda_{10}^3}\Big)t
  \Big]:=i\tilde\varphi_1.
\end{equation}
Inserting $\lambda_1=\lambda_{10}$, $\mu_1=\mu_{10}$ and $\varphi_1=i\tilde\varphi_1$
from (\ref{phase_periodic_solution2}) into (\ref{real_condition_sec5}) one obtains the the relation:
\begin{equation}
\Big(\frac{\lambda_1}{\mu_1}-\frac{\mu_1}{\lambda_1}\Big)\big[a_1 e^{i\tilde\varphi_1} + \overline{a}_1 e^{-i\tilde\varphi_1}\big]
\Big[1 - |a_1|^2\Big(\frac{\lambda_1+\mu_1}{\lambda_1-\mu_1}\Big)^2\Big] = 0,
\end{equation}
which nontrivially satisfies for
\begin{equation}\label{Periodic_solutions_condition2}
    |a_1|=\pm \frac{\lambda_{10}-\mu_{10}}{\lambda_{10}+\mu_{10}}.
\end{equation}
The solution $u(\xi,\eta,t)$ of the NVN equation (\ref{NVN}) due to (\ref{u_reconstructFormulae}),
 (\ref{real_condition_sec5}), (\ref{phase_periodic_solution2}) and (\ref{Periodic_solutions_condition2})
is given by expression:
\begin{equation}\label{periodic_solution_2}
    u=-\epsilon-\epsilon
    \frac{(\lambda_{10}-\mu_{10})^2}{2\lambda_{10}\mu_{10}}\frac{1}{\cos^2({\frac{\tilde\varphi_1+\arg{a_1}}{2}}\mp\frac{\pi}{4})},
\end{equation}
where $\mp\pi/4$ corresponds to $\pm$ signs in (\ref{Periodic_solutions_condition2}).
%


${\overline{\partial}}$-dressing in  present paper is carried out for the fixed nonzero value of parameter $\epsilon$.
Nevertheless as in subsections 4.1 and 5.1 one can correctly consider the limit $\epsilon \rightarrow 0$, for this one can set $\epsilon=c_{k}\mu_{k0}, (k=1,2)$ ($c_{k}$-arbitrary real constant) and take the limit $\epsilon=c_{k}\mu_{k0}\rightarrow 0, (k=1,2)$ in all derived formulas. Limiting procedure can be correctly performed by the following settings in all required formulas:  $\epsilon \rightarrow 0$ and $\mu_{k0}\rightarrow 0$ in cases when uncertainty is absent, but $\frac{\mu_{20}}{\mu_{10}}=-\frac{\lambda_{10}}{\lambda_{20}}\rightarrow \frac{c_{1}}{c_{2}}$ in accordance with the relations $\epsilon=c_{k}\mu_{k0}$ and $\mu_{10}\lambda_{10}+\mu_{20}\lambda_{20}=0$ (\ref{EquivSeparatCondit}); the last relation is assumed to be valid in considered limit.
The periodic solution (\ref{uTwoSolGen}) in the limit $\epsilon \rightarrow 0$ takes the form:
\begin{equation}\label{NVNE0LimitPeriodic}
u=-\frac{c_{1}\lambda_{10}}{2\cos^2(\frac{\tilde\varphi_1+\arg{a_1}}{2}-\frac{\pi}{4})} -\frac{c_{2}\lambda_{20}}{2\cos^2(\frac{\tilde\varphi_2+\arg{a_2}}{2}-\frac{\pi}{4})},
\end{equation}
where the phases  $\tilde\varphi_{k}(\xi,\eta,t)$ due to (\ref{phase_periodic_solution2}) are given in considered limit by the expressions:
\begin{equation}\label{NVNPhazesE0LimitPeriodic}
\tilde\varphi_k(\xi,\eta,t) =(-\lambda_{k0}\xi - c_k\eta -
\kappa_1 \lambda_{k0}^{3}t-\kappa_2 c_{k}^{3}t).
\end{equation}
One can check by direct substitution that NVN-II equation (\ref{NVN})
with $\sigma=1$ satisfies by $u$
given by (\ref{NVNE0LimitPeriodic}), it satisfies also by each item
\begin{equation}\label{uE0LimitSolsNVNPeriodic}
u^{(k)}=-\frac{c_{k}\lambda_{k0}}{2\cos^2(\frac{\tilde\varphi_1+\arg{a_k}}{2}-\frac{\pi}{4})} , \quad (k=1,2)
\end{equation}
of the sum (\ref{NVNE0LimitPeriodic}). So in considered case the linear principle
of superposition $u=u^{(1)}+u^{(2)}$ for such special solutions $u^{(1)},u^{(2)}$  (\ref{uE0LimitSolsNVNPeriodic}) is valid.
%

\emph{\bf{The elliptic case}}. For elliptic version of NVN equation (\ref{NVN}) the condition of
imaginary  phase $\varphi_1 = - \overline{\varphi_1}$ given by (\ref{phase_VN_sec5}) leads to the relation:
\begin{eqnarray}\label{Periodic VN phi = -bar phi}
\fl    \varphi_1 = i\Big[(\mu_1 - \lambda_1)z - \Big(\frac{\epsilon}{\mu_1} - \frac{\epsilon}{\lambda_1}\Big)\overline{z} + \kappa(\mu_1^3 - \lambda_1^3)t - \overline{\kappa}\Big(\frac{\epsilon^3}{\mu_1^3} - \frac{\epsilon^3}{\lambda_1^3}\Big)t\Big] = \nonumber \\
\fl    = i\Big[(\overline{\mu}_1 - \overline{\lambda}_1)\overline{z} - \Big(\frac{\epsilon}{\overline{\mu}_1} - \frac{\epsilon}{\overline{\lambda}_1}\Big)z + \overline{\kappa}(\overline{\mu}_1^3 - \overline{\lambda}_1^3)t - \kappa\Big(\frac{\epsilon^3}{\overline{\mu}_1^3} - \frac{\epsilon^3}{\overline{\lambda}_1^3}\Big)t\Big].
\end{eqnarray}
From the space-dependent part of (\ref{Periodic VN phi = -bar phi}) follows the system of equations:
\begin{equation}\label{Periodic VN relation mu_1 and lambda_1}
  \mu_1 - \lambda_1 = -\frac{\epsilon}{\overline{\mu}_1} + \frac{\epsilon}{\overline{\lambda}_1}, \quad \overline{\mu}_1 - \overline{\lambda}_1 = -\frac{\epsilon}{\mu}_1 + \frac{\epsilon}{\lambda_1}.
\end{equation}
The solution $\mu_1=\lambda_1$ of (\ref{Periodic VN relation mu_1 and lambda_1})
leads to lumps solutions $u(\xi,\eta,t)$ of NVN equation (\ref{NVN}), which are not considered here (see the papers \cite{DubrForm_01}, \cite{DubrForm_03}).
Excluding parameter $\epsilon$ from (\ref{Periodic VN relation mu_1 and lambda_1}) one obtains the relations:
\begin{equation}\label{Periodic VN epsilon}
  \epsilon = \overline{\mu}_1\overline{\lambda}_1\frac{\mu_1 - \lambda_1}{\overline{\mu}_1 - \overline{\lambda}_1} = \mu_1\lambda_1\frac{\overline{\mu}_1 - \overline{\lambda}_1}{\mu_1 - \lambda_1},
\end{equation}
and their consequence:
\begin{equation}\label{Periodic VN without epsilon}
  (|\mu_1|^2 - |\lambda_1|^2)(\mu_1\overline{\lambda}_1 - \overline{\mu}_1\lambda_1) = 0.
\end{equation}
Due to (\ref{Periodic VN epsilon}) and (\ref{Periodic VN without epsilon}) the system (\ref{Periodic VN relation mu_1 and lambda_1}) has the solutions:
\begin{equation}\label{Periodic VN two solution, epsilon}
  1.\: \epsilon = -|\mu_1|^2 = -|\lambda_1|^2, \quad 2.\: \epsilon = \overline{\lambda}_1\mu_1 = \lambda_1\overline{\mu}_1.
\end{equation}
One can show that time-dependent part of (\ref{Periodic VN phi = -bar phi}) satisfies by solutions (\ref{Periodic VN two solution, epsilon}) of the system (\ref{Periodic VN relation mu_1 and lambda_1}).
For both solutions of the system (\ref{Periodic VN relation mu_1 and lambda_1})
the pure imaginary $\varphi_1$ given by (\ref{Periodic VN phi = -bar phi}) takes the form:
\begin{equation}\label{Periodic VN solution 1 phi}
\fl  \varphi_1(z,\bar{z},t) = i[(\mu_1 - \lambda_1)z + (\overline{\mu}_1 - \overline{\lambda}_1)\bar{z} + \kappa(\mu_1^3 - \lambda_1^3)t + \overline{\kappa}(\overline{\mu}_1^3 - \overline{\lambda}_1^3)t] := i\tilde\varphi_1(z,\bar{z},t)
\end{equation}
 The condition (\ref{real_condition_sec5}) of reality of $u$ for the first case
in (\ref{Periodic VN two solution, epsilon}) gives the  relation:
%
%
\begin{equation}\label{Periodic VN solution 1 real condition}
\Big(\frac{\lambda_1}{\mu_1}-\frac{\mu_1}{\lambda_1}\Big)\big[a_1 e^{i\tilde\varphi_1} - \overline{a}_1 e^{-i\tilde\varphi_1}\big]
\Big[1 + |a_1|^2\Big(\frac{\lambda_1+\mu_1}{\lambda_1-\mu_1}\Big)^2\Big] = 0,
\end{equation}
which nontrivially satisfies for the following choice of amplitude $a_1$
\begin{equation}\label{Periodic VN solution 1 |a_1|}
  |a_1| = \pm\frac{\lambda_1-\mu_1}{\lambda_1+\mu_1}= \pm \tan\frac{\delta}{2};\quad  \delta_1:=\arg(\mu_1) - \arg(\lambda_1).
\end{equation}
For $|a_1| = \tan\frac{\delta}{2}$ due to (\ref{u_reconstructFormulae}) and
(\ref{real_condition_sec5}), (\ref{Periodic VN two solution, epsilon}) - (\ref{Periodic VN solution 1 |a_1|})
one obtains the periodic  solution $u$ with constant asymptotic
values $-\epsilon$ at infinity of elliptic NVN equation:
\begin{equation}\label{Periodic VN solution 1b u}
\fl  u(z,\bar{z},t) = -\epsilon -\frac{|\lambda_{1}-\mu_{1}|^2}{2}\frac{1}{\cos^2\Big(\frac{\tilde\varphi_1 + \arg(a_1)}{2}\Big)} = -\epsilon + \frac{2\epsilon\sin^2\frac{\delta}{2}}{\cos^2\Big(\frac{\tilde\varphi_1 + \arg(a_1)}{2}\Big)},
\end{equation}
and for $|a_1| = - \tan\frac{\delta}{2}$  another periodic solution
\begin{equation}\label{Periodic VN solution 1a u}
\fl  u(z,\bar{z},t) = -\epsilon -\frac{|\lambda_{1}-\mu_{1}|^2}{2}\frac{1}{\sin^2\Big(\frac{\tilde\varphi_1 + \arg(a_1)}{2}\Big)} = -\epsilon + \frac{2\epsilon\sin^2\frac{\delta}{2}}{\sin^2\Big(\frac{\tilde\varphi_1 + \arg(a_1)}{2}\Big)}.
\end{equation}

The condition (\ref{real_condition_sec5}) of reality of $u$ for the second case
in (\ref{Periodic VN two solution, epsilon}) gives the  relation:
\begin{equation}
\Big(\frac{\lambda_1}{\mu_1}-\frac{\mu_1}{\lambda_1}\Big)\big[a_1 e^{i\tilde\varphi_1} + \overline{a}_1 e^{-i\tilde\varphi_1}\big]
\Big[1 - |a_1|^2\Big(\frac{\lambda_1+\mu_1}{\lambda_1-\mu_1}\Big)^2\Big] = 0
\end{equation}
%
which satisfies for
\begin{equation}\label{Periodic_solutions_condition2VN}
|a_1|=\pm\frac{\lambda_{1}-\mu_{1}}{\lambda_{1}+\mu_{1}}.
\end{equation}
For the second case in (\ref{Periodic VN two solution, epsilon}) periodic solution $u(\xi,\eta,t)$ for the NVN equation (\ref{NVN}) due to (\ref{u_reconstructFormulae}), (\ref{Periodic VN solution 1 phi}), (\ref{Periodic_solutions_condition2VN}) has the form:
\begin{equation}\label{periodic_solution_2VN}
\fl   u=-\epsilon-
    \frac{|\lambda_{1}-\mu_{1}|^2}{2}\frac{1}{\cos^2({\frac{\tilde\varphi_1+\arg{a_1}}{2}}\mp\frac{\pi}{4})},\quad \epsilon=\lambda_1\bar\mu_1=\bar\lambda_1\mu_1,
\end{equation}
where $\mp \pi/4$ corresponds to $\pm$ signs in (\ref{Periodic_solutions_condition2VN}).

${\overline{\partial}}$-dressing in  present paper is carried out for the fixed nonzero value of parameter $\epsilon$ or, in  context of  present section, for nonzero energy
$E\neq  0$.
%
Nevertheless as in subsections 4.1 and 5.1 one can correctly  consider the limit $\epsilon \rightarrow 0$ in all derived formulas and obtain some interesting results also for the case of zero energy $E=-2\epsilon=0$. Limiting procedure  $E=-2\epsilon=-\mu_k\bar\lambda_k-\bar\mu_k\lambda_k \rightarrow 0, (k=1,2)$  can be correctly performed by the following settings in all required formulas:  $\epsilon \rightarrow 0$ and $\mu_k \rightarrow 0$ in cases when uncertainty is absent, but ${\frac{\epsilon}{\mu_k}}\rightarrow \bar\lambda_k$ in accordance with the relation $\epsilon=\mu_k\bar\lambda_k$; in addition the formula $\lambda_2=i\tau^{-1}\lambda_1$ (followed from the relations
$\bar\mu_k\lambda_k=\mu_k\bar\lambda_k$ and $\mu_1\lambda_1+\mu_2\lambda_2=0$) with arbitrary real constant $\tau$  is assumed to be valid. The periodic solution due to (\ref{uTwoSolGen}) in considered limit has the form:
\begin{equation}\label{VSchrE0LimitPeriodic}
u=-\frac{|\lambda_1|^2}{2\cos^2(\frac{\tilde\varphi_1+\arg{a_1}}{2}-\frac{\pi}{4})} -\frac{|\lambda_2|^2}{2\cos^2(\frac{\tilde\varphi_2+\arg{a_2}}{2}-\frac{\pi}{4})} ,
\end{equation}
 the phases  $\tilde\varphi_{k}(z,\bar{z},t)$  due to (\ref{Periodic VN solution 1 phi}) have in considered limit the forms:
\begin{equation}\label{PhazesE0LimitPeriodic}
\tilde\varphi_k(z,\bar{z},t) = \big(-\lambda_kz - \overline{\lambda}_k\overline{z} -
\kappa \lambda_{k}^{3}t -\overline{\kappa} \overline{\lambda}_{k}^{3}t\big).
\end{equation}
One can check by direct substitution that NVN-I equation (\ref{NVN})
with $\sigma=i$ satisfies by $u=\tilde u=-V_{Schr}/2$
given by (\ref{VSchrE0LimitPeriodic}), but it also satisfies by each item
\begin{equation}\label{uE0LimitSolsPeriodic}
u^{(k)}=-\frac{|\lambda_k|^2}{2\cos^2(\frac{\tilde\varphi_k+\arg{a_k}}{2}-\frac{\pi}{4})}, \quad (k=1,2)
\end{equation}
of the sum (\ref{VSchrE0LimitPeriodic}). Thus, in considered case the linear principle
of superposition $u=u^{(1)}+u^{(2)}$ for such special periodic solutions $u^{(1)},u^{(2)}$  (\ref{uE0LimitSolsPeriodic}) is valid. One can show using relation $\lambda_2=i\tau^{-1}\lambda_1$, (\ref{PhazesE0LimitPeriodic})  that periodic solutions $u^{(1)}$ and $u^{(2)}$ are propagate in the plane $(x,y)$ in perpendicular to each other directions.
%

\begin{figure}[h]
\begin{center}
a\includegraphics[width=0.50\textwidth,keepaspectratio]{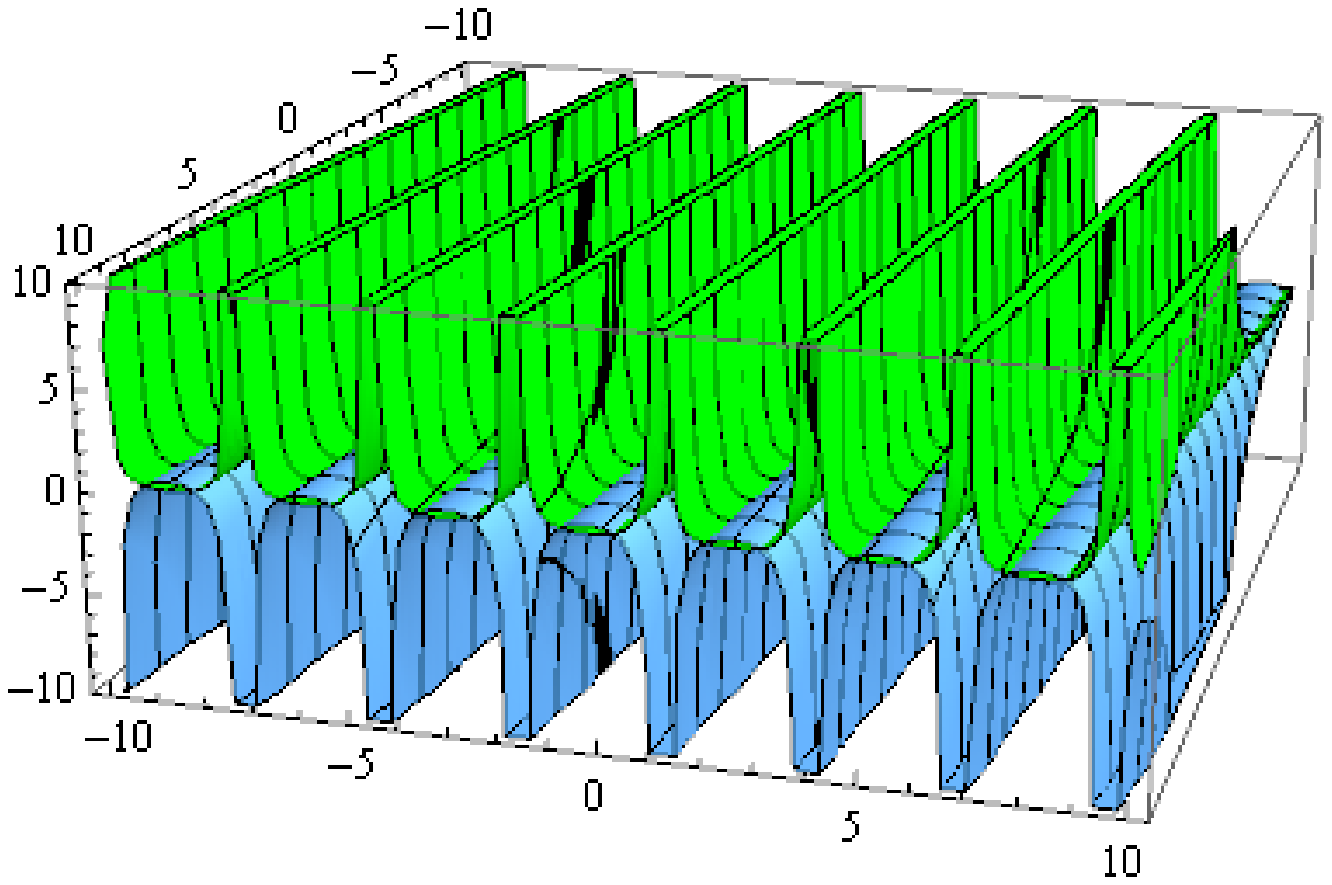}\hfill
b\includegraphics[width=0.45\textwidth,keepaspectratio]{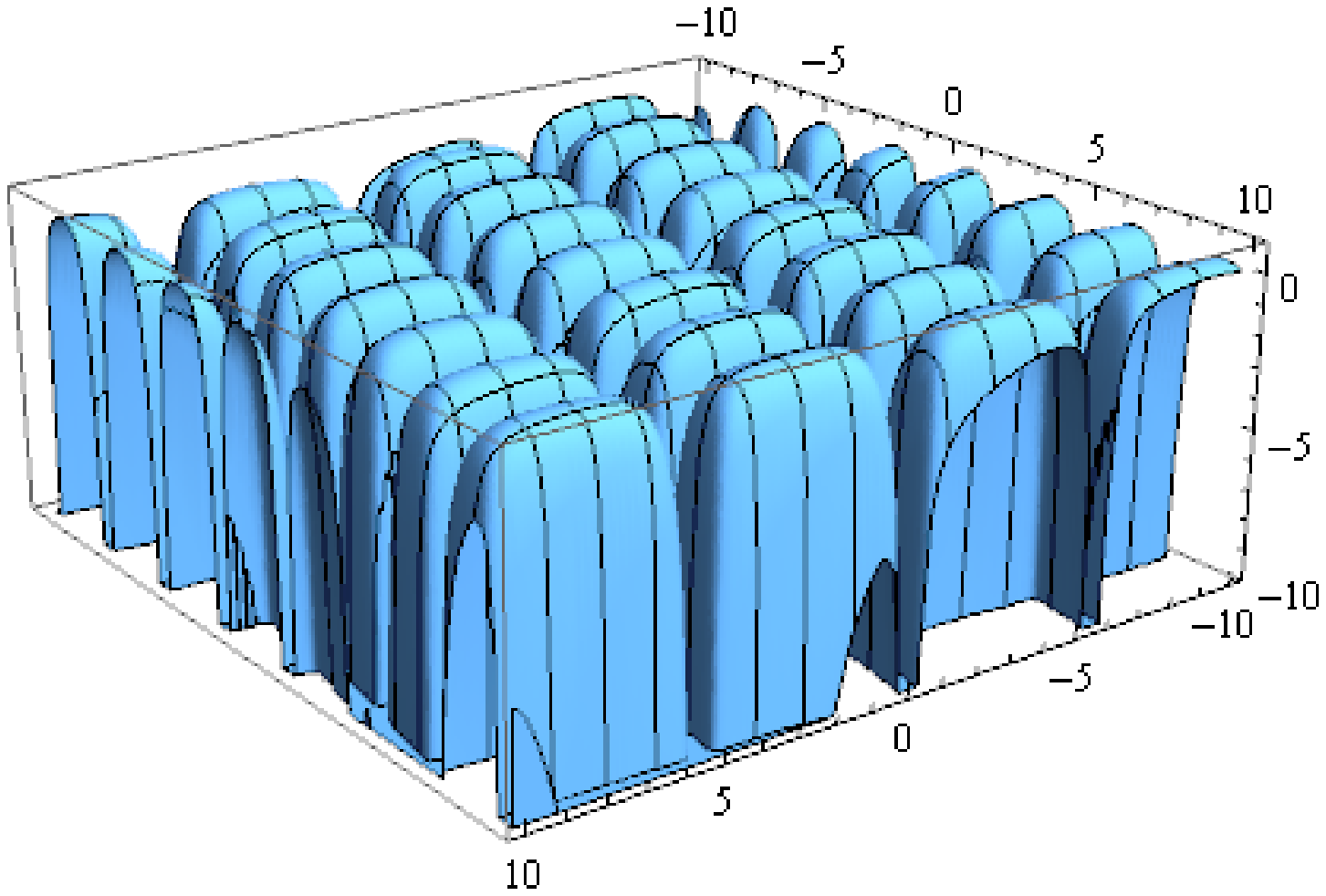}\\
\end{center}
\end{figure}

\begin{figure}[h]
\begin{center}
\fl\parbox[t]{1\textwidth}{\caption{
a)Periodic solution $u(x,y,t=0)$ (\ref{Periodic VN solution 1b u}) (blue) and the squared absolute value of corresponding wave functions $|\psi(\mu_1)|^2=|\psi(-\lambda_{1})|^2$ (\ref{PsiWFNVNN=1,1}) (green) with parameters $\arg(a_1)=\frac{\pi}{5},\delta_1=\frac{\pi}{3},\lambda_{1}=1-0.5i,\epsilon=1.25$,
b)Two-periodic solution $u(x,y,t=0)$ (\ref{uTwoSolGen}) with parameters $\arg(a_1)=\frac{\pi}{3},\delta_1=\frac{\pi}{3},\lambda_{1}=1-0.5i$;
$\arg(a_2)=\frac{\pi}{3},\delta_2=\frac{\pi}{6},\lambda_{2}=0.1-1.11355i,\epsilon=1.25$.}\label{PeriodicSolutions}}
\end{center}
\end{figure}

Last two figures, Fig.\ref{PeriodicSolutions} a) and Fig.\ref{PeriodicSolutions} b), demonstrate the simplest one - (N=1 in kernel $R_0$ (\ref{sum delta_kernel_satisfied_potent})) and two-periodic (N=2 in kernel $R_0$ (\ref{sum delta_kernel_satisfied_potent})) solutions of NVN equation (\ref{NVN}) calculated by the formulas (\ref{Periodic VN solution 1b u}) and  (\ref{uTwoSolGen}) under certain values of corresponding parameters. It is assumed also that for two-periodic solution the condition (\ref{EquivSeparatCondit}) of splitting the solution (\ref{uN=2Gen}) into two terms is fulfilled.
All constructed in the present section periodic solutions evidently are singular.
The further study of periodic solutions of NVN equation in the framework of $\bar{\partial}$-dressing
method will be continued elsewhere.

\section{Solutions of NVN equation with functional parameters}
\label{Section_7}
\setcounter{equation}{0}

Constructed in the previous sections multi line soliton and periodic solutions can be embedded into  more general class
of exact solutions with functional parameters. Such solutions correspond to degenerate
 kernel $R_0(\mu,\overline{\mu};\lambda,\overline{\lambda})$ of $\overline{\partial}$-problem
(\ref{dibar_problem})
\begin{equation}\label{Kernel_VN_FP}
R_{0}(\mu,\overline{\mu},\lambda,\overline{\lambda})=
\pi\sum\limits_{k=1}^{N}f_k(\mu,\overline{\mu})g_k(\lambda,\overline{\lambda}).
\end{equation}
As in section 2 one can easily derive general determinant formula for the class of exact solutions $u(\xi,\eta,t)$
with constant asymptotic value $-\epsilon$ at infinity with functional parameters of the NVN equation (\ref{NVN}).
Indeed, inserting (\ref{Kernel_VN_FP}) into (\ref{di_problem1}) and integrating one obtains
\begin{equation}\label{chi_FP}
\chi (\lambda) = 1 + \pi\sum\limits_{k=1}^{N}h_k(\xi,\eta,t)
\int\int\limits_C {\frac{d{\lambda }'\wedge
d{\overline {\lambda }}'}{2\pi i(\lambda'-\lambda)}}
g_k(\lambda',\overline{\lambda'})e^{-F(\lambda')}
\end{equation}
where
\begin{equation}\label{h_FP}
h_k(\xi,\eta,t):=\int\int \limits_C \chi(\mu,\overline{\mu}) e^{F(\mu)}f_k(\mu,\overline{\mu})d\mu\wedge
d\overline{\mu}.
\end{equation}
From (\ref{chi_FP}), (\ref{h_FP}) follows the system of linear algebraic equations for the quantities
$h_k$:
\begin{equation}\label{system_FP}
\sum\limits_{k=1}^{N}A_{lk} h_k =\alpha_l,\quad
(l=1,\cdots,N)
\end{equation}
with
\begin{equation}\label{alpha_FP}
 \alpha_l(\xi,\eta,t): = \int\int\limits_C f_l(\mu,\overline{\mu})e^{F(\mu)}d\mu\wedge d\overline{\mu}
\end{equation}
and matrix $A$ is given by expression:
\begin{equation}\label{matrix_FP}
\fl A_{lk}:=\delta_{lk}+\pi\int\int\limits_C d\lambda\wedge d\overline{\lambda}\int\int\limits_C \frac{d\lambda'\wedge d\overline{\lambda'}}{2\pi i}\frac{e^{F(\lambda)-F(\lambda')}}{\lambda-\lambda'} f_l(\lambda,\overline{\lambda})g_k(\lambda',\overline{\lambda'}).
\end{equation}
Introducing the quantities
\begin{equation}\label{beta_FP}
\beta_l(\xi,\eta,t): = \int \int\limits_C g_l(\lambda,\overline{\lambda})e^{-F(\lambda)}d\lambda \wedge d\overline{\lambda}
\end{equation}
one can rewrite the matrix $A_{lk}$ (\ref{matrix_FP}) in the following form:
\begin{equation}\label{matrix1_FP}
A_{lk}=\delta_{lk}+\frac{1}{2}\partial_{\xi}^{-1}\alpha_l \beta_k.
\end{equation}
The functions $\alpha_k(\xi,\eta,t)$, $\beta_k(\xi,\eta,t)$  given by (\ref{alpha_FP}) and (\ref{beta_FP}) are known as functional parameters. By the definitions (\ref{F_formula}) and (\ref{alpha_FP}), (\ref{beta_FP}) the functional parameters $\alpha_n$ and $\beta_n$ to the following linear equations are satisfy:
\begin{equation}\label{linearDiffEqForParam_FP}
\alpha_{n\xi\eta} = \epsilon\alpha_n,\quad
\alpha_{nt}+\kappa_1\alpha_{n\xi\xi\xi}+\kappa_2\alpha_{n\eta\eta\eta}=0,
\end{equation}
\begin{equation}\label{linearDiffEqForParam_FP}
\beta_{n\xi\eta} = \epsilon\beta_n, \quad
\beta_{nt}+\kappa_1\beta_{n\xi\xi\xi}+\kappa_2\beta_{n\eta\eta\eta}=0.
\end{equation}

From (\ref{di_problem_chi_-1}) and (\ref{system_FP})-(\ref{beta_FP}) follows compact formula
for the coefficient $\chi_{-1}$ of the expansion (\ref{series of chi})
\begin{eqnarray}\label{chi_-1_FP}
 \chi_{-1}=-\frac{1}{2i}\sum\limits_{k=1}^{N}h_k\beta_k= -\frac{1}{2i}\sum\limits_{l,k=1}^{N}A^{-1}_{kl}\alpha_l\beta_k=i
\sum\limits_{k,l=1}^{N}A^{-1}_{kl}\frac{\partial A_{lk}}{\partial \xi}= \nonumber \\
=i Tr(A^{-1}\frac{\partial A}{\partial \xi})=
i\,\partial_{\xi}(\ln \det A).
\end{eqnarray}
Here and below useful determinant identities
\begin{equation}\label{useful_identities_FP}
Tr(\frac{\partial A}{\partial \xi}A^{-1})=\frac{\partial}{\partial \xi} \ln (\det A),
\quad 1+trB=det(1+B)
\end{equation}
are used.
The matrix $B$ in the last identity of (\ref{useful_identities_FP}) is degenerate with rank 1.
Using reconstruction formula (\ref{u_reconstructFormulae}) and the expression (\ref{chi_-1_FP}) one obtains general determinant formula for the solution $u$ with constant asymptotic values $-\epsilon$ at infinity with functional parameters $\alpha_k(\xi,\eta,t)$, $\beta_k(\xi,\eta,t)$ (given by (\ref{alpha_FP}),(\ref{beta_FP}))  of the NVN equation (\ref{NVN}):
\begin{equation}\label{Solution_general_formula_FP}
u(\xi,\eta,t)=-\epsilon-i\chi_{-1\eta}=-\epsilon+\frac{\partial^2}{\partial\xi\partial\eta}
\ln \det A.
\end{equation}

Potentiality condition (\ref{potencial_condition}) due to (\ref{Kernel_VN_FP}), (\ref{h_FP})-(\ref{beta_FP}) also can be
expressed in terms of functional parameters
\begin{equation}\label{potent_cond_FP}
\fl \chi_0-1=-\frac{1}{2\epsilon}\sum\limits_{k=1}^{N} h_k
\beta_{k\eta}=-\frac{1}{2\epsilon}\sum\limits_{k,m=1}^{N} A_{km}^{-1}
\alpha_m \beta_{k\eta}=-\frac{1}{2\epsilon}\sum\limits_{k,m=1}^{N} A_{km}^{-1} B_{mk}=0
\end{equation}
where degenerate matrix $B$ with rank 1 is defined by the formula
\begin{equation}\label{matrixB_FP}
B_{mk}=\alpha_m \beta_{k\eta}.
\end{equation}
Due to (\ref{potencial_condition}) and (\ref{matrixB_FP})  potentiality
condition (\ref{potent_cond_FP}) takes the form
\begin{equation}\label{potent_cond1_FP}
0=\sum\limits_{k,m=1}^{N} A_{km}^{-1} B_{mk}=tr(A^{-1}B)=det(B A^{-1}+1)-1,
\end{equation}
here matrix $BA^{-1}$ is degenerate of rank 1 and in deriving the last equality
in (\ref{potent_cond1_FP}) 
second matrix identity (\ref{useful_identities_FP}) is used. So due to (\ref{potent_cond1_FP}) the potentiality condition takes the following convenient form:
\begin{equation}\label{potent_cond2_FP}
\det(A+B)=\det{A}.
\end{equation}

Important class of exact multi line soliton solutions of the NVN equation (\ref{NVN}) can be obtained from solutions
with functional parameters by the following choice of the functions $f_k(\mu,\overline{\mu})$,
$g_k(\lambda,\overline{\lambda})$ in the kernel $R_0$ (\ref{Kernel_VN_FP}):
\begin{equation}\label{delta_func_FP}
f_k(\mu,\overline{\mu})=\delta (\mu-M_k), \quad
g_k(\lambda,\overline{\lambda})=A_k \delta (\lambda-\Lambda_k).
\end{equation}
Inserting (\ref{delta_func_FP}) into (\ref{matrix_FP}) one obtains
\begin{equation}\label{matrix_soliton_FP}
A_{lk}=\delta_{lk}+ 2i\frac{A_{k}}{M_{l}-\Lambda_{k}}e^{F(M_{l})-F(\Lambda_{k})}.
\end{equation}
For the matrix $B$ due to (\ref{Kernel_VN_FP}), (\ref{beta_FP}) and (\ref{matrixB_FP}), (\ref{delta_func_FP}) one derives
the expression:
\begin{equation}\label{matrixB_soliton_FP}
B_{lk}=\alpha_l \beta_{k\eta}=-\frac{4i\epsilon}{\Lambda_k}A_k e^{F(M_l)-F(\Lambda_k)}.
\end{equation}

The main problem in construction of exact solutions of the NVN equation (\ref{NVN}) is an "effectivization" of general determinant formula
(\ref{Solution_general_formula_FP}) by satisfying to the conditions (\ref{real_condition_NVN}), (\ref{real_condition_VN})
of reality and to the condition of potentiality (\ref{potencial_condition}) or (\ref{potent_cond2_FP})
of operator $L_1$ in (\ref{NVN dressing auxiliary problems}).
In order to satisfy to the condition of potentiality (\ref{potencial_condition}) the terms in the sum (\ref{Kernel_VN_FP}) for the kernel $R_0$ can be grouped by pairs. Indeed, inserting the expression $R_0=\pi p_1(\mu,\overline{\mu})q_1(\lambda,\overline{\lambda})+
\pi p_2(\mu,\overline{\mu})q_2(\lambda,\overline{\lambda})$ into (\ref{potencial_condition}) and performing the change of variables $\mu\leftrightarrow-\lambda$ in the second term one obtains in the limit of weak fields ($\chi=1$ in the equality (\ref{potencial_condition})):
\begin{equation}\label{potent_cond_in_weak_field}
 \fl   \int\int\limits_C\int\int\limits_C\Big[\frac{p_1(\mu,\overline{\mu})q_1(\lambda,\overline{\lambda})}{\lambda}-
    \frac{p_2(-\lambda,-\overline{\lambda})q_2(-\mu,-\overline{\mu})}{\mu}\Big]e^{F(\mu)-F(\lambda)}
    d\mu\wedge d\overline{\mu}\,d\lambda\wedge d\overline{\lambda}=0.
\end{equation}
The relation (\ref{potent_cond_in_weak_field}) will be satisfied  if $\frac{1}{\lambda}p_1(\mu,\overline{\mu})q_1(\lambda,\overline{\lambda})=
    \frac{1}{\mu}p_2(-\lambda,-\overline{\lambda})q_2(-\mu,-\overline{\mu})$, or separating variables, if
\begin{equation}\label{potent_cond_in_weak_field_in_sep}
    \frac{q_1(\lambda,\overline{\lambda})}{\lambda p_2(-\lambda,-\overline{\lambda})}=
   \frac{q_2(-\mu,-\overline{\mu})}{\mu p_1(\mu,\overline{\mu})}=c
\end{equation}
where $c$ is some constant.
Due to (\ref{potent_cond_in_weak_field_in_sep}) $p_2$ and $q_2$ through $q_1$ and $q_1$ are expressed
\begin{equation}\label{p2&q2_through_p1&q1}
p_2(\lambda,\overline{\lambda})=\frac{-1}{c\lambda}q_1(-\lambda,-\overline{\lambda}),\quad
q_2(\mu,\overline{\mu})=-c\mu p_1(-\mu,-\overline{\mu}).
\end{equation}
So to the potentiality condition (\ref{potencial_condition}) due to (\ref{p2&q2_through_p1&q1}) is satisfied the following kernel \begin{equation}\label{kernel_satify_potent_cond}
\fl R_{0}(\mu,\overline{\mu},\lambda,\overline{\lambda})=
\pi\sum\limits_{k=1}^{N}\Big(p_k(\mu,\overline{\mu})q_k(\lambda,\overline{\lambda})+
\frac{q_k(-\mu,-\overline{\mu})}{\mu}\lambda p_k(-\lambda,-\overline{\lambda})\Big)
\end{equation}
$R_0$  of the $\overline{\partial}$-problem (\ref{dibar_problem})
with $N$ pairs of correlated with each other terms.

The conditions (\ref{real_condition_NVN}) and (\ref{real_condition_VN}) of reality $u=\overline{u}$ give further restrictions on the functions $p_k$ and $q_k$ in the sum (\ref{kernel_satify_potent_cond}). It is convenient to perform the calculations of these restrictions and exact solutions $u(\xi,\eta,t)$  separately for Nizhnik $\sigma^2=1$, $\xi=x+y$, $\eta=x-y$ and Veselov-Novikov $\sigma^2=-1$, $\xi=z=x+iy$, $\eta=\overline{z}=x-iy$ versions of the NVN equation (\ref{NVN}).

\section{Exact solutions with functional parameters of  NVN-II equation}
\label{Section_8}
\setcounter{equation}{0}

Let us consider at first the case $\sigma^2=1$ of real space variables $\xi=x+y$, $\eta=x-y$ or hyperbolic version of the NVN equation (\ref{NVN}). To the condition (\ref{real_condition_NVN}) of reality $u=\overline{u}$ one can satisfy imposing on each pair of terms in the sum (\ref{kernel_satify_potent_cond}) the following restriction:
\begin{eqnarray}\label{kernel_reality_cond}
p_n(\mu,\overline{\mu})q_n(\lambda,\overline{\lambda})+\frac{1}{\mu}q_n(-\mu,-\overline{\mu}){\lambda} p_n(-\lambda,-\overline{\lambda})&=&\\ \nonumber
=\overline{p_n(-\overline{\mu},-\mu)}\,\overline{q_n(-\overline{\lambda},-\lambda)}+
\frac{1}{\mu}\overline{q_n(\overline{\mu},\mu)}{\lambda} \overline{p_n(\overline{\lambda},\lambda)}.
\end{eqnarray}
Due to (\ref{kernel_reality_cond}) two cases are possible
\begin{equation}\label{kernel_real_case_A}
8.A.\quad p_n(\mu,\overline{\mu})q_n(\lambda,\overline{\lambda})=\overline{p_n(-\overline{\mu},-\mu)}\,
\overline{q_n(-\overline{\lambda},-\lambda)},
\end{equation}
\begin{equation}\label{kernel_real_case_B}
8.B.\quad p_n(\mu,\overline{\mu})q_n(\lambda,\overline{\lambda})=\frac{1}{\mu}\overline{q_n(\overline{\mu},\mu)}{\lambda} \overline{p_n(\overline{\lambda},\lambda)}.
\end{equation}
In the case $8.A$ by separating variables
\begin{equation}\label{caseA_separate}
\frac{p_n(\mu,\overline{\mu})}{\overline{p_n(-\overline{\mu},-\mu)}}=
\frac{\overline{q_n(-\overline{\lambda},-\lambda)}}{q_n(\lambda,\overline{\lambda})}=c_n
\end{equation}
one obtains the following restrictions on the functions $p_n(\mu,\overline{\mu})$ and $q_n(\lambda,\overline{\lambda})$:
\begin{equation}\label{caseA_p&q}
p_n(\mu,\overline{\mu})=c_n\overline{p_n(-\overline{\mu},-\mu)},\quad
q_n(\lambda,\overline{\lambda})=\frac{1}{c_n}\overline{q_n(-\overline{\lambda},-\lambda)}.
\end{equation}
Constants $c_n$ in (\ref{caseA_p&q}) without restriction of generality can be chosen equal to unity.
In the case $8.B$ by separating variables
\begin{equation}\label{caseB_separate}
\frac{\mu p_n(\mu,\overline{\mu})}{\overline{q_n(\overline{\mu},\mu)}}=
\frac{\lambda\overline{p_n(\overline{\lambda},\lambda)}}{q_n(\lambda,\overline{\lambda})}=
c^{-1}_n
\end{equation}
one obtains the another restrictions on the functions $p_n(\mu,\overline{\mu})$ and $q_n(\lambda,\overline{\lambda})$:
\begin{equation}\label{caseB_p&q}
q_n(\lambda,\overline{\lambda})=\lambda c_n\overline{p_n(\overline{\lambda},\lambda)}.
\end{equation}
The constants $c_n$ in (\ref{caseB_p&q}) due to (\ref{caseB_separate}) are real.

In applying general determinant formula (\ref{Solution_general_formula_FP}) for exact solutions $u$ one must to identify the corresponding kernels (\ref{Kernel_VN_FP}) and (\ref{kernel_satify_potent_cond}). For the case $8.A$   taking into account (\ref{kernel_satify_potent_cond}) and (\ref{caseA_p&q}) one has:
\begin{eqnarray}\label{kernel_satify_potent_cond&potential}
 R_{0}(\mu,\overline{\mu},\lambda,\overline{\lambda})=\pi\sum\limits_{n=1}^{N}f_n(\mu,\overline{\mu})
g_n(\lambda,\overline{\lambda})=\nonumber \\
=\pi\sum\limits_{n=1}^{N}\Big(p_n(\mu,\overline{\mu})q_n(\lambda,\overline{\lambda})+
\frac{1}{\mu}\overline{q_n(\overline{\mu},\mu)}\lambda \overline{p_n(\overline{\lambda},\lambda)}\Big)
\end{eqnarray}
and from (\ref{kernel_satify_potent_cond&potential}) one can choose  the following convenient sets $f$ and $g$ of functions $f_n$, $g_n$:
\begin{equation}\label{set_of_f}
\fl f:=(f_1,\ldots,f_{2N})=(p_1(\mu,\overline{\mu}),\ldots, p_N(\mu,\overline{\mu});\frac{1}{\mu}\overline{q_1(\overline{\mu},\mu)},\ldots,\frac{1}{\mu}
\overline{q_N(\overline{\mu},\mu)}),
\end{equation}
\begin{equation}\label{set_of_g}
\fl g:=(g_1,\ldots,g_{2N})=(q_1(\lambda,\overline{\lambda}),\ldots, q_N(\lambda,\overline{\lambda});\lambda \overline{p_1(\overline{\lambda},\lambda)},\ldots,\lambda \overline{p_N(\overline{\lambda},\lambda)}).
\end{equation}
Due to definitions (\ref{alpha_FP}), (\ref{beta_FP}) and (\ref{set_of_f}), (\ref{set_of_g}) taking into account (\ref{caseA_p&q}) one can derive the following interrelations between different functional parameters:
\begin{equation}\label{alpha&beta_FP}
\fl \alpha_n: = \int\int\limits_C p_n(\mu,\overline{\mu})e^{F(\mu)}d\mu\wedge d\overline{\mu}=\overline{\alpha_n},\quad
\beta_n: = \int\int \limits_C q_n(\lambda,\overline{\lambda})e^{-F(\lambda)}d\lambda \wedge d\overline{\lambda}=\overline{\beta_n},
\end{equation}
\begin{equation}\label{a2&b1_FP}
\alpha_{N+n}: = \int\int\limits_C \frac{1}{\mu}\overline{q_n(\overline{\mu},\mu)}e^{F(\mu)}d\mu\wedge d\overline{\mu}=\frac{i}{\epsilon}{\beta}_{n\eta},
\end{equation}
\begin{equation}\label{a1&b2_FP}
\beta_{N+n}: = \int\int \limits_C \lambda \overline{p_n(\overline{\lambda},\lambda)}e^{-F(\lambda)}d\lambda \wedge d\overline{\lambda}=i\,{\alpha}_{n\xi}, \quad (n=1,\ldots,N).
\end{equation}
So due to (\ref{alpha&beta_FP})-(\ref{a1&b2_FP})  the sets of functional parameters have the following structure:
\begin{equation}\label{set_of_alpha}
(\alpha_1,\ldots,\alpha_{2N}):=(\alpha_1,\ldots, \alpha_N;\frac{i}{\epsilon}\beta_{1\eta},\ldots,
\frac{i}{\epsilon}\beta_{N\eta})
\end{equation}
\begin{equation}\label{set_of_beta}
(\beta_1,\ldots,\beta_{2N}):=(\beta_1,\ldots, \beta_N;i\,\alpha_{1\xi},\ldots,i\,\alpha_{N\xi})
\end{equation}
i.e. both sets express through $2N$ independent real functional parameters $(\alpha_1,\ldots, \alpha_N)$ and $(\beta_1,\ldots, \beta_N)$.

General determinant formula (\ref{Solution_general_formula_FP}) with matrix $A$ (\ref{matrix1_FP}) corresponding to the kernel $R_0$ (\ref{kernel_satify_potent_cond&potential}) of the $\overline{\partial}$-problem (\ref{dibar_problem}) gives the class of exact solutions $u$ with constant asymptotic value $-\epsilon$ at infinity of hyperbolic version of the NVN equation (\ref{NVN}). By construction these solutions depend on $2N$ real functional parameters $(\alpha_1,\ldots,\alpha_N)$ and $(\beta_1,\ldots, \beta_N)$ given by (\ref{set_of_alpha}),(\ref{set_of_beta}). In the simplest case $N=1$ $(\alpha_1,\alpha_{2}):=(\alpha_1,\frac{i}{\epsilon}{\beta}_{1\eta}),\quad
(\beta_1,\beta_{2}):=(\beta_1,i\,{\alpha}_{1\xi})$
the determinant of $A$ due to (\ref{matrix1_FP}) is given by expression
\begin{eqnarray}\label{determinant_FP_NVN1}
&\det{A}=\big(1+\frac{1}{2}\partial_{\xi}^{-1}\alpha_1\beta_1\big)
\big(1-\frac{1}{2\epsilon}\partial_{\xi}^{-1}\alpha_{1\xi}\beta_{1\eta}\big)+
\frac{1}{8\epsilon}\alpha_1^2\partial_{\xi}^{-1}\beta_{1}\beta_{1\eta}=\nonumber\\
&=\Big(1+\frac{1}{2}\partial_{\xi}^{-1}
\alpha_1\beta_{1}-\frac{\alpha_{1}\beta_{1\eta}}{4\epsilon}\Big)^2
=\Delta^2.
\end{eqnarray}
The corresponding solution $u$ due to (\ref{Solution_general_formula_FP}) and (\ref{determinant_FP_NVN1}) has the form:
\begin{equation}\label{Solution_FP_NVN1}
\fl u(\xi,\eta,t)=-\epsilon+\frac{1}{2\Delta}(\alpha_{1\eta}\beta_{1}-
\frac{1}{\epsilon}\alpha_{1\xi}\beta_{1\eta\eta})-\frac{1}{8\Delta^2\epsilon}
(\alpha_{1}\beta_{1}-\frac{1}{\epsilon}\alpha_{1\xi}\beta_{1\eta})
(\alpha_{1\eta}\beta_{1\eta}-\alpha_1\beta_{1\eta\eta}).
\end{equation}

For the delta-functional kernel $R_0$ (\ref{kernel_satify_potent_cond}) of the type (\ref{kernel_satify_potent_cond&potential}) with
\begin{equation}\label{p&q_soliton_NVN1}
\fl p_n(\mu,\overline{\mu})=\delta(\mu-i\mu_{n0}),
\quad q_n(\lambda,\overline{\lambda})=a_n\lambda_{n0}\delta(\lambda-i\lambda_{n0}),
\quad n=1,\ldots,N
\end{equation}
the general determinant formula (\ref{Solution_general_formula_FP}) leads to corresponding exact multisoliton solutions. In the simplest case of $N=1$ from (\ref{alpha&beta_FP}) one obtains the functional parameters $\alpha_{1}=-2ie^{F(i\mu_{10})},\quad \beta_{1}=-2ia_1\lambda_{10}e^{-F(i\lambda_{10})}$
and from (\ref{Solution_FP_NVN1}), under the condition $\frac{a_1(\lambda_{10}+\mu_{10})}{\lambda_{10}-\mu_{10}}=-e^{\varphi_0}<0$, the exact nonsingular line soliton solution of the hyperbolic NVN equation:
\begin{equation}\label{one-soliton N a1 u}
    u(\xi,\eta,t) = -\epsilon - \frac{\epsilon(\lambda_{10}-\mu_{10})^2}{2\lambda_{10}\mu_{10}}
    \frac{1}{\cosh^2{\frac{\varphi(\xi,\eta,t)+\varphi_0}{2}}}
\end{equation}
where the phase $\varphi$ has the form
\begin{eqnarray}\label{one-soliton N a phi}
    \varphi(\xi,\eta,t):= F(i\mu_{10})-F(i\lambda_{10})=\nonumber \\
\fl = (\lambda_{10}-\mu_{10})\xi+\Big(\frac{\epsilon}{\lambda_{10}}-
    \frac{\epsilon}{\mu_{10}}\Big)\eta-\kappa_1\big(\lambda_{10}^3-\mu_{10}^3\big)t-
    \kappa_2\Big(\frac{\epsilon^3}{\lambda_{10}^3}-\frac{\epsilon^3}{\mu_{10}^3}\Big)t.
\end{eqnarray}

For the case $8.B$ taking into account  (\ref{caseB_p&q}) and identifying expressions for $R_0$ given by (\ref{Kernel_VN_FP}) and (\ref{kernel_satify_potent_cond}) one obtains
\begin{eqnarray}\label{kernel_satify_potent_cond&real2}
R_{0}(\mu,\overline{\mu},\lambda,\overline{\lambda})=\pi\sum\limits_{n=1}^{N}f_n(\mu,\overline{\mu})
g_n(\lambda,\overline{\lambda})=\nonumber \\
=\pi\sum\limits_{n=1}^{N}\Big(c_n p_n(\mu,\overline{\mu})\lambda \overline{p_n(\overline{\lambda},\lambda)}-c_n
\overline{p_n(-\overline{\mu},-\mu)}\lambda{p_n(-\lambda,-\overline{\lambda})}\Big).
\end{eqnarray}
From (\ref{kernel_satify_potent_cond&real2}) one can choose  the following convenient sets $f$, $g$ of functions $f_n$, $g_n$:
\begin{equation}\label{set_of_f2}
\fl f:=(f_1,\ldots,f_{2N})=(p_1(\mu,\overline{\mu}),\ldots, p_N(\mu,\overline{\mu});\overline{p_1(-\bar
{\mu},-\mu)},\ldots,
\overline{p_N(-\overline{\mu},-\mu)}),
\end{equation}
\begin{eqnarray}\label{set_of_g2}
 g:=(g_1,\ldots,g_{2N})=\nonumber \\
 \fl (c_1\lambda \overline{p_1(\overline{\lambda},\lambda)},\ldots, c_N\lambda \overline{p_N(\overline{\lambda},\lambda)};-c_1\lambda {p_1(-\lambda,-\overline{\lambda})},\ldots,-c_N\lambda
{p_N(-\lambda,-\overline{\lambda})}).
\end{eqnarray}
Due to the definitions (\ref{alpha_FP}), (\ref{beta_FP}) and (\ref{set_of_f2}), (\ref{set_of_g2}) one derives the interrelations between different functional parameters:
\begin{equation}\label{relations_between_alpha_n&beta_n_NVN1_FP}
\fl \alpha_n: = \int\int\limits_C p_n(\mu,\overline{\mu})e^{F(\mu)}d\mu\wedge d\overline{\mu},
\beta_{n}: = \int\int \limits_C c_n\lambda \overline{p_n(\overline{\lambda},\lambda)}e^{-F(\lambda)}d\lambda \wedge d\overline{\lambda}=ic_n\overline{\alpha}_{n\xi},
\end{equation}
\begin{equation}\label{relations_between_alpha_Nn_NVN1_FP}
\alpha_{N+n}=\int\int\limits_C \overline{p_{n}(-\overline{\mu},-\mu)}e^{F(\mu)}d\mu\wedge d\overline{\mu}=
\overline{\alpha_{n}},
\end{equation}
\begin{equation}\label{relations_between_beta_Nn_NVN1_FP}
\fl \beta_{N+n}=-\int\int\limits_C c_n\lambda p_{n}(-\lambda,-\overline{\lambda})e^{-F(\lambda)}d\lambda\wedge d\overline{\lambda}=
\overline{\beta}_n, \quad (n=1,\cdots, N).
\end{equation}

So due to (\ref{relations_between_alpha_n&beta_n_NVN1_FP}) and (\ref{relations_between_alpha_Nn_NVN1_FP}), (\ref{relations_between_beta_Nn_NVN1_FP}) the sets $\alpha$, $\beta$ of functional parameters
\begin{equation}\label{alpha_sets_NVN1}
\alpha:=(\alpha_1,\alpha_2,\ldots,\alpha_{2N})=(\alpha_1,\ldots,\alpha_N;
\overline{\alpha_1},\ldots,\overline{\alpha_N}),
\end{equation}
\begin{equation}\label{beta_sets_NVN1}
\fl \beta:=(\beta_1,\beta_2,\ldots,\beta_{2N})=
(ic_1\overline{\alpha}_{1\xi},
\ldots,ic_N\overline{\alpha}_{N\xi};
-i c_1\alpha_{1\xi},\ldots,ic_N{\alpha}_{N\xi})
\end{equation}
express through the $N$ independent complex parameters $(\alpha_1,\ldots,\alpha_N)$.

General determinant formula (\ref{Solution_general_formula_FP}) with matrix $A$ given by (\ref{matrix1_FP}) with kernel $R_0$ (\ref{set_of_g2}) of the $\overline{\partial}$-problem (\ref{dibar_problem}) gives another class of exact solutions with constant asymptotic value at infinity of the hyperbolic version of the NVN equation (\ref{NVN}). By construction these solutions depend on $N$ independent complex parameters ($\alpha_1,\ldots,\alpha_N$) given by (\ref{alpha_sets_NVN1}), (\ref{beta_sets_NVN1}). In the simplest case $N=1$ $(\alpha_1,\alpha_{2}):=(\alpha_1,\overline{\alpha}_1),\quad
(\beta_1,\beta_{2}):=(ic_1\overline{\alpha}_{1\xi},-ic_1\alpha_{1\xi})$
the determinant of $A$ due to (\ref{matrix1_FP}) is given by expression
\begin{eqnarray}\label{determinant_FP_NVN2}
\det{A}=(1+\frac{ic_1}{2}\partial_{\xi}^{-1}\alpha_1\overline{\alpha}_{1\xi})
(1-\frac{ic_1}{2}\partial_{\xi}^{-1}\alpha_{1\xi} \overline{\alpha_1})-
\frac{c_1^2|\alpha_1|^4}{16}&=&\\
=(1+\frac{ic_1}{2}\partial_{\xi}^{-1}
(\alpha_1\overline{\alpha}_{1\xi}-\alpha_{1\xi} \overline{\alpha}_1))^2=\Delta^2.\nonumber
\end{eqnarray}
The corresponding solution $u$ due to (\ref{Solution_general_formula_FP}) and (\ref{determinant_FP_NVN2}) has the form:
\begin{equation}\label{Solution_FP_NVN2}
\fl u(\xi,\eta,t)=-\epsilon+\frac{ic_1}{2\Delta}(\alpha_{1\eta}\overline{\alpha}_{1\xi}-
\overline{\alpha}_{1\eta}\alpha_{1\xi})+
\frac{c_1^2}{8\Delta^2}(\alpha_{1}\overline{\alpha}_{1\xi}-\overline{\alpha}_{1}\alpha_{1\xi})
(\alpha_{1\eta}\overline{\alpha}_{1}-
\overline{\alpha}_{1\eta}\alpha_{1}).
\end{equation}

For the delta-functional kernel of the type (\ref{kernel_satify_potent_cond&real2}) with
\begin{equation}\label{p&q_soliton_NVN2}
p_n(\mu,\overline{\mu})=\delta(\mu-i\overline{\lambda}_n), \quad n=1,\ldots,N
\end{equation}
general determinant formula (\ref{Solution_general_formula_FP}) taking into account (\ref{set_of_f2})-(\ref{beta_sets_NVN1}) leads to corresponding exact multi line soliton solutions. In the simplest case of $N=1$ from (\ref{relations_between_alpha_n&beta_n_NVN1_FP})
one obtains the functional parameter $\alpha_{1}=-2ie^{F({\lambda}_1)}$
and due to (\ref{Solution_FP_NVN2})  corresponding exact solution $u$, under the condition $\frac{c_1 \lambda_R}{\lambda_I}=e^{\varphi_0}>0$, is the one line nonsingular soliton:
\begin{equation}\label{one-soliton NVN2 a1 u}
\fl u(\xi,\eta,t) =-\epsilon+\frac{8\epsilon c_1\lambda_{R}\lambda_{I}e^{\varphi(\xi,\eta,t)}}
{|\lambda|^2(1+\frac{c_1 \lambda_R}{\lambda_I}e^{\varphi(\xi,\eta,t)})^2}= -\epsilon+\frac{2\epsilon \lambda_{I}^2}{|\lambda|^2}\frac{1}
{\cosh^2\frac{\varphi(\xi,\eta,t)+\varphi_0}{2}}
\end{equation}
where the phase $\varphi$ has the form
\begin{equation}\label{one-soliton NVN2 a phi}
\fl \varphi(\xi,\eta,t) = i\Big[(\overline{\lambda}-\lambda)\xi-\Big(\frac{\epsilon}{\overline{\lambda}}-
\frac{\epsilon}{\lambda}\Big)\eta+\kappa_1\big(\overline{\lambda}^3-\lambda^3\big)t-
\kappa_2\Big(\frac{\epsilon^3}{\overline{\lambda}^3}-\frac{\epsilon^3}{\lambda^3}\Big)t\Big].
\end{equation}

\section{Exact solutions with functional parameters of  NVN-I equation}
\label{Section_9}
\setcounter{equation}{0}

Let us consider also the case $\sigma^2=-1$ of complex space variables $\xi=z=x+iy$, $\eta=\bar
{z}=x-iy$ or elliptic version of the NVN equation (\ref{NVN}). To the condition (\ref{real_condition_VN}) of reality $u=\overline{u}$ one can satisfy imposing on each pair of terms in the sum (\ref{kernel_satify_potent_cond}) the following restriction:
\begin{eqnarray}\label{kernel_reality_cond_VN}
\fl p_n(\mu,\overline{\mu})q_n(\lambda,\overline{\lambda})+\frac{1}{\mu}q_n(-\mu,-\overline{\mu}){\lambda} p_n(-\lambda,-\bar
{\lambda})=\nonumber \\
\fl =\frac{\epsilon^3}{|\lambda|^2|\mu|^2\overline{\lambda}\overline{\mu}}
\overline{p_n\Big(-\frac{\epsilon}{\overline{\lambda}},-\frac{\epsilon}{{\lambda}}\Big)}
\,\overline{q_n\Big(-\frac{\epsilon}{\overline{\mu}},-\frac{\epsilon}{{\mu}}\Big)}+
\frac{\epsilon^3}{|\lambda|^2|\mu|^2\overline{\lambda}\overline{\mu}}
\lambda\overline{q_n\Big(\frac{\epsilon}{\overline{\lambda}},\frac{\epsilon}{{\lambda}}\Big)}\frac{1}{\mu} \overline{p_n\Big(\frac{\epsilon}{\overline{\mu}},\frac{\epsilon}{{\mu}}\Big)}.
\end{eqnarray}
Due to (\ref{kernel_reality_cond_VN}) two cases are possible
\begin{equation}\label{kernel_real_case_A_VN1}
\fl 9.A.\quad p_n(\mu,\overline{\mu})q_n(\lambda,\overline{\lambda})=
\frac{\epsilon^3}{|\lambda|^2|\mu|^2\overline{\lambda}\overline{\mu}}
\overline{p_n\Big(-\frac{\epsilon}{\overline{\lambda}},-\frac{\epsilon}{{\lambda}}\Big)}
\,\overline{q_n\Big(-\frac{\epsilon}{\overline{\mu}},-\frac{\epsilon}{{\mu}}\Big)},
\end{equation}
\begin{equation}\label{kernel_real_case_B_VN2}
\fl 9.B.\quad p_n(\mu,\overline{\mu})q_n(\lambda,\overline{\lambda})=
\frac{\epsilon^3\lambda}{|\lambda|^2|\mu|^4\overline{\lambda}}
\overline{q_n\Big(\frac{\epsilon}{\overline{\lambda}},\frac{\epsilon}{{\lambda}}\Big)}\, \overline{p_n\Big(\frac{\epsilon}{\overline{\mu}},\frac{\epsilon}{{\mu}}\Big)}.
\end{equation}
In the case $9.A$ separating in (\ref{kernel_real_case_A_VN1}) the variables
\begin{equation}\label{caseA_separate_VN1}
\fl \frac{p_n(\mu,\overline{\mu})|\mu|^2\overline{\mu}}{\overline{q_n(-\frac{\epsilon}{\overline{\mu}},
-\frac{\epsilon}{{\mu}})}}=
\frac{\epsilon^3}{|\lambda|^2\overline{\lambda}}
\frac{\overline{p_n(-\frac{\epsilon}{\overline{\lambda}},-\frac{\epsilon}{{\lambda}})}}
{q_n(\lambda,\overline{\lambda})}=c_n^{-1}
\end{equation}
one obtains the following relations on the functions $q_n$
 and $p_n$:
\begin{equation}\label{caseA_p&q_VN1a}
\fl p_n(\mu,\overline{\mu})=\frac{1}{c_n|\mu|^2\overline{\mu}}
\overline{q_n\Big(-\frac{\epsilon}{\overline{\mu}},-\frac{\epsilon}{{\mu}}\Big)}, \quad q_n(\lambda,\overline{\lambda})=\frac{\epsilon^3 c_n}{|\lambda|^2\overline{\lambda}}
\overline{p_n\Big(-\frac{\epsilon}{\overline{\lambda}},-\frac{\epsilon}{{\lambda}}\Big)}.
\end{equation}
Comparing two relations in (\ref{caseA_p&q_VN1a}) one concludes that constant $c_n$ are pure imaginary: $c_n=i\,a_n$. In applying general determinant formula (\ref{Solution_general_formula_FP}) for exact solutions $u$ one must to identify the corresponding expressions (\ref{Kernel_VN_FP}) and (\ref{kernel_satify_potent_cond}) for the kernel $R_0$, due to relations (\ref{caseA_p&q_VN1a}) one obtains
\begin{eqnarray}\label{kernel_satify_potent_cond&real_VN1}
\fl R_{0}(\mu,\overline{\mu},\lambda,\overline{\lambda})=\pi\sum\limits_{n=1}^{N}f_n(\mu,\overline{\mu})
g_n(\lambda,\overline{\lambda})=\nonumber \\
\fl=\pi\sum\limits_{n=1}^{N}\Big(p_n(\mu,\overline{\mu})\frac{i\,a_n \epsilon^3}{|\lambda|^2\overline{\lambda}}
\overline{p_n\Big(-\frac{\epsilon}{\overline{\lambda}},-\frac{\epsilon}{{\lambda}}\Big)}-
\frac{i\,a_n \epsilon^3}{|\mu|^4}
\overline{p_n\Big(\frac{\epsilon}{\overline{\mu}},\frac{\epsilon}{{\mu}}\Big)}\lambda {p_n(-\lambda,-\overline{\lambda})}\Big).
\end{eqnarray}
From (\ref{kernel_satify_potent_cond&real_VN1}) one can choose  the following convenient sets $f$, $g$ of functions $f_n$, $g_n$:
\begin{equation}\label{set_of_f_VN1}
\fl f:=(f_1,\ldots,f_{2N})=\Big(p_1(\mu,\overline{\mu}),\ldots, p_N(\mu,\overline{\mu});\frac{\epsilon^3}{|\mu|^4}
\overline{p_1\Big(\frac{\epsilon}{\overline{\mu}},\frac{\epsilon}{{\mu}}\Big)},\ldots,\frac{\epsilon^3}{|\mu|^4}
\overline{p_N\Big(\frac{\epsilon}{\overline{\mu}},\frac{\epsilon}{{\mu}}\Big)}\Big),
\end{equation}
\begin{eqnarray}\label{set_of_g_VN1}
\fl g:=(g_1,\ldots,g_{2N})=\Big(i\frac{\epsilon^3 a_1}{|\lambda|^2\overline{\lambda}}\overline{p_1\Big(-\frac{\epsilon}{\overline{\lambda}},
-\frac{\epsilon}{{\lambda}}\Big)},
\ldots, i\frac{\epsilon^3 a_N}{|\lambda|^2\overline{\lambda}}\overline{p_N\Big(-\frac{\epsilon}{\overline{\lambda}},
-\frac{\epsilon}{{\lambda}}\Big)};\nonumber \\
-ia_1\lambda {p_1(-\lambda,-\overline{\lambda})},\ldots,-ia_N\lambda {p_N(-\lambda,-\overline{\lambda})}\Big).
\end{eqnarray}

Due to definitions (\ref{alpha_FP}), (\ref{beta_FP}) and (\ref{set_of_f_VN1}), (\ref{set_of_g_VN1}) taking into account (\ref{caseA_p&q_VN1a}) one can derive the interrelations between different functional parameters:
\begin{equation}\label{alpha_FP_VN1}
\alpha_n: = \int\int\limits_C p_n(\mu,\overline{\mu})e^{F(\mu)}d\mu\wedge d\overline{\mu},
\end{equation}
\begin{eqnarray}\label{beta_FP_VN1}
\beta_n: = i\,a_n\epsilon^3\int\int \limits_C \frac{1}{|\lambda|^2\overline{\lambda}} \overline{p_n(-\frac{\epsilon}{\overline{\lambda}},-\frac{\epsilon}{{\lambda}})}e^{-F(\lambda)}
d\lambda \wedge d\overline{\lambda}=\\
=-a_n\epsilon\partial_{z}\int\int \limits_C \overline{p_n(\lambda,\overline{\lambda})}e^{\overline{F(\lambda)}}
\overline{d\lambda \wedge d\overline{\lambda}}=-\epsilon a_n \overline{\alpha}_{nz},\nonumber
\end{eqnarray}
\begin{equation}\label{a2&b1_FP_VN1a}
\fl \alpha_{N+n}: = \int\int\limits_C \frac{\epsilon^3}{|\mu|^4}\overline{p_n(\frac{\epsilon}{\overline{\mu}},\frac{\epsilon}{{\mu}})}
e^{F(\mu)}d\mu\wedge d\overline{\mu}=\epsilon\int\int\limits_C \overline{p_n(\mu,\overline{\mu})}
e^{F(\frac{\epsilon}{\overline{\mu}})}\overline{d\mu\wedge d\overline{\mu}}=\epsilon\overline{\alpha}_{n},
\end{equation}
\begin{equation}\label{a1&b2_FP_VN1a}
\fl \beta_{N+n}: = -i\int\int \limits_C \lambda a_n p_n(-\lambda,-\overline{\lambda})e^{-F(\lambda)}d\lambda \wedge d\overline{\lambda}=a_n\alpha_{nz}, \qquad (n=1,\ldots,N).
\end{equation}
So due to (\ref{alpha_FP_VN1})-(\ref{a1&b2_FP_VN1a}) the sets of functional parameters
\begin{equation}\label{alpha_sets_VN1}
\fl (\alpha_1,\ldots,\alpha_{2N}):=(\alpha_1,\ldots,\alpha_N,
\epsilon\overline{\alpha_1},\ldots,\epsilon\overline{\alpha_N})
\end{equation}
\begin{equation}\label{beta_sets_VN1}
\fl (\beta_1,\beta_2,\ldots,\beta_{2N})=
(-\epsilon a_1\overline{\alpha}_{1z},\ldots,-\epsilon a_N\overline{\alpha}_{Nz};a_1\alpha_{1z},\ldots,a_N\alpha_{Nz}).
\end{equation}
are express through $N$ independent complex functional parameters $(\alpha_1,\ldots,\alpha_N)$.

General determinant formula (\ref{Solution_general_formula_FP}) with matrix $A$ (\ref{matrix1_FP}) corresponding to the kernel $R_0$ (\ref{kernel_satify_potent_cond&real_VN1}) of the $\overline{\partial}$-problem (\ref{dibar_problem}) gives the class of exact solutions $u$ with constant asymptotic value $-\epsilon$ at infinity of the elliptic version of the NVN equation (\ref{NVN}).
By construction these solutions depends on $N$ complex functional parameters $\alpha_1,\ldots,\alpha_N$. In the simplest case $N=1$ $(\alpha_1,\alpha_{2}):=(\alpha_1,\epsilon\overline{\alpha}_{1}),\quad
(\beta_1,\beta_{2}):=(-\epsilon a_1\overline{\alpha}_{1z}, a_1{\alpha}_{1z})$
and  due to (\ref{matrix1_FP})  the determinant of $A$ is given by expression:
\begin{eqnarray}\label{determinant_FP_VN1}
\fl \det{A}=(1-\frac{a_1\epsilon}{2}\partial_{z}^{-1}(\alpha_1\overline{\alpha}_{1z}))
(1+\frac{a_1\epsilon}{2}\partial_{z}^{-1}(\overline{\alpha}_1 \alpha_{1z}))+
\frac{a_1^2\epsilon^2}{16}|\alpha_1|^4=\nonumber \\
=(1-\frac{a_1\epsilon}{2}\partial_{z}^{-1}
(\alpha_1\overline{\alpha}_{1z})+
\frac{a_1\epsilon}{4}|\alpha_1|^2)^2=\Delta^2.
\end{eqnarray}
The corresponding solution $u$ due to (\ref{Solution_general_formula_FP}) and (\ref{determinant_FP_VN1}) has the form:
\begin{equation}\label{Solution_FP_VN1}
\fl u(z,\bar{z},t)=-\epsilon+\frac{a_1\epsilon}{2\Delta}(|\alpha_{1z}|^2-|\alpha_{1\overline{z}}|^2)-
\frac{a_1^2 \epsilon^2}{8\Delta^2}
|\alpha_{1}\overline{\alpha}_{1\overline{z}}-\overline{\alpha}_{1}\alpha_{1\overline{z}}|^2.
\end{equation}

For the delta-functional kernel of the type (\ref{kernel_satify_potent_cond&real_VN1}) with
\begin{equation}\label{p&q_soliton_VN1}
p_n(\mu,\overline{\mu})=\delta(\mu-\mu_n), \quad n=1,\ldots,N
\end{equation}
and $\lambda_n\overline{\mu}_n=\mu_n\overline{\lambda}_n=-\epsilon$,
general determinant formula (\ref{Solution_general_formula_FP}) taking into account (\ref{set_of_f_VN1})-(\ref{beta_sets_VN1}) leads to corresponding exact multi line soliton solutions.
In the simplest case of $N=1$ from (\ref{alpha_FP_VN1}) one obtains the functional parameter $\alpha_{1}=-2ie^{F({\mu}_1)}$
and due to (\ref{Solution_FP_VN1})
corresponding exact solution $u$, under the condition $\epsilon a_1\frac{\mu_1+\lambda_1}{\lambda_1-\mu_1}=-e^{\varphi_0}<0$, is the nonsingular one line soliton:
\begin{equation}\label{one-soliton VN2 a1 u}
u(z,\overline{z},t) = -\epsilon + \frac{|\lambda_1-\mu_1|^2}{2}\frac{1}{\cosh^2{\frac{\varphi(z,\overline{z},t)+\varphi_0}{2}}}
\end{equation}
where the phase $\varphi$ has the form
\begin{equation}\label{one-soliton VN2 a phi}
\fl \varphi(z,\overline{z},t) = i[(\mu_1-\lambda_1)z-(\overline{\mu}_1-\overline{\lambda}_1)\overline{z}+
    \kappa(\mu_{1}^{3}-\lambda_{1}^{3})t-\overline{\kappa}(\overline{\mu}_{1}^{3}
    -\overline{\lambda}_{1}^{3})t].
\end{equation}

In the case $9.B$ separating in (\ref{kernel_real_case_B_VN2}) the variables
\begin{equation}\label{caseB_separate_VN2}
\frac{p_n(\mu,\overline{\mu})|\mu|^4}{\epsilon^2\overline{p_n(\frac{\epsilon}{\overline{\mu}})}}=
\frac{\epsilon}{\overline{\lambda}^2}
\frac{\overline{q_n(\frac{\epsilon}{\overline{\lambda}},\frac{\epsilon}{\lambda})}}
{q_n(\lambda,\overline{\lambda})}=c_n
\end{equation}
one obtains the following relations on the functions $q_n(\lambda,\overline{\lambda})$
and $p_n(\mu,\overline{\mu})$:
\begin{equation}\label{caseB_p&q_VN2a}
p_n(\mu,\overline{\mu})=c_n\frac{\epsilon^2}{|\mu|^4}
\overline{p_n\Big(\frac{\epsilon}{\overline{\mu}},\frac{\epsilon}{{\mu}}\Big)},
\quad
q_n(\lambda,\overline{\lambda})=\frac{\epsilon}{c_n\overline{\lambda}^2}
\overline{q_n\Big(\frac{\epsilon}{\overline{\lambda}},\frac{\epsilon}{{\lambda}}\Big)}.
\end{equation}
The constants $c_n$ in (\ref{caseB_separate_VN2}), (\ref{caseB_p&q_VN2a}) without loss of generality  can be choosen equal to unity.
In applying general determinant formula (\ref{Solution_general_formula_FP}) for exact solutions $u$ one must to identify the corresponding expressions (\ref{Kernel_VN_FP}) and (\ref{kernel_satify_potent_cond}) for the kernel $R_0$ $\overline{\partial}$-problem (\ref{dibar_problem}). In the considered $9.B$ case taking into account (\ref{caseB_p&q_VN2a}) one obtains from (\ref{Kernel_VN_FP}) and (\ref{kernel_satify_potent_cond}):
\begin{eqnarray}\label{kernel_satify_potent_cond&real_VN2}
R_{0}(\mu,\overline{\mu},\lambda,\overline{\lambda})=\pi\sum\limits_{n=1}^{N}f_n(\mu,\overline{\mu})
g_n(\lambda,\overline{\lambda})=\nonumber \\
\fl =\pi\sum\limits_{n=1}^{N}\Big(p_n(\mu,\overline{\mu})
q_n(\lambda,\overline{\lambda})+\frac{\epsilon}{\mu \overline{\mu}^2}
\overline{q_n\Big(-\frac{\epsilon}{\overline{\mu}},-\frac{\epsilon}{{\mu}}\Big)}\lambda \frac{\epsilon^2}{|\lambda|^4}
\overline{p_n\Big(-\frac{\epsilon}{\overline{\lambda}},-\frac{\epsilon}{{\lambda}}\Big)}\Big).
\end{eqnarray}
From (\ref{kernel_satify_potent_cond&real_VN2}) one can choose  the following convenient sets $f$, $g$ of functions $f_n$, $g_n$:
\begin{eqnarray}\label{set_of_f_VN2}
f:=(f_1,\ldots,f_{2N})=\Big(p_1(\mu,\overline{\mu}),\ldots, p_N(\mu,\overline{\mu});\nonumber \\
\frac{\epsilon}{\mu \overline{\mu}^2}
\overline{q_1\Big(-\frac{\epsilon}{\overline{\mu}},-\frac{\epsilon}{{\mu}}\Big)},\ldots,
\frac{\epsilon}{\mu \overline{\mu}^2}\overline{q_N\Big(-\frac{\epsilon}{\overline{\mu}},-\frac{\epsilon}{{\mu}}\Big)}\Big),
\end{eqnarray}
\begin{eqnarray}\label{set_of_g_VN2}
g:=(g_1,\ldots,g_{2N})=\Big(q_1(\lambda,\overline{\lambda}),
\ldots, q_N(\lambda,\overline{\lambda});\nonumber \\
\frac{\epsilon^2}{\overline{\lambda}^2\lambda}
\overline{p_1\Big(-\frac{\epsilon}{\overline{\lambda}},-\frac{\epsilon}{{\lambda}}\Big)},
\ldots,\frac{\epsilon^2}{\overline{\lambda}^2\lambda}
\overline{p_N\Big(-\frac{\epsilon}{\overline{\lambda}},-\frac{\epsilon}{{\lambda}}\Big)}\Big).
\end{eqnarray}

Due to definitions (\ref{alpha_FP}), (\ref{beta_FP}) and (\ref{set_of_f_VN2}), (\ref{set_of_g_VN2}) taking into account (\ref{caseB_p&q_VN2a}) one can derive the interrelations between different functional parameters:
\begin{equation}\label{alpha_FP_VN2}
\fl\alpha_n: = \int\int\limits_C \frac{\epsilon^2}{|\mu|^4}\overline{p_n\Big(\frac{\epsilon}{\overline{\mu}},\frac{\epsilon}{\mu}\Big)}
e^{F(\mu)}d\mu\wedge d\overline{\mu}=\int\int\limits_C \overline{p_n(\mu,\overline{\mu})}
e^{\overline{F(\mu)}}\overline{d\mu\wedge d\overline{\mu}}=\overline{\alpha}_n,
\end{equation}
\begin{equation}\label{beta_FP_VN2}
\fl \beta_n: = \int\int \limits_C  \frac{\epsilon}{\overline{\lambda}^2}
\overline{q_n\Big(\frac{\epsilon}{\overline{\lambda}},\frac{\epsilon}{\lambda}\Big)}e^{-F(\lambda)}
d\lambda \wedge d\overline{\lambda}=\int\int \limits_C  \frac{\epsilon}{\overline{\lambda}^2}
\overline{q_n({\lambda},\overline{\lambda})}e^{-F(\frac{\epsilon}{\overline{\lambda}})}
\overline{d\lambda \wedge d\overline{\lambda}}=-\frac{1}{\epsilon}\overline{\beta}_{nzz},
\end{equation}
\begin{eqnarray}\label{alphaN+n_FP_VN2}
\fl \alpha_{N+n}: = \int\int\limits_C \frac{\epsilon}{\mu \overline{\mu}^2}\overline{q_n\Big(-\frac{\epsilon}{\overline{\mu}},-\frac{\epsilon}{{\mu}}\Big)}
e^{F(\mu)}d\mu\wedge d\overline{\mu}=-\frac{i}{\epsilon}\overline{\beta}_{nz},
\end{eqnarray}
\begin{equation}\label{a1&b2_FP_VN2a}
\fl \beta_{N+n}: = \int\int \limits_C \frac{\epsilon^2}{\overline{\lambda}^2\lambda}
\overline{p_n\Big(-\frac{\epsilon}{\overline{\lambda}},-\frac{\epsilon}{{\lambda}}\Big)}e^{-F(\lambda)}
d\lambda \wedge d\overline{\lambda}=i\overline{\alpha}_{nz},\quad (n=1,\ldots,N).
\end{equation}
From (\ref{beta_FP_VN2}) it follows that
\begin{equation}\label{beta1_FP_VN2}
\fl \beta_{n\overline{z}}=-\overline{\beta}_{nz}=-\overline{\beta_{n\overline{z}}},\quad (n=1,\ldots,N).
\end{equation}
So due to (\ref{alpha_FP_VN2})-(\ref{beta1_FP_VN2}) the sets of functional parameters
\begin{equation}\label{alpha_sets_VN2}
\fl (\alpha_1,\ldots,\alpha_{2N}):=(\alpha_1,\ldots,\alpha_N;
\frac{i}{\epsilon}\beta_{1\overline{z}},\ldots,\frac{i}{\epsilon}\beta_{N\overline{z}})
\end{equation}
\begin{equation}\label{beta_sets_VN2}
\fl (\beta_1,\beta_2,\ldots,\beta_{2N})=
(\beta_1,\ldots,\beta_N;i\,\alpha_{1z},\ldots,i\,\alpha_{Nz}).
\end{equation}
are express through $2N$ independent  functional parameters $(\alpha_1,\ldots,\alpha_N)$ and ($\beta_1,\ldots,\beta_N$) given by (\ref{alpha_FP_VN2})-(\ref{a1&b2_FP_VN2a}).

General determinant formula (\ref{Solution_general_formula_FP}) with matrix $A$ (\ref{matrix1_FP}) corresponding to the kernel $R_0$ (\ref{kernel_satify_potent_cond&real_VN2}) of the $\overline{\partial}$-problem (\ref{dibar_problem}) gives the class of exact solutions $u$ with constant asymptotic value $-\epsilon$ at infinity of the elliptic version of the NVN equation (\ref{NVN}).
By construction these solutions depend in fact due to (\ref{beta1_FP_VN2}) on $N$ real functional parameters $\alpha_1,\ldots,\alpha_N$ and $N$ real functional parameters  $i\beta_{1\overline{z}},\ldots,i\beta_{N\overline{z}}$. In the simplest case $N=1$ $(\alpha_1,\alpha_{2}):=(\alpha_1,\frac{i}{\epsilon}\beta_{1\overline{z}}),\quad
(\beta_1,\beta_{2}):=(\beta_1, i\,\alpha_{1z})$
the determinant of $A$ due to (\ref{matrix1_FP}) and (\ref{alpha_sets_VN2}), (\ref{beta_sets_VN2}) is given by expression
\begin{eqnarray}\label{determinant_FP_VN2}
\det{A}=(1+\frac{1}{2}\partial_{z}^{-1}\alpha_1\beta_1)
(1-\frac{1}{2\epsilon}\partial_{z}^{-1}{\alpha}_{1z} \beta_{1\overline{z}})+
\frac{\alpha_1^2\beta_{1\overline{z}}^2}{16\epsilon^2}=\nonumber \\
=(1+\frac{1}{2}\partial_{z}^{-1}(\alpha_1\beta_1)
-\frac{1}{4\epsilon}\alpha_1\beta_{1\overline{z}})^2=\Delta^2.
\end{eqnarray}
Using identity $\partial_{z}^{-1}(\alpha_1\beta_1)-
\partial_{\overline{z}}^{-1}(\overline{\alpha_1}\overline{\beta_1})=\frac{1}{\epsilon}\alpha_1\beta_{1\overline{z}}$
(which is valid due to the relations (\ref{alpha_FP_VN2})-(\ref{beta1_FP_VN2})) one obtains explicitly real expression for $\det{A}$:
\begin{equation}\label{determinant1_FP_VN2}
\det{A}
=(1+\frac{1}{4}\partial_{z}^{-1}(\alpha_1\beta_1)
+\frac{1}{4}\partial_{\overline{z}}^{-1}(\overline{\alpha_1}\overline{\beta_1}))^2=\Delta^2.
\end{equation}
Using (\ref{determinant1_FP_VN2}) one calculates by (\ref{Solution_general_formula_FP}) the corresponding
exact solution
\begin{equation}\label{Solution_FP_VN2}
\fl u=-\epsilon+\frac{1}{2\Delta}\Big((\alpha_1\beta_1)_{\overline{z}}+(\overline{\alpha_1}
\overline{\beta_1})_{z}\Big)-
\frac{1}{8\Delta^2}\Big(\alpha_1\beta_1+
\frac{1}{\epsilon}\overline{\alpha}_{1z}
\overline{\beta}_{1z}\Big)\Big(\overline{\alpha_1}
\overline{\beta_1}+
\frac{1}{\epsilon}\alpha_{1\overline{z}}\beta_{1\overline{z}}\Big).
\end{equation}

For the delta-functional kernel of the type (\ref{kernel_satify_potent_cond&real_VN2}) with
\begin{equation}\label{p&q_soliton_VN2}
\fl p_n(\mu,\overline{\mu})=i\delta(\mu-\mu_n), \quad q_n(\lambda,\overline{\lambda})=-ia_n\lambda_n\delta(\lambda-\lambda_n),
(n=1,\ldots,N)
\end{equation}
with $|\mu_n|^2=|\lambda_n|^2=\epsilon$ and real constants $a_n=\overline{a_n}$
general determinant formula (\ref{Solution_general_formula_FP}) taking into account (\ref{set_of_f_VN2})-(\ref{beta_sets_VN2}) leads to corresponding exact multi line soliton solutions. In the simplest case of $N=1$ from (\ref{alpha_FP_VN2})-(\ref{beta_FP_VN2}) one obtains the functional parameters $\alpha_{1}=2e^{F(\mu_1)}$, $\beta_{1}=-2 a_1 \lambda_1 e^{-F(\lambda_1)}$
and due to (\ref{Solution_FP_VN2})
corresponding exact solution $u$, under the condition $i\,a\frac{\mu_1+\lambda_1}{\mu_1-\lambda_1}=-e^{\varphi_0}<0$, is nonsingular line soliton:
\begin{equation}\label{one-soliton VN2 a1 u}
\fl u(z,\overline{z},t) =-\epsilon+\epsilon\frac{2\sin^{2}(\frac{\delta}{2})}{\cosh^{2}(\frac{\varphi(z,\overline{z},t)+\varphi_{0}}{2})}
\end{equation}
where  $\delta=\arg{\mu_1}-\arg{\lambda_1}$ and the phase $\varphi$ has the form
\begin{equation}\label{one-soliton VN2 a phi}
\fl \varphi(z,\overline{z},t) = i[(\mu_1-\lambda_1)z-(\overline{\mu}_1-\overline{\lambda}_1)\overline{z}+
\kappa(\mu_{1}^{3}-\lambda_{1}^{3})t-\overline{\kappa}(\overline{\mu}_{1}^{3}
-\overline{\lambda}_{1}^{3})t].
\end{equation}

\section{Conclusions and Acknowledgments}
The powerful $\bar{\partial}$-dressing method of Zakharov and Manakov, discovered a quarter of century ago, continues to develop and successfully apply  for construction of exact solutions of multidimensional integrable nonlinear equations. The realization of the method goes due to basic idea of IST through the careful study of auxiliary linear problems by the methods of modern theory of functions of complex variables.  Following  this way one constructs exact complex wave functions (with rich analytical structure) of linear auxiliary problems and by using  the wave functions, via reconstruction formulas, exact (or solvable) potentials - exact solutions of integrable nonlinear equations.

Constructed in the paper exact solutions of hyperbolic and elliptic versions of NVN equation (\ref{NVN}) as exact potentials for one-dimensional perturbed telegraph (or perturbed string) and 2D stationary Schr\"{o}dinger equations (\ref{The_first_aux_probl}) respectively together with calculated exact wave functions may find an applications in modern differential geometry of surfaces and in solid state physics of planar nanostructures. Interesting problem of quantum mechanics of particle in the field of multi line soliton potentials  will be discussed elsewhere.

This work was supported by: 1. scientific Grant for fundamental researches of Novosibirsk State Technical University  (2009); 2. by the Grant  of Ministry of Science and Education of Russia Federation (registration number 2.1.1/1958) via analytical departmental special programm "Development of potential of High School (2009-2010)"; 3. by the international RFFI and Italy Grant for scientific research  (2009).
\setcounter{equation}{0}

%
%
%
{\bf {References}}


\end{document}